\documentclass[12pt]{article}
\usepackage{amsfonts,amssymb,epsfig,amsmath,mathtools}
\usepackage{color}

\usepackage{mathrsfs}
\usepackage{graphicx}
\usepackage{subcaption}
\renewcommand{\baselinestretch}{1.2}

\begin{document}

\makeatletter \@addtoreset{equation}{section} \makeatother
\renewcommand{\theequation}{\thesection.\arabic{equation}}
\renewcommand{\thefootnote}{\alph{footnote}}

\begin{titlepage}

\begin{center}
\hfill {\tt KIAS-P23007}\\
\hfill {\tt SNUTP23-001}\\

\vspace{2cm}

{\Large\bf Towards quantum black hole microstates}

\vspace{2cm}

\renewcommand{\thefootnote}{\alph{footnote}}

{\large Sunjin Choi$^1$, Seok Kim$^2$, Eunwoo Lee$^2$, Siyul Lee$^3$ and
Jaemo Park$^4$}

\vspace{0.7cm}

\textit{$^1$School of Physics, Korea Institute for Advanced Study, Seoul 02455, Korea.}\\

\vspace{0.2cm}

\textit{$^2$Department of Physics and Astronomy \& Center for
Theoretical Physics,\\
Seoul National University, Seoul 08826, Korea.}\\

\vspace{0.2cm}

\textit{$^3$Leinweber Center for Theoretical Physics, University of Michigan,\\ 
Ann Arbor, MI 48109, USA.}\\

\vspace{0.2cm}

\textit{$^4$ Department of Physics, Pohang University of Science and Technology (POSTECH)\\
Pohang 37673, Korea.}\\

\vspace{0.7cm}

E-mails: {\tt sunjinchoi@kias.re.kr, seokkimseok@gmail.com, eunwoo42@snu.ac.kr\\
siyullee@umich.edu, jaemo@postech.ac.kr
}

\end{center}

\vspace{1cm}

\begin{abstract}

We study the cohomology of local BPS operators in $\mathcal{N}=4$ Yang-Mills
theory. The finite $N$ cohomologies consist of the graviton part (subject to the
stringy exclusion principle) and the rest which may describe black hole
microstates in quantum AdS/CFT. We construct an infinite tower of non-graviton
cohomologies in the $SU(2)$ theory and study to what extent they simulate 
quantum black holes. We find signals for partial no-hair behaviors by showing that  
certain gravitons are forbidden to dress these cohomologies. This is in qualitative 
agreement with the perturbative hairs allowed around black holes, which
also leads us to a natural setup to construct hairy BPS black holes.
The cohomologies are simpler to study in the BMN matrix model truncation of the 
classical field theory. 

\end{abstract}

\end{titlepage}

\renewcommand{\thefootnote}{\arabic{footnote}}

\setcounter{footnote}{0}

\renewcommand{\baselinestretch}{1}

\tableofcontents

\renewcommand{\baselinestretch}{1.2}

\section{Introduction}

Black holes exhibit extreme properties with the information they carry.  For instance, 
the Bekenstein-Hawking entropy \cite{Bekenstein:1973ur,Hawking:1975vcx} of black holes is
supposed to represent the maximal amount of information allowed per size \cite{Bekenstein:1980jp}. So regarding this entropy as the most fine-grained information 
of Nature, we can learn from it the fundamental degrees of freedom of quantum gravity
\cite{Witten:1998zw,Sundborg:1999ue,Aharony:2003sx}. To explore other extreme behaviors 
of black holes, we want access to the individual black hole microstates which 
account for this entropy.
In this paper, we `construct' (in certain sense) and study the local BPS operators 
which can describe the microstates of BPS black holes in quantum AdS/CFT
\cite{Maldacena:1997re}. The entropies of BPS black holes in AdS$_{D>3}$ 
were recently computed from the dual field theories: see
\cite{Cabo-Bizet:2018ehj,Choi:2018hmj,Benini:2018ywd} and references thereof. 
We shall construct representations of some of these microstates and study their
properties, hoping that we can better address and exactly solve interesting black 
hole information problems with this knowledge.

We shall study the 4d $\mathcal{N}=4$ Yang-Mills theory dual to type IIB string theory 
in $AdS_5\times S^5$. We are interested in
%Our
BPS states that preserve $2$ of the $32$ supersymmetries
\cite{Romelsberger:2005eg,Kinney:2005ej}, called $\frac{1}{16}$-BPS 
states. The spectrum of these BPS states may change as the 
coupling constant $g_{\rm YM}$ is varied, and in fact they are
different between the free and 1-loop levels. 
The 1-loop BPS states of the $SU(N)$ theory were studied in the past \cite{Kinney:2005ej,Berkooz:2006wc,Janik:2007pm,Grant:2008sk,Chang:2013fba}.
They can be reformulated as classical cohomologies with respect to a nilpotent
supercharge $Q$. It has been conjectured that 
the spectrum of 1-loop BPS states remains unchanged at general nonzero coupling  
\cite{Kinney:2005ej,Berkooz:2006wc,minwalla}. This conjecture is partly proved in 
perturbation theory \cite{Chang:2022mjp} with certain assumptions. 
The discussions in this paper will often assume this conjecture. 
Although we shall construct representatives of new cohomologies, they are generally 
not equal to the BPS operators. 
The two in general differ by certain $Q$-exact terms. So what we achieve is weaker than
`constructing' the black hole microstates. It remains to be seen what kind of questions 
can be addressed just from cohomologies. (See section 3 for an example.)

Our precise goal is to construct cohomologies at finite $N$ which are not of
the graviton type. One may wonder what gravitons mean at finite $N$. 
First, single-particle graviton cohomologies are
constructed with single trace operators only. Multi-gravitons are simply defined as 
products of the single-particle gravitons.\footnote{While products of 
BPS operators are generally not BPS, cohomologies multiply to yield cohomologies.}
At large $N$, all these operators are independent because there are no trace relations 
of matrix fields. More precisely, an operator is free of trace relations when 
the number of fields appearing in the operator is no greater than $N$. When 
the number of fields is larger than $N$, trace relations may apply. 
On the gravity side, trace relations are realized by gravitons polarizing into D3-brane giant gravitons \cite{McGreevy:2000cw,Grisaru:2000zn,Hashimoto:2000zp}, after which fewer 
states are allowed than the naive estimate. This is 
called the `stringy exclusion principle.' The physical mechanism of this principle is 
same at all $N\geq 2$. So it makes physical sense to define finite $N$ graviton 
cohomologies as multiplications of single-trace cohomologies modulo 
trace relations. These cohomologies are fully understood.

Unless there are exotic microstates that neither qualify to be called gravitons nor 
black holes, we expect the remaining cohomologies to describe the black hole microstates 
in AdS. We shall therefore call them black hole cohomologies for simplicity of nomenclature, 
but also having in mind that `black hole' could broadly mean all possible novelties 
beyond gravitons. There may be two viewpoints on these cohomologies at finite $N$. First, 
they are intermediate steps to the cohomologies at parametrically large $N$. Second, 
more progressively, one may regard the finite $N$ Yang-Mills theory as a model of 
quantum AdS/CFT at finite Newton constant. The `black hole cohomologies' at finite $N$ may
simulate quantum black holes therein. Not all interesting questions on semiclassical black
holes survive in these finite $N$ models, but some questions do. For instance, the Cardy limit
\cite{Choi:2018hmj,ArabiArdehali:2019tdm,Honda:2019cio,Kim:2019yrz,Cabo-Bizet:2019osg} 
exhibits universal deconfining behaviors at large charges, naturally generalizing the large
black hole physics to all finite $N$.

The progress in this paper is all with the $SU(2)$ theory, the most quantum version 
of AdS/CFT. Already in this model, the new cohomologies exhibit some qualitative features
reminiscent of black holes. Although this problem has been discussed since 2005, 
not a single black hole cohomology was found until last year. In \cite{Chang:2022mjp}, 
it was shown that the $SU(2)$ theory has the lowest black hole cohomology at energy 
$E=\frac{19}{2}$, R-charges $R_1=R_2=R_3=\frac{3}{2}$ and angular 
momenta $J_1=J_2=\frac{5}{2}$. A simple representative of this cohomology 
was constructed in \cite{Choi:2022caq}. This is the primary of a superconformal
representation of $PSU(1,2|3)$ $\subset PSU(2,2|4)$. There are infinitely many descendants 
that one can trivially construct from this primary.

We construct an infinite tower of new black hole primaries in the $SU(2)$ theory. 
For technical reasons, we often focus on the `BMN sector' of the Yang-Mills 
theory on $S^3\times \mathbb{R}$ \cite{Berenstein:2002jq,Kim:2003rza}. This is a 
consistent truncation of the classical theory, also yielding a consistent truncation of 
our cohomology problem. The superconformal index
\cite{Romelsberger:2005eg,Kinney:2005ej} of the Yang-Mills theory can also be restricted 
to the BMN sector. For the $SU(2)$ theory in the BMN 
sector, we find infinitely many new cohomologies which 
saturate the index. More concretely, the energy $E_{(n)}$, 
R-charge $R_{(n)}(=R_1=R_2=R_3)$ and the angular momentum $J_{(n)}(=J_1=J_2)$ of our 
$n$'th `core' black hole primary $O_n$ are given by
\begin{equation}
  (E_{(n)},R_{(n)},J_{(n)})=({\textstyle \frac{19}{2}+4n,\frac{3}{2},\frac{5}{2}+2n})\ ,\ \ 
  %j_n\equiv 6(Q_n+J_n)=24+12n\ ,\ \ 
  n=0,1,2,\cdots\ .
\end{equation}
The operator found in \cite{Chang:2022mjp} corresponds to $n=0$. 
One can exactly compute the full index $Z$ and the graviton index $Z_{\rm grav}$ in 
the BMN sector. Their difference is given by
\begin{equation}\label{bmn-bh}
  Z-Z_{\rm grav}=\left[-\frac{e^{-4(\Delta_1+\Delta_2+\Delta_3)}}{1-e^{-2(\Delta_1+\Delta_2+\Delta_3)}}\right]\cdot 
  \left[\prod_{I=1}^3(1-e^{-\Delta_I})\right]\cdot \left[\prod_{I=1}^3
  \frac{1}{1-e^{-\Delta_I}e^{-\Delta_1-\Delta_2-\Delta_3}}\right]\ ,
\end{equation}
where $\Delta_I$ is the chemical potential conjugate to the charge $R_I+J$.
The three factors in the square brackets come respectively from: 
(left) the tower of core black hole primaries $O_n$, 
(middle) the Fock space of their superconformal descendants, (right) 
the hairs by a limited subset of gravitons. The BMN sector in the 
$SU(2)$ theory does not show large enough entropy of BPS states even at large 
charges. However, we expect that the BPS entropy in the BMN sector should exhibit 
a black hole like growth at large $N$.

Since cohomologies are multiplicative, we can consider the product cohomologies 
of gravitons and core black hole cohomologies. We find that a surprisingly large set 
of them does not appear in the index. The most natural interpretation of this phenomenon
is that these product operators are $Q$-exact (i.e. absent in the BPS Hilbert space), 
which we prove explicitly at some low orders. We view this as a partial 
no-hair theorem of black holes in the $SU(2)$ 
model, in that our black hole cohomologies abhor the dressing by some gravitons. 
This phenomenon is most clearly visible for the conformal primary states of gravitons. 
In (\ref{bmn-bh}), the index only captures $3$ out of $17$ species of BMN graviton particles  
dressing our black holes. We also studied the general $SU(2)$ index for black holes up to $40$'th order in the charge $j=6(R+J)$. Till this order, the index captures only $3$ out 
of $32$ conformal primary gravitons dressing the core black holes.
(See sections 3.2 and 4 for the conformal descendants.) 
Since AdS black holes are expected to allow certain graviton hairs 
\cite{Basu:2010uz,Bhattacharyya:2010yg,Markeviciute:2018yal,Markeviciute:2018cqs}, 
partial no-hair phenomenon seems to be the right behavior in AdS/CFT 
models. In section 4 we perturbatively study the black hole 
hairs in the BPS limit to clarify the similar behaviors in the gravity dual.

We also estimate when new black hole cohomologies should appear in the 
$SU(2)$ theory beyond $O_n$. By studying the index and all possibilities of  
finite $N$ graviton hairs, we show that new black hole core primaries
should appear at or before the $j=39$ order.

Note that black holes in the BMN model were studied recently 
\cite{Pateloudis:2022ijr,Biggs:2023sqw,Maldacena:2023acv}. As emphasized in 
\cite{Biggs:2023sqw,Maldacena:2023acv}, microscopic detection of the quasinormal modes 
is a signature of seeing black holes since these modes are falling into the horizon. 
In our stationary (BPS) setup, a direct consequence is the no-hair theorem for
the corresponding modes: we try to perturb a black hole by multiplying
certain gravitons and find that they do not exist in the BPS sector after a long time.

The remaining part of this paper is organized as follows. In section 2, we explain 
the cohomology problem and the graviton cohomologies at finite $N$. 
In section 3.1, we study the BMN sector and find an infinite class 
of new black hole cohomologies for $SU(2)$. In section 3.2, we study 
the general $SU(2)$ cohomologies up to  $j\leq 40$ and find 
no-hair behaviors. In section 4, we study how perturbative black hole hairs 
behave in the BPS limit and comment on similarities with the results in section 3. 
In section 5, we conclude with remarks.

\section{The cohomology problem}

We review the problem of local BPS operators in $\mathcal{N}=4$ Yang-Mills theory on 
$\mathbb{R}^4$ and its cohomological formulation. By operator-state map, they 
map to BPS states of the Yang-Mills theory on $S^3\times\mathbb{R}$. 
We shall discuss the $SU(N)$ theory. 
We sketch the problem briefly before explaining the details, to emphasize the nature of the 
problem. We start by selecting two supercharges among $32$ of them, which will annihilate 
our BPS operators. We first consider all gauge-invariant local BPS 
operators in the free limit. Then we turn 
on small coupling $g_{\rm YM}\neq 0$ and see how many of the free BPS operators 
remain BPS, order by order in perturbation theory.

The $\mathcal{N}=4$ Yang-Mills theory consists of the following fields in 
the adjoint representation of $SU(N)$ ($N\times N$ traceless matrices):
\begin{eqnarray}
  \textrm{vector}&:&A_\mu\sim A_{\alpha\dot\beta}\ ,\ \mu=1,2,3,4\ ,\ 
  \alpha=\pm\ ,\ \dot\beta=\dot{\pm}\nonumber\\
  \textrm{scalar}&:&\Phi_{ij}(=-\Phi_{ji})\ ,\ \overline{\Phi}^{ij}\sim
  {\textstyle \frac{1}{2}}\epsilon^{ijkl}\Phi_{kl}
  \ ,\ \ i,j,k,l=1,2,3,4\nonumber\\
  \textrm{fermion}&:&\Psi_{i\alpha}\ ,\ \overline{\Psi}^i_{\dot\alpha}\ .
\end{eqnarray}
We consider the Euclidean CFT on $\mathbb{R}^4$, which is 
related to the CFT on $S^3\times\mathbb{R}$ by radial quantization.
$\alpha,\dot\alpha$ are the doublet indices of $SU(2)_L\times SU(2)_R\sim SO(4)$ 
which rotate the $S^3$, and $\mu$ is the vector index. Uppercase $i,j$ are the 
indices for the fundamental representation of $SU(4)$ R-symmetry, while the 
lowercase indices are for the anti-fundamental representation. Hermitian conjugations 
of these fields should be understood with a bit care in radial quantization. 
See, for instance, \cite{Grant:2008sk} for the details. 
It will be useful to decompose these fields in the 
$\mathcal{N}=1$ language as follows, with manifest $SU(3)\times U(1)\subset SU(4)$ symmetry:
\begin{eqnarray}
  \textrm{vector multiplet}&:&A_{\alpha\dot\beta}\ ,\ 
  \lambda_\alpha=\Psi_{4\alpha}\ ,\ \bar\lambda_{\dot\alpha}=\overline{\Psi}^4_{\dot\alpha}\\
  \textrm{chiral multiplets}&:&\phi_m=\Phi_{4m}\ ,\ \bar\phi^m=\overline{\Phi}^{4m}\ ,\ 
  \psi_{m\alpha}=-i\Psi_{m\alpha}\ ,\ \bar\psi^m_{\dot\alpha}=i\overline{\Psi}^m_{\dot\alpha}
  \nonumber\ .
\end{eqnarray}
$m=1,2,3$ is the $SU(3)$ index, either for the fundamental/anti-fundamental representation. 
The system carries a continuous real coupling $g_{\rm YM}$, and 
enjoys $\mathcal{N}=4$ superconformal symmetry at any value of $g_{\rm YM}$. 
(The theta angle will not be relevant in our discussions.)
The supercharges consist of $16$ Poincare supercharges $Q^i_{\alpha}$, 
$\overline{Q}_{i\dot\alpha}$ and $16$ conformal supercharges $S_{i\alpha}$, 
$\overline{S}^i_{\dot\alpha}$. In radial quantization, $Q$ and $S$ are 
Hermitian conjugate to each other: $(Q^i_\alpha)^\dag=S_i^{\alpha}$, 
$(\overline{Q}_{i\dot\alpha})^\dag=\overline{S}^{i\dot\alpha}$.  
The quantum supercharges may depend on the coupling $g_{\rm YM}$, which is hard to 
write down in general. The supercharges of the interacting classical field theory 
contain terms up to half-loops, i.e. $\mathcal{O}(g_{\rm YM}^1)$ order. 
The transformations for the classical Poincare supercharges are listed in Appendix A.

The BPS operators of our interest are defined with 
$Q\equiv Q^4_-$ and $S\equiv S_{4}^-=Q^\dag$. 
We are interested in gauge-invariant local operators $O$ located at the origin 
$x^\mu=0$ of $\mathbb{R}^4$, satisfying
\begin{equation}
  [Q,O\}=0\ ,\ [Q^\dag,O\}=0\ .
\end{equation}
The part of the superconformal algebra which is important to us is
\begin{equation}
    Q^2=0\ ,\ (Q^\dag)^2=0 \ ,\ \{Q,Q^\dag\}\sim H-\sum_{I=1}^3R_I-\sum_{i=1}^2J_i\ .
\end{equation}
$H$ is the scaling dimension of the local operators, or the energy of the corresponding 
states times the radius of $S^3$. $R_I$ are the three Cartan charges of $SO(6)\sim SU(4)$ 
which rotate three orthogonal 2-planes in $\mathbb{R}^6$. $J_i$ are 
the two angular momenta on $\mathbb{R}^4$ rotating two orthogonal 
2-planes. The eigenvalues of $R_I$ and $J_i$ are integers for bosons and half of odd integers 
for fermions. The BPS operators $O$ annihilated by $Q$ and $Q^\dag$ commute with 
$\{Q,Q^\dag\}$. They can be arranged to be the eigenstates of 
$H,R_I,J_i$, whose eigenvalues $E,R_I,J_i$ satisfy 
\begin{equation}\label{BPS-relation}
  E=\sum_I R_I+\sum_i J_i\ .
\end{equation}

\begin{table}[t]
\begin{center}
\begin{tabular}{c|c|c|c}
	\hline
    field & $E$ & ($R_1$, $R_2$, $R_3$)  & ($J_1$, $J_2$) \\
	\hline $\phi_m$ & $1$ & ($-1,0,0$), ($0,-1,0$), ($0,0,-1$) & ($0,0$)\\
    $\bar\phi^m$ & $1$ & ($1,0,0$), ($0,1,0$), ($0,0,1$) & ($0,0$)\\
    $\psi_{m\pm}$ & $\frac{3}{2}$ & ($-\frac{1}{2},\frac{1}{2},\frac{1}{2}$),
    ($\frac{1}{2},-\frac{1}{2},\frac{1}{2}$), ($\frac{1}{2},\frac{1}{2},-\frac{1}{2}$)&
    ($\pm\frac{1}{2},\pm\frac{1}{2}$)\\
    $\bar\psi^m_{\dot\pm}$ & $\frac{3}{2}$ & ($\frac{1}{2},-\frac{1}{2},-\frac{1}{2}$),
    ($-\frac{1}{2},\frac{1}{2},-\frac{1}{2}$), ($-\frac{1}{2},-\frac{1}{2},\frac{1}{2}$)&
    ($\pm\frac{1}{2},\mp\frac{1}{2}$)\\
    \hline
    $A_{+\dot\pm}$ & $1$ & $(0,0,0)$ & $(1,0)$, $(0,1)$\\
    $A_{-\dot\pm}$ & $1$ & $(0,0,0)$ & $(-1,0)$, $(0,-1)$\\
    $f_{++},f_{+-},f_{--}$ & $2$ & ($0,0,0$) & ($1,1$), ($0,0$), ($-1,-1$)\\
    $f_{\dot{+}\dot{+}},f_{\dot{+}\dot{-}},f_{\dot{-}\dot{-}}$ & 
    $2$ & ($0,0,0$) & ($1,-1$), ($0,0$), ($-1,1$)\\
    $\lambda_\pm$ & $\frac{3}{2}$ & $(-\frac{1}{2},-\frac{1}{2},-\frac{1}{2})$ & 
    ($\pm\frac{1}{2},\pm\frac{1}{2}$)\\
    $\bar{\lambda}_{\dot\pm}$ & $\frac{3}{2}$ & $(\frac{1}{2},\frac{1}{2},\frac{1}{2})$ & 
    ($\pm\frac{1}{2},\mp\frac{1}{2}$)\\
    \hline
    $\partial_{+\dot\alpha}$ & $1$ &($0,0,0$) & ($1,0$), ($0,1$)\\
    $\partial_{-\dot\alpha}$ & $1$ &($0,0,0$) & ($-1,0$), ($0,-1$)\\
    \hline
\end{tabular}
\end{center}
\caption{The charges of elementary fields. Charges in the parenthesis 
are listed in the order of $m=1,2,3$ or $\dot\alpha=\dot{+},\dot{-}$ or in the order 
of the fields listed in the first column.}\label{charges-fields}
\end{table}

In the free limit, the operators satisfying the BPS relation (\ref{BPS-relation}) 
can be easily constructed using the BPS elementary fields satisfying the same relation. 
The charges for the fields and the derivative $\partial_\mu\sim \partial_{\alpha\dot\beta}$ 
are listed in Table \ref{charges-fields}. The field strengths in the bispinor basis 
are defined by $f_{\alpha\beta}\sim(\sigma^{\mu\nu})_{\alpha\beta}F_{\mu\nu}$ and 
$f_{\dot\alpha\dot\beta}\sim(\bar{\sigma}^{\mu\nu})_{\dot\alpha\dot\beta}F_{\mu\nu}$. 
Among them, the gauge-covariant BPS fields and derivatives which satisfy the BPS relation 
are given by
\begin{equation}\label{BPS-fields}
  \bar\phi^m\ ,\ \ \psi_{m+}\ ,\ \ f_{++}\ ,\ \ 
  \bar\lambda_{\dot\alpha}\ ,\ \ \partial_{+\dot\alpha}\ .
\end{equation}
With these, we construct independent `letters' for the gauge invariant operators. 
Basically, acting derivatives $\partial_{+\dot\alpha}$ to a BPS field 
forms a letter. However, the equation of motion operator is zero and should  
not be included. The only equation of motion constructed using 
(\ref{BPS-fields}) is 
\begin{equation}
  \partial_{+\dot\alpha}\bar\lambda^{\dot\alpha}=0\ \Leftrightarrow\ 
  \partial_{+[\dot\alpha}\bar\lambda_{\dot{\beta}]}=0\ .
\end{equation}
So the $SU(2)_R$ indices carried by the derivatives and the gaugino have to be 
symmetrized in a letter. Also, in the free theory, the derivatives
$\partial_{+\dot\alpha}$ acting on the same field commute. Therefore, all $SU(2)_R$
indices appearing in a letter should be symmetrized. So we find the following independent
letters (all  at $x^\mu=0$),
\begin{eqnarray}
  \partial_{+(\dot\alpha_1}\cdots\partial_{+\dot\alpha_n)}\varphi
  &\sim&(\partial_{+\dot{+}})^{n_1}(\partial_{+\dot{-}})^{n_2}\varphi\ \ \ 
  (n_1,n_2\geq 0,n\geq 0)\\
  \partial_{+(\dot\alpha_1}\cdots\partial_{+\dot\alpha_{n-1}}\bar\lambda_{\dot\alpha_{n})}
  &\sim& n_1(\partial_{+\dot{+}})^{n_1-1}(\partial_{+\dot{-}})^{n_2}\bar{\lambda}_{\dot{+}}+
  n_2(\partial_{+\dot{+}})^{n_1}(\partial_{+\dot{-}})^{n_2-1}\bar{\lambda}_{\dot{-}}\ \ \ 
  (n_1,n_2\geq 0,n\geq 1)
  \nonumber
\end{eqnarray}
where $n_1+n_2=n$.
$\varphi$ denotes any field except $\bar\lambda_{\dot\alpha}$, i.e. chosen 
among $\bar\phi^m$, $\psi_{m+}$, $f_{++}$. Multiplying these letters and 
pairwise contracting the $SU(N)$ indices, one can construct general 
gauge-invariant BPS operators in the free theory.

Now we consider the 1-loop BPS operators, i.e. operators 
which diagonalize $H$ and saturate the BPS relation till $O(g_{\rm YM}^2)$ order. 
For very small $g_{\rm YM}$, we can start from the free BPS operators 
explained in the previous paragraph and study which of them remain BPS at the 
1-loop level. The only 
modification needed is to replace $\partial_{+\dot\alpha}$ and 
$f_{++}$ by the covariant expressions by adding $\mathcal{O}(g_{\rm YM}^1)$ corrections. 
The covariant letters used in the free theory should be modified as follows:
\begin{equation}\label{letter-interact}
  D_{+(\dot\alpha_1}\cdots D_{+\dot\alpha_n)} \varphi\ ,\ \
 D_{+(\dot\alpha_1}\cdots D_{+\dot\alpha_{n-1}}\bar\lambda_{\dot\alpha_{n})}\ .
\end{equation}
Here $D_{+\dot\alpha}$ is the covariant derivative defined by 
$D_{+\dot\alpha}=\partial_{+\dot\alpha}-ig_{\rm YM}[A_{+\dot\alpha},\ \ ]$, related 
to $f_{++}$ by $[D_{+\dot\alpha},D_+^{\ \dot\alpha}]\sim 
[D_{+[\dot\alpha},D_{+\dot\beta]}]\sim g_{\rm YM} f_{++}$. 
Although two different $D_{+\dot\alpha}$'s do not commute anymore, we still completely 
symmetrize the derivatives in a letter. This is to avoid introducing redundant 
letters: if some $D$'s are antisymmetrized in (\ref{letter-interact}), 
it can be rewritten as a sum of products of letters involving $f_{++}$'s. 
We also symmetrize the $SU(2)_R$ indices in $D_{+\dot\alpha}$ and $\bar\lambda_{\dot\alpha}$ 
to avoid redundant letters, because the gaugino equation is given by 
$g_{\rm YM}[\bar\phi^m,\psi_{m+}]\sim D_{+\dot\alpha}\bar\lambda^{\dot\alpha}
\sim D_{+[\dot\alpha}\bar\lambda_{\dot\beta]}$. 
Among all possible gauge-invariant operators $O$ constructed from (\ref{letter-interact}), 
we would like to find the combinations satisfying 
\begin{equation}
  0=\left[\frac{}{}\!\!\right.H-\sum_IR_I-\sum_i J_i,O\left.\!\!\frac{}{}\right]\sim \left[\{Q,Q^\dag\},O\right]\ .
\end{equation}
The last equation holds if and only if $[Q,O\}=0$ and $[Q^\dag,O\}=0$. 
Note that the coupling dependence is only in $H,Q,Q^\dag$, since 
the quantized non-Abelian charges $R_I$, $J_i$ cannot depend on $g_{\rm YM}$. At the 1-loop level, 
$H-\sum_IR_I-\sum_i J_i$ acting on the BPS letters (\ref{letter-interact}) 
is at $\mathcal{O}(g_{\rm YM}^2)$ order, while $Q,Q^\dag$ acting on 
(\ref{letter-interact}) are at $\mathcal{O}(g_{\rm YM}^1)$ order. Their 
complete forms are given in \cite{Beisert:2004ry}. In particular, 
$Q$ at the half-loop order is simply that of the classical field theory at nonzero 
$g_{\rm YM}$, which transform the BPS fields by
\begin{eqnarray}\label{half-loop-Q}
  &&[Q,\bar\phi^m]=0\ ,\ \{Q,\bar\lambda_{\dot\alpha}\}=0\ ,\ 
  \{Q,\psi_{m+}\}=-ig_{\rm YM}\epsilon_{mnp}[\bar\phi^n,\bar\phi^p]\\
  &&[Q,f_{++}]=-ig_{\rm YM}[\psi_{m+},\bar\phi^m]\ ,\ 
  [Q,D_{+\dot\alpha}](\cdots)=-ig_{\rm YM}[\bar\lambda_{\dot\alpha},(\cdots)\}\ .
  \nonumber
\end{eqnarray}
$S=Q^\dag$ acts on pairs of letters \cite{Beisert:2004ry}.
From here, we absorb $g_{\rm YM}$ into the normalization of fields.

Instead of directly constructing the BPS operators, 
we study the cohomology classes which are in 1-to-1 map to the BPS operators. 
The cohomology class is defined by the set of operators $O$ made of (\ref{letter-interact}) 
that are closed under the action of $Q$, i.e. $[Q,O\}=0$, with
the equivalence relation $O\sim O+[Q,\Lambda\}$. 
Here, $\Lambda$ is also an operator constructed with (\ref{letter-interact}). 
We can call this a cohomology because of the nilpotency 
$Q^2=0$. This cohomology class is in 1-to-1 map to the BPS operators 
$O_{\rm BPS}$ satisfying $[Q,O_{\rm BPS}\}=0$ and $[Q^\dag,O_{\rm BPS}\}=0$, because 
the latter is basically a harmonic form in this language \cite{Grant:2008sk}. 
The cohomology problem is defined using $Q$ without referring to $S=Q^\dag$.
Note also that this problem is completely classical, 
since the transformation (\ref{half-loop-Q}) is 
that of the classical field theory.

It has been conjectured implicitly/explicitly 
(especially in \cite{minwalla}) that the 1-loop BPS states remain BPS at general nonzero coupling. 
However, even if this conjecture is true,  it does not mean that the form 
of the BPS operator $O_{\rm BPS}$ takes the same form at different values of $g_{\rm YM}$. 
The conjecture just claims that the operators will survive to be BPS with $g_{\rm YM}$-dependent 
deformations, being in 1-to-1 map with the 1-loop BPS states without any pairwise jumps. 
Recently, the validity of this conjecture was argued at all orders 
in perturbation theory \cite{Chang:2022mjp}. 

We shall construct the representatives of new cohomology classes. 
Our interest is those which have chances to describe black holes rather than 
BPS gravitons. The BPS graviton cohomologies are completely 
understood, even at finite $N$ subject to the stringy exclusion principle. 
We shall explain them in section 2.1. However, enumerating them without overcounting 
is technically quite tricky, due to the trace relations. We shall present 
a strategy which allows us to solve the trace relations in terms of diagonal matrix fields. 
The strategy applies to arbitrary $N$, but is particularly useful for $N=2$. (At $N=2$, this 
idea was employed in \cite{Grant:2008sk} in special subsectors.) We shall 
analytically/numerically implement this idea in section 3 (and appendix C) 
to efficiently count graviton-type cohomologies.

\subsection{The finite $N$ gravitons}

We first explain the cohomologies for the BPS gravitons in the large $N$ limit, 
and then explain how to define and understand the finite $N$ graviton cohomologies.

We first take $N\gg 1$ and the number of fields in the cohomologies 
much smaller than $N$ (say at order $1$). In this limit, the mixing of
single- and multi-trace operators by the dilatation operator $H$ is suppressed by 
$\frac{1}{N}$. So one can first construct the BPS operators using single trace operators only, 
and then multiply them to obtain multi-trace BPS operators. The former and latter are 
the single- and multi-particle graviton states, respectively.

\begin{table}[t]
\begin{center}
\begin{tabular}{c|c|c|c|c}
	\hline
    $(-1)^F E^\prime$ & $J^\prime$ & $R_1^\prime$ & $R_2^\prime$ & construction\\
	\hline $n$ & $0$ & $n$ & $0$ & $|n\rangle$ \\
    $-(n+\frac{1}{2})$ & $\frac{1}{2}$ & $n-1$ & $0$ &
    $\overline{Q}_{m\dot\alpha}|n\rangle$ \\
    $n+1$ & $0$ & $n-2$ & $0$
    & $\overline{Q}_{m\dot{+}}\overline{Q}_{n\dot{-}}|n\rangle$ \\
	$-(n+1)$ & $0$ & $n-1$ & $1$ &
    $Q^m_+|n\rangle$\\
    $n+\frac{3}{2}$ & $\frac{1}{2}$ & $n-2$ & $1$ &
    $Q^m_+\overline{Q}_{n\dot\alpha}|n\rangle$ \\
    $-(n+2)$ & $0$ & $n-3$ & $1$ &
    $Q^m_+\overline{Q}_{n\dot{+}}\overline{Q}_{p\dot{-}}|n\rangle$ \\
    $n+2$ & $0$ & $n-1$ & $0$ & $Q^m_+Q^n_+|n\rangle $ \\
    $-(n+\frac{5}{2})$ & $\frac{1}{2}$ & $n-2$ & $0$ &
    $Q^m_+Q^n_+\overline{Q}_{p\dot\alpha}|n\rangle$ \\
    $n+3$ & $0$ & $n-3$ & $0$ &
    $Q^m_+Q^n_+\overline{Q}_{p\dot{+}}\overline{Q}_{q\dot{-}}|n\rangle$ \\
    \hline
\end{tabular}
\caption{The state contents of the $PSU(1,2|3)$ multiplet $S_n$. 
For low $n$'s, the rows with negative $R_1^\prime$ are absent. $|n\rangle$ schematically 
denotes the superconformal primaries.}\label{graviton-multiplet}
\end{center}
\end{table}

We consider the single-trace BPS operators from the viewpoint of
the $Q$-cohomology problem. The cohomology problem is well-defined in 
the single trace sector, since the action of $Q$ does not change trace numbers.
This problem is completely solved \cite{Janik:2007pm,Chang:2013fba}. 
The single trace cohomologies can be arranged into  
the supermultiplets of $PSU(1,2|3)\subset PSU(2,2|4)$ which commute with 
our $Q,Q^\dag$. In \cite{Kinney:2005ej}, 
the short multiplets for the single trace cohomologies are called 
$S_n$, with $n=2,3,\cdots$.\footnote{The $S_1$ multiplet 
comes from the overall $U(1)$ mode of the $U(N)$ theory, which is absent 
for $SU(N)$.} In this paper, we shall call the multiplets/representations of $PSU(1,2|3)$ 
`$\frac{1}{16}$-BPS multiplets/representations' and those of 
$PSU(2,2|4)$ `$\mathcal{N}=4$ multiplets/representations,' respectively.
The ways in which $\frac{1}{16}$-BPS representations sit in the $\mathcal{N}=4$ representations 
are explained in the appendix B.3 of \cite{Kinney:2005ej}, which we review in  
appendix B. The $\frac{1}{16}$-BPS multiplet $S_n$ is contained in the 
short $\mathcal{N}=4$ multiplet $B\overline{B}[0;0]^{[0,n,0]}_n$ in the notation of \cite{Cordova:2016emh}, 
where the superscripts/subscript are the $SU(4)$ Dynkin labels and the scaling dimension of 
the primaries, respectively, and $[J_1+J_2,J_1-J_2]=[0,0]$ 
denotes the spins of the primaries. (This is a multiplet of ${\bf bb}$ type in the notation of \cite{Kinney:2005ej}.) Following \cite{Kinney:2005ej}, 
we define $E^\prime=E+\frac{J_1+J_2}{2}$, $J^\prime=\frac{J_1-J_2}{2}$, and 
$[R_1^\prime,R_2^\prime]$ is the $SU(3)$ highest weight. 
The states in $S_n$ obtained by acting 
$Q^m_+$ and $\overline{Q}_{m\dot\alpha}$ generators of $PSU(1,2|3)$ are listed 
in Table \ref{graviton-multiplet}. These are obtained by assuming that all supercharges
anticommute. In fact some of the anticommutators 
yield $\{Q^m_+,\overline{Q}_{n\dot\alpha}\}\sim P_{+\dot\alpha}$, but they 
generate the conformal descendants can be generated later. In the gravity dual language, 
the table lists the Kaluza-Klein field contents in AdS$_5$. 
The conformal descendants are obtained by acting $P_{+\dot\alpha}$ on 
the `fields' in the table, which generate the wavefunctions in $AdS_5$.

The superconformal primaries $|n\rangle$ of Table \ref{graviton-multiplet} are 
given by the following operators,
\begin{equation}\label{scalar-primary}
  {\rm tr}[\bar\phi^{(m_1}\cdots\bar\phi^{m_n)}]\ .
\end{equation}
From the cohomology viewpoint, (\ref{scalar-primary}) can be 
understood as follows. All gauge-invariant operators made of BPS scalars are $Q$-closed 
since $Q\bar\phi^m=0$.
However, the presence of antisymmetrized pairs of scalars 
will make the operator $Q$-exact because $[\bar\phi^m,\bar\phi^n]\sim \epsilon^{mnp}Q\psi_{p+}$ from (\ref{half-loop-Q}).\footnote{$\bar\phi^{m_1}\bar\phi^{n_1}\cdots \bar\phi^{n_i}\bar\phi^{m_2}-
\bar\phi^{m_2}\bar\phi^{n_1}\cdots \bar\phi^{n_i}\bar\phi^{m_1}$ is also a 
linear combination of terms involving commutators.} So among the single-trace operators 
made of $n$ scalars, only (\ref{scalar-primary}) represent nontrivial 
cohomology classes. All the other states in the table are obtained by acting
$Q^m_+,\overline{Q}_{m\dot\alpha}$'s on  (\ref{scalar-primary}). 
Multiplying them yields independent multi-particle cohomologies at large $N$.
Cohomologies in $S_2$  are given in Appendix A.

Now we discuss the finite $N$ cohomologies of graviton type. 
Finite $N$ gravitons are nothing but the `graviton cohomologies' that we have explained 
so far. Various steps that we explained above go through for finite $N$ cohomologies. 
For instance, constructing cohomologies within single trace operators 
is consistent because $Q$ does not change the trace number.
Also, despite the existence of trace relations at finite $N$ 
which may relate some single trace cohomologies to multi-trace ones, 
it is just a matter of using redundant basis and we can eliminate them later.
(Of course, it will be convenient later to use only a subset of a such single trace 
cohomologies, as we will explain.)
Next we consider the step of multiplying them to make multi-trace operators. 
Due to the Leibniz rule of $Q$, cohomologies multiply to yield a new cohomology. 
Therefore, even at finite $N$, multiplying the single trace cohomologies yields 
multi-trace cohomologies. Again due to trace relations, there might be some multiplications 
which yield a trivial cohomology, i.e. $Q$-exact operator. However, this is again a 
question of redundancy.

The only new issue to consider is that the number of independent 
cohomologies reduce because of the trace relation. Even at large $N$, 
one should take into account the trace relations if the number of fields becomes 
larger than $N$. This is the regime in which semi-classical giant gravitons can 
account for the trace relations as the stringy exclusion principle. Even at 
finite $N$, it makes good sense to regard these cohomologies as the graviton cohomologies 
in quantum AdS gravity. In other words, these BPS states can be completely 
understood by knowing the (giant) graviton physics.\footnote{Historically, giant gravitons 
are discovered as objects realizing the trace relations, reducing the number of states 
than the naive estimates. This is the context in which we quote giant gravitons 
here. However, note that it has been shown \cite{Choi:2022ovw} recently that more 
complicated giant graviton states (with brane intersections and light open 
strings stretched between the branes) represent black hole microstates.}

On the other hand, these operators cannot account for certain 
finite $N$ physics. For instance, in the high-temperature Cardy limit 
\cite{Choi:2018hmj}, charges/energy are taken to be $N^2$ times a   
large number independent of $N$. The entropy in this limit scales like 
$N^2$ times a universal large number for the whole 
sequence of $U(N)$ gauge theories, from small to large $N$. Although this is not large $N$ at all, such a deconfining behavior $\propto N^2$ is universal at large energy/temperature.
This generalizes the large black hole physics straightforwardly to finite $N$. (The black 
hole like growth of the entropy at finite $N$ was also confirmed by numerical studies \cite{Murthy:2020rbd,Agarwal:2020zwm}.) Such behaviors do not happen with the finite $N$
graviton cohomologies. All these make it natural to study novel finite $N$
cohomologies which are not of the graviton type as defined above.

In section 3, an important technical step of identifying the black hole cohomologies 
will be counting finite $N$ gravitons and subtracting them from the full degeneracy 
(index). Once we know the overcomplete set for gravitons as described above, it is in principle 
straightforward to identify the redundant operators due to trace relation.
In practice, even if one uses a computer, this becomes quickly cumbersome as the charges 
and $N$ grow. In the remaining part of this section, we will explain some structures which 
make this estimate easier. We explain two tricks, both related to the fact that 
the primaries (\ref{scalar-primary}) are given by symmetrized scalars.

The single-trace cohomologies in the $S_n$ multiplets are the generators of the graviton
cohomologies, i.e. we multiply them to generate multi-trace graviton cohomologies. 
If some of them decompose to products of other generators in $S_n$'s, clearly 
they need not be used as generators. For $SU(N)$, one can show that all elements of 
$S_{n\geq N+1}$ decompose to the those of $S_{n\leq N}$. To see this, we first show that 
any primary of the form (\ref{scalar-primary}) at $n\geq N+1$ decomposes to 
those at $n\leq N$. This can be shown using the Cayley-Hamilton equation,
\begin{equation}
  M^N+c_{N-1}(M)M^{N-1}+\cdots c_1(M) M+c_0(M){\bf 1}_{N\times N}=0\ ,
\end{equation}
where $M$ is an $N\times N$ matrix, and the coefficients $c_n(M)$'s are explicitly known 
in terms of ${\rm tr}(M^k)$ with $k\leq N-n$. (In particular, $c_0(M)=(-1)^N\det(M)$.) 
Using this identity, one can express the matrix 
$M^N$ (the first term) as a linear combination of various multi-trace operators 
times the matrix $M^n$, where $0\leq n\leq N-1$. 
Inserting $M=M_1+M_2$, $M=M_1+M_2+M_3$, $\cdots$, $M=M_1+\cdots +M_N$ 
with independent matrices and applying the Cayley-Hamilton equations repeatedly
to various parts, one obtains a more generalized identity for the symmetrized product  
$M_{(1}\cdots M_{N)}$ of $N$ matrices $M_i$. For traceless 
$M_i$, some examples at low $N$ are 
\begin{eqnarray}
  N=2&:&M_{(1}M_{2)}={\textstyle \frac{1}{2}}{\rm tr}(M_1M_2){\bf 1}_{2\times 2}\\
  N=3&:&M_{(1}M_2M_{3)}={\textstyle \frac{1}{2}}{\rm tr}(M_{(1}M_2)M_{3)}
  +{\textstyle \frac{1}{3}}{\rm tr}(M_{(1}M_2M_{3)}){\bf 1}_{3\times 3}
  \nonumber\\
  N=4&:&M_{(1}M_2M_3M_{4)}={\textstyle \frac{1}{2}}{\rm tr}(M_{(1}M_2)M_3M_{4)}
  +{\textstyle \frac{1}{3}}{\rm tr}(M_{(1}M_2M_3)M_{4)}\nonumber\\
  &&\hspace*{3cm} 
  +{\textstyle \frac{1}{4}}{\rm tr}(M_{(1}M_2M_3M_{4)}){\bf 1}_{4\times 4}
  -{\textstyle \frac{1}{2\cdot 2\cdot 2!}}
  {\rm tr}(M_{(1}M_2){\rm tr}(M_3M_{4)}){\bf 1}_{4\times 4}\nonumber\\
  N=5&:&M_{(1}\cdots M_{5)}=-{\textstyle \frac{1}{2\cdot 2\cdot 2!}}
  {\rm tr}(M_{(1}M_2){\rm tr}(M_3M_4)M_{5)}
  -{\textstyle \frac{1}{2\cdot 3}}{\rm tr}(M_{(1}M_2){\rm tr}(M_3M_4M_{5)}){\bf 1}_{5\times 5}\nonumber\\
  &&\hspace*{2.6cm}+{\textstyle \frac{1}{2}}{\rm tr}(M_{(1}M_2)M_3M_4M_{5)}
  +{\textstyle \frac{1}{3}}{\rm tr}(M_{(1}M_2M_3)M_4M_{5)}\nonumber\\
  &&\hspace*{2.6cm}+{\textstyle \frac{1}{4}}{\rm tr}(M_{(1}M_2M_3M_4)M_{5)}
  +{\textstyle \frac{1}{5}}{\rm tr}(M_{(1}\cdots M_{5)}){\bf 1}_{5\times 5}\ .
  \nonumber
\end{eqnarray}
The factors on the right-hand sides are the symmetry factors for the cyclicity of 
matrices inside a trace and also for exchanging identical single-trace operators. The 
sign is $-1$ if the term involves even number of single-trace operators. 
Inserting this expression into (\ref{scalar-primary})
with $n\geq N+1$, many times if necessary, the single trace operators can be written 
as sums of products of operators of the form (\ref{scalar-primary}) with $n\leq N$. 
In other words, the chiral primaries of $S_{n\geq N+1}$ can be written in terms 
of those of $S_{n\leq N}$. Acting $Q^m_+$, $\overline{Q}_{m\dot\alpha}$, 
$P_{+\dot\alpha}$ to these expressions, all cohomologies in $S_{n\geq N+1}$ can be 
written in terms of those in $S_{n\leq N}$. So finite $N$ gravitons can be constructed 
with the single-trace generators in $S_{n\leq N}$.

Even if the single-trace generators $S_{n\leq N}$ are independent in the sense that 
none of them can be written as multi-trace operators, still there are new trace relations 
among multi-traces operators after multiplying them. 
For example, consider the $SU(2)$ theory and the following primaries 
of the $S_2$ multiplet:
\begin{equation}\label{primary-S2}
  {\rm tr}(\bar\phi^{(m}\bar\phi^{n)})\sim
  {\rm tr}(X^2)\ ,\ {\rm tr}(Y^2)\ ,\ {\rm tr}(Z^2)\ ,\
  {\rm tr}(XY)\ ,\ {\rm tr}(YZ)\ ,\ {\rm tr}(ZX)\ .
\end{equation}
Here and below, we shall often use the notation 
$(X,Y,Z)\equiv(\bar\phi^1,\bar\phi^2,\bar\phi^3)$.
Since we consider cohomologies, we study the trace relations  
up to $Q$-exact terms. One finds the following relations at the double trace order, 
\begin{eqnarray}
  &&{\rm tr}(X^2){\rm tr}(Y^2)-[{\rm tr}(XY)]^2\sim {\rm tr}([X,Y][X,Y])\sim
  Q{\rm tr}(\psi_3[X,Y])\\
  &&{\rm tr}(X^2){\rm tr}(YZ)-{\rm tr}(XY){\rm tr}(XZ)\sim
  {\rm tr}([X,Y][X,Z])\sim Q{\rm tr}(\psi_3[X,Z]-[X,Y]\psi_2)\ ,\nonumber
\end{eqnarray}
and $4$ more relations obtained by permuting $X,Y,Z$.
More covariantly, one can write them as
\begin{equation}
  \epsilon_{mab}\epsilon_{ncd}{\rm tr}(\bar\phi^{(a}\bar\phi^{c)})
  {\rm tr}(\bar\phi^{(b}\bar\phi^{d)})=Q\textrm{-exact}\ ,
\end{equation}
in the representation $\bar{\bf 6}$, or $[0,2]$, of $SU(3)$. They are the primaries 
of a constraint supermultiplet of $PSU(1,2|3)$, contained in $B\overline{B}[0;0]_4^{[2,0,2]}$ 
of $PSU(2,2|4)$. These constraints are not independent in the sense that 
there are `relations of relations' at higher orders. See, for instance, \cite{Grant:2008sk} 
on these relations of relations in special sectors. Unfortunately, we do not 
know a simple algorithm to identify all such trace relations 
in terms of the generators.

The strategy we explained so far, using single-trace generators, is analogous or dual to the 
 counting of BPS states  from giant gravitons (D3-branes expanded in $S^5$) . There is another 
way of counting graviton-like BPS states using dual giant gravitons, which are D3-branes
expanded in $AdS_5$. In the field theory, this is dual to counting operators in terms of 
the eigenvalues of diagonal matrices, as we shall explain in a moment. For 
the chiral primaries (\ref{scalar-primary}), the two complementary methods are fully 
explored and yield the same result \cite{Biswas:2006tj,Mandal:2006tk}. For  
general graviton-like BPS states, neither method has been fully developed. In a sense, 
we shall now provide a hybrid method of the two.

The multiplet $B\overline{B}[0;0]^{[0,n,0]}_n$ for single-particle gravitons 
is absolutely protected, meaning that it cannot combine with 
other short multiplets to form a long multiplet. BPS operators in this 
multiplet never acquire anomalous dimensions as the coupling changes. Therefore, 
one may count the graviton cohomologies in the free limit $g_{\rm YM}\rightarrow 0$.

We start by considering the chiral primaries (\ref{scalar-primary}). 
Since all the scalars are symmetrized in the trace, we can regard all the fields 
as diagonal matrices for the purpose of enumerating cohomologies. Each scalar contains 
$N-1$ eigenvalues, or $N$ eigenvalues whose sum is zero. The generators (\ref{scalar-primary}) are 
Weyl-invariant polynomials of the eigenvalues. Then we consider the superconformal  
descendants of (\ref{scalar-primary}) in the free theory limit. In free theory, supersymmetry 
transformation of diagonal fields only involves diagonal components of the superpartner 
fields. So we can restrict all the elementary fields appearing in $S_{n\leq N}$ to be 
diagonal. The covariant derivatives on the fields also reduce to ordinary 
derivatives since $g_{\rm YM}=0$. 
In particular, trace relations are now relations of eigenvalue polynomials, 
including the ordinary derivatives acting on the eigenvalues.  
So we should count the polynomials which can be written as sums of products of 
the generators in $S_{n\leq N}$. The reduction to the polynomial counting will ease 
numerical studies, and sometimes will admit analytic counting.

This procedure becomes simpler in the $SU(2)$ theory, because each
elementary field is represented by one eigenvalue. We denote by 
$x,y,z,\psi_m,\lambda_\alpha,f$ the eigenvalues of the fields 
$X,Y,Z,\psi_{m+},\bar\lambda_{\dot\alpha},f_{++}$, respectively,
e.g. $X = {\rm diag}(x,-x)$. The BPS derivatives 
$D_{+\dot\alpha}$ can also be replaced by ordinary derivatives, 
which we call $\partial_\alpha$. We should consider polynomials of these 
eigenvalues and derivatives acting on them, 
\begin{equation}
  \partial_+^{n_1}\partial_-^{n_2}(x,y,z,\psi_m,f)\ \ ,\ \ 
  n_1\partial_+^{n_1-1}\partial_-^{n_2}\lambda_++n_2\partial_+^{n_1}\partial_-^{n_2-1}\lambda_-
\end{equation}
with $n_1,n_2\geq 0$.
In order for a polynomial to be a BPS graviton,
it must be a sum of products of the generators in $S_2$.
In terms of eigenvalues, those generators are given by
\begin{eqnarray}
  &&\partial_+^{n_1}\partial_-^{n_2}(x^2,\ y^2,\ z^2,\ xy,\ yz,\ zx)\\
  &&\partial_+^{n_1}\partial_-^{n_2}
  (y\psi_1,\ z\psi_1,\ x\psi_2,\ z\psi_2,\ x\psi_3,\ y\psi_3,\
  x\psi_1-y\psi_2,\ y\psi_2-z\psi_3)\ \nonumber\\
  &&\partial_+^{n_1}\partial_-^{n_2}(x\lambda_\pm,\ y\lambda_\pm,\ z\lambda_\pm)
  \ \ ,\ \ {\textstyle 
  \partial_+^{n_1}\partial_-^{n_2}(xf-\frac{1}{2}\psi_2\psi_3,yf-\frac{1}{2}\psi_3\psi_1,zf-\frac{1}{2}\psi_1\psi_2)}\nonumber\\
  &&\partial_+^{n_1}\partial_-^{n_2}(\psi_1\lambda_\pm+2y\partial_\pm z,
  \psi_2\lambda_\pm+2z\partial_\pm x,\psi_3\lambda_\pm+2x\partial_\pm y)\nonumber\\
  &&\partial_+^{n_1}\partial_-^{n_2}(\lambda_+\lambda_-)\ ,\ \ 
  {\textstyle 
  \partial_+^{n_1}\partial_-^{n_2}(\lambda_\pm f-\frac{2}{3}\psi_m\partial_\pm \phi^m+\frac{1}{3}\phi^m\partial_\pm \psi_m)}\ .\nonumber
\end{eqnarray}
If a polynomial is expressed as a sum of the products of these generators, 
it must be invariant under the Weyl reflection of the 
eigenvalues:
\begin{equation}
  (x,y,z,\psi_m,\lambda_\alpha,f)\rightarrow -(x,y,z,\psi_m,\lambda_\alpha,f)\ .
\end{equation}
In other words it must be an even polynomial.
We used this approach to numerically count the $SU(2)$ graviton operators till certain 
order in the charge expansion. The studies of section 3.2 will be based on this calculation.
In a simple subsector of the $SU(2)$ theory, we can analytically count independent 
graviton cohomologies using this approach as explained in Appendix C.

\section{The black hole cohomologies for $SU(2)$}

\cite{Chang:2022mjp} systematically constructed all cohomologies till certain order.
We shall employ a more streamlined approach in this section. We 
will first compute the index over the black hole cohomologies by subtracting 
the contributions from finite $N$ gravitons. 
Then with some guesses if necessary, we shall explicitly 
construct the black hole cohomologies which account for this index.  
We find that this approach is much more efficient in detecting new cohomologies.

The index of the $\mathcal{N}=4$ Yang-Mills theory is defined as 
\cite{Romelsberger:2005eg,Kinney:2005ej}
\begin{equation}\label{index}
  Z(\Delta_I,\omega_i)={\rm Tr}\left[(-1)^Fe^{-\sum_{I=1}^3\Delta_I R_I}
  e^{-\sum_{i=1}^2\omega_i J_i}\right]
\end{equation}
where the chemical potentials should satisfy $\sum_I\Delta_I-\sum_i\omega_i=0$ 
for this quantity to be an index. For the $SU(N)$ theory, this is given by
the following integral
\begin{equation}\label{index-integral}
  Z(\Delta_I,\omega_i)=
  \frac{1}{N!}\int_0^{2\pi}\prod_{a=1}^N\frac{d\alpha_a}{2\pi}
  \exp\left[\sum_{a\neq b}\sum_{n=1}^\infty\frac{f(n\Delta_I,n\omega_i)-1}{n}
  e^{in(\alpha_a-\alpha_b)}\right]
  \exp\left[(N\!-\!1)\sum_{n=1}^\infty\frac{f(n\Delta_I,n\omega_i)}{n}\right]\ ,
\end{equation}
where $f(\Delta_I,\omega_i)\equiv 1-\frac{(1-e^{-\Delta_1})(1-e^{-\Delta_2})
(1-e^{-\Delta_3})}{(1-e^{-\omega_1})(1-e^{-\omega_2})}$ is the single letter index. 
Since this index is independent of the coupling $g_{\rm YM}$, 
we can regard it as the index over our 1-loop cohomologies.

\subsection{The BMN sector}

The radially quantized QFT lives on $S^3\times\mathbb{R}$. 
The fields are expanded in spherical harmonics of $SO(4)$.
It was shown in \cite{Kim:2003rza} that the \textit{classical} $\mathcal{N}=4$ 
Yang-Mills theory has a consistent truncation which keeps finite degrees of freedom, 
described by the BMN matrix model \cite{Berenstein:2002jq}. 
The modes kept after the truncation are given by: 
(1) s-wave modes $\phi_m(t)$, $\bar\phi^m(t)$ of the scalars, 
(2) lowest spinor harmonics modes $\psi_{m\alpha}(t)$, 
$\lambda_\alpha(t)$ (the spinor indices 
are defined using the labels of Killing spinor fields \cite{Kim:2003rza}), (3) vector 
potential 1-form restricted to $A=A_0(t)dt+A_i(t)\sigma_i$ where $\sigma_i$ with 
$i=1,2,3$ are the right-invariant 1-forms on $S^3$ in our 
convention. This is a 
consistent truncation of the nonlinear equations of motion, and not a quantum 
reduction in any sense. So the full quantum BMN theory is 
a priori unrelated to the 4d Yang-Mills theory. However, since our 1-loop cohomology problem 
uses classical supercharge $Q$ only, it can be truncated to the BMN model. If 
the conjecture of \cite{minwalla} is true, the whole BPS cohomology problem would 
have a quantum truncation to this model.

In general, the BMN theory and the full Yang-Mills theory behave differently in many ways. 
The difference starts from the number of ground states. The Yang-Mills 
theory on $S^3\times\mathbb{R}$ has a unique vacuum, while the BMN model has 
many ground states labeled semiclassically by the discrete values of $A_i$. 
In the quantum BMN theory, viewed as M-theory in the plane wave background, these ground 
states describe various M2/M5-brane configurations with zero lightcone energies 
\cite{Maldacena:2002rb}. In the Yang-Mills theory, however, there are large gauge
transformations on $S^3$ which can gauge away these ground states to $A_i=0$. 
So if one wishes to study the Yang-Mills theory using this matrix model, it suffices to 
consider the physics around $A_i=0$.

Recall that our cohomology problem is completely classical, using the classical
supercharge $Q$ at half-loop order. Therefore, this problem should have a truncation to 
the BMN matrix model. This turns out to be the cohomology problem defined using
\begin{equation}\label{letter-bmn}
  \bar\phi^m\ ,\ \ \psi_{m+}\ ,\ \ f_{++}\ ,
\end{equation}
without using any gauginos $\bar\lambda_{\dot\alpha}$ or derivatives $D_{+\dot\alpha}$.
These operators close by the action of $Q$: 
$[Q,\bar\phi^m]=0$, $\{Q,\psi_{m+}\}=-i\epsilon_{mnp}[\bar\phi^m,\bar\phi^n]$, 
$[Q,f_{++}]=-i[\psi_{m+},\bar\phi^m]$. So it is possible to restrict the cohomology 
problem by using operators constructed using the letters (\ref{letter-bmn}).
Note that the truncation is also applied to the operator $\Lambda$ when one 
identifies two operators $O_1$ and $O_2$ related as
$O_2-O_1=[Q,\Lambda\}$. This is why the gauginos $\bar\lambda_{\dot\alpha}$ cannot be 
included in this truncation. Although it is $Q$-closed by itself, $\bar\lambda_{\dot\alpha}$ 
can be obtained by acting $Q$ on the covariant derivative, 
$[Q,D_{+\dot\alpha}]=-i[\bar\lambda_{\dot\alpha},\ \}$.
So if one had tried to include $\bar\lambda_{\dot\alpha}$ into the truncation and construct 
operators like $O_1$, $O_2$, $\Lambda$, one may incorrectly conclude that certain 
$O_1$ and $O_2$ are different by not including derivatives in $\Lambda$. This truncation of 
the cohomology problem was known in \cite{Chang:2022mjp,Choi:2022caq}, 
although the relation to the BMN truncation was not explicitly addressed.\footnote{We thank 
Nakwoo Kim for first pointing this out to us.} Notice also that this truncation is not 
kinematic, i.e. cannot be inferred without knowing the dynamical information of 
the classical theory. 

The BMN truncation is the $SU(2)_R$ invariant truncation. 
In our cohomology problem, this means that no ingredients include the $\dot\alpha$ 
indices for $SU(2)_R$. This is why $\bar\lambda_{\dot\alpha}$ and $D_{+\dot\alpha}$ 
are excluded. Similarly, in the representation theory, only a small subset of 
$PSU(1,2|3)$ generators can be used to generate a multiplet. Among the $PSU(1,2|3)$ 
generators $Q^m_+$, $\overline{Q}_{m\dot\alpha}$ and $P_{+\dot\alpha}$, only the 
three supercharges $Q^m_+$ which belong to $SU(1|3)$ act 
within BMN cohomologies.

We also consider the index over the BMN cohomologies. Keeping the letters 
(\ref{letter-bmn}) only, the letter index is given by
\begin{equation}
  \tilde{f}=\sum_{I=1}^3e^{-\Delta_I}-\sum_{I<J}e^{-\Delta_I-\Delta_J}
  +e^{-\Delta_1-\Delta_2-\Delta_3}=
  1-(1-e^{-\Delta_1})(1-e^{-\Delta_2})(1-e^{-\Delta_3})\ .
\end{equation}
The first three terms come from $\bar\phi^m$, the next three terms from 
$\psi_{m+}$, and the last term from $f_{++}$. $\Delta_I$ is the chemical 
potential conjugate to $R_I+\frac{J_1+J_2}{2}\equiv R_I+J$. The fourth 
chemical potential $\omega_1-\omega_2$ of (\ref{index}) does 
not appear since BMN truncation is $SU(2)_R$ invariant.
The matrix integral expression for the BMN index is 
\begin{eqnarray}\label{index-bmn}
  Z(\Delta_I)&=&
  \frac{1}{N!}\int_0^{2\pi}\prod_{a=1}^N\frac{d\alpha_a}{2\pi}
  \exp\left[\sum_{a\neq b}\sum_{n=1}^\infty\frac{\tilde{f}(n\Delta_I)-1}{n}
  e^{in(\alpha_a-\alpha_b)}\right]
  \exp\left[(N\!-\!1)\sum_{n=1}^\infty\frac{\tilde{f}(n\Delta_I)}{n}\right]\nonumber\\
  &=&\frac{1}{N!}\int_0^{2\pi}\prod_{a=1}^N\frac{d\alpha_a}{2\pi}\cdot 
  \prod_{a\neq b}\frac{(1-e^{i\alpha_{ab}})\prod_{I<J}
  (1-e^{-\Delta_I-\Delta_J}e^{i\alpha_{ab}})}
  {(1-e^{-\Delta_1-\Delta_2-\Delta_3}e^{i\alpha_{ab}})
  \prod_{I=1}^3(1-e^{-\Delta_I}e^{i\alpha_{ab}})}\nonumber\\
  &&\hspace*{3cm}\cdot\left[\frac{\prod_{I<J}
  (1-e^{-\Delta_I-\Delta_J})}{(1-e^{-\Delta_1-\Delta_2-\Delta_3})
  \prod_{I=1}^3(1-e^{-\Delta_I})}\right]^{N-1}
\end{eqnarray}
where $\alpha_{ab}\equiv\alpha_a-\alpha_b$.
This expression cannot be obtained from the original index (\ref{index-integral}) 
by taking limits of the chemical potentials, because the BMN truncation is 
not kinematic. To ease the calculations, 
one may replace the Haar measure part of the integrand by half of it 
\cite{Hanany:2008sb},
\begin{equation}\label{half-haar}
  \frac{1}{N!}\prod_{a\neq b}(1-e^{i\alpha_{ab}})\ \rightarrow\ 
  \prod_{a<b}(1-e^{i\alpha_{ab}})\ ,
\end{equation}
without changing the integral.

(\ref{index-bmn}) can be computed easily for low $N$'s because the 
integrand is a finite product. One can evaluate it by a residue sum. 
For instance, for the $SU(2)$ index, the only nontrivial integral variable is 
$e^{i\alpha_{12}}$ along the unit circle. We evaluate 
the integral by the residue sum by shrinking this unit circle.
After the replacement (\ref{half-haar}) in the integrand, we should sum over four 
residues at the following poles:
\begin{equation}
  e^{i\alpha_{12}}=e^{-\Delta_1}\ ,\ \ e^{-\Delta_2}\ ,\ \ e^{-\Delta_3}\ ,\ \ 
  e^{-\Delta_1-\Delta_2-\Delta_3}\ .
\end{equation}
We have a general result, but just to illustrate it at 
$e^{-\Delta_1}=e^{-\Delta_2}=e^{-\Delta_3}\equiv t^2$, one obtains
\begin{eqnarray}
  Z&=&\left[\frac{}{}\!\!\right.
  1+\!3t^2+\!12t^4 + \!20t^6 + \!42t^8 + \!48t^{10} + \!75t^{12} + \!66t^{14} 
  + \!81t^{16} \\
  &&\hspace*{.3cm}+ \!55t^{18} + \!54t^{20} + \!27t^{22} + \!19t^{24} + \!6t^{26} + 
  \!3t^{28}\left.\!\!\frac{}{}\right]\frac{(1-t^2)^3}{ (1-t^{12})(1 - t^8 )^3 }\ .
  \nonumber
\end{eqnarray}
This index contains both black hole and graviton cohomologies 
made with (\ref{letter-bmn}).

To compute the index over black hole cohomologies in this sector, 
we need to enumerate the graviton cohomologies and subtract 
their index from the full index. This can be done analytically 
for the $SU(2)$ gravitons in the BMN sector, 
employing the eigenvalue counting explained in section 2. 
We need to consider even polynomials of the following seven eigenvalues:
\begin{equation}
  x\ ,\ y\ ,\ z\ ,\ f\ ,\ \psi_1\ ,\ \psi_2\ ,\ \psi_3\ ,
\end{equation}
where one should remember that the last three are Grassmannian numbers. We 
can write all possible even monomials of these eigenvalues, and find the 
combinations which can be written as sums of products of the following BMN
generators in $S_2$:
\begin{eqnarray}\label{bmn-SU(2)-generator}
  {\bf 6}&:& x^2\ ,\ y^2\ ,\ z^2\ ,\ xy\ ,\ yz\ ,\ zx\\
  {\bf 8}&:&\psi_1\cdot(y,z)\ ,\ \psi_2\cdot(z,x)\ ,\ \psi_3\cdot(x,y)\ ,\ 
  \psi_1 x-\psi_2 y\ ,\ \psi_2 y-\psi_3 z\nonumber\\
  {\bf 3}&:&xf-\textstyle{\frac{1}{2}}\psi_2\psi_3\ ,\ yf-\textstyle{\frac{1}{2}}\psi_3\psi_1\ ,\ zf-\textstyle{\frac{1}{2}}\psi_1\psi_2\ .
  \nonumber
\end{eqnarray}
Counting these polynomials in the right order, one can count them basically 
at the monomial levels. The procedure 
is explained in Appendix B. We can compute the full partition function for 
these cohomologies. Giving $-1$ weights to fermions and unrefining some fugacities, 
we can compute the index $Z_{\rm grav}(\Delta_I)$ over gravitons.

Subtracting this from the full index, 
the difference between $Z$ and $Z_{\rm grav}$ is given by 
\begin{equation}\label{bh-index-bmn}
  Z-Z_{\rm grav}=-\frac{e^{-4(\Delta_1+\Delta_2+\Delta_3)}}{1-e^{-2(\Delta_1+\Delta_2+\Delta_3)}}
  \cdot \prod_{I=1}^3(1-e^{-\Delta_I})\cdot 
  \prod_{I=1}^3\frac{1}{(1-e^{-\Delta_I}e^{-\Delta_1-\Delta_2-\Delta_3})}\ .
\end{equation}
Unrefining as $e^{-\Delta_I}=t^2$, 
one obtains a series in $t$ given by
\begin{equation}
  Z-Z_{\rm grav}=-t^{24}+3t^{26}-3t^{28}+t^{30}-3t^{32}+\cdots\ .
\end{equation}
$t$ is conjugate to  $j\equiv 6(R+J)$, where 
$R\equiv\frac{R_1+R_2+R_3}{3}$, $J\equiv\frac{J_1+J_2}{2}$.
From this formula, one finds the first black hole cohomology at 
$j=24$. This `threshold' black hole cohomology was already identified 
in \cite{Chang:2022mjp,Choi:2022caq}, as we shall review and rewrite in a more compact 
form below. It may look like there are many black hole states beyond this threshold, 
but most of them are rather trivial. To make this point clear, we would like 
to first interpret various factors of  
(\ref{bh-index-bmn}), which will be extensively justified later.

(\ref{bh-index-bmn}) is a multiplication of three factors, divided by the $\cdot$ products.
We interpret the first factor as the `core' black hole primary operators. Constructing this 
part of the cohomologies will be the goal of this subsection. The second factor 
comes from the $SU(1|3)$ descendants obtained from the first factor by acting $Q^m_+$. 
The supercharge $Q^m_+$ carries charges $R_I=\delta_{I,m}-\frac{1}{2}$ and $J=\frac{1}{2}$, so is 
weighted by  $e^{-\Delta_I}$. So the second factor comes from the Fock space 
obtained by acting three $Q^m_+$'s. Finally, the third factor comes from the multiplications of 
certain multi-gravitons to the core black hole cohomologies. Among the $17$ graviton states listed in  
(\ref{bmn-SU(2)-generator}), only $3$ types on the third line can contribute. 
The remaining $14$ gravitons of (\ref{bmn-SU(2)-generator}) multiplying the core 
black hole operators do not appear in the index. This aspect will be discussed further 
in section 3.2.

For $SU(2)$, the BMN index (\ref{bh-index-bmn}) 
does not show very large entropy even at large charges. From an inspection 
of the integral (\ref{index-bmn}) and having in mind applying the techniques developed in 
\cite{Choi:2021lbk}, we expect that at large enough $N$ and charges one will 
be able to show that the entropies will scale like those for black holes.

Now we construct the cohomologies that will account for the first factor 
of (\ref{bh-index-bmn}),
\begin{equation}\label{index-core-primary}
  -\frac{t^{24}}{1-t^{12}}=-t^{24}-t^{36}-t^{48}-t^{60}-\cdots\ ,
\end{equation}
where $t^6=e^{-\Delta_1-\Delta_2-\Delta_3}$. The index predicts unique fermionic 
cohomology at $j=24+12n$ ($n=0,1,2,\cdots$), all singlets of
$SU(3)\subset SU(4)$.  For $SU(2)$ gauge group, we use the 3-dimensional vector notation for 
the adjoint fields. In the remaining part of this subsection, 
$\phi^m=(X,Y,Z)$, $\psi_{m},f$ would mean 3 dimensional vectors, and 
inner/outer products will replace the trace/commutators. 
The $Q$-transformations of these $3$-vectors are given by
\begin{equation}\label{Q-bmn-SU(2)}
  Q\phi^m=0\ ,\ \ Q\psi_m={\textstyle \frac{1}{2}}\epsilon_{mnp}\phi^n\times \phi^p
  \ ,\ Qf=\phi^m\times\psi_m\ .
\end{equation}
See \cite{Choi:2022caq} and appendix A for our normalization.

\hspace*{-0.65cm}\underline{\bf $O_0$ operator at $t^{24}$}

\hspace*{-0.65cm}This operator has charges 
$E=\frac{19}{2}$, $R_1=R_2=R_3=\frac{3}{2}$, $J_1=J_2=\frac{5}{2}$. 
A representative of this cohomology \cite{Choi:2022caq} is given by
\begin{eqnarray}\label{n=0-old}
  O_0^\prime&=&(X\cdot \psi_1-Y\cdot \psi_2)(X\cdot \psi_3)(\psi_2\cdot \psi_1\times\psi_1)
  +(Y\cdot \psi_2-Z\cdot \psi_3)(Y\cdot \psi_1)(\psi_3\cdot \psi_2\times\psi_2)\nonumber\\
  &&+(Z\cdot \psi_3-X\cdot \psi_1)(Z\cdot \psi_2)(\psi_1\cdot \psi_3\times\psi_3)\ .
\end{eqnarray} 
Note that the second and third terms are obtained by making cyclic permutations 
of $(X,\psi_1)$, $(Y,\psi_2)$, $(Z,\psi_3)$ on the first term. The cyclic permutations 
are part of the $SU(3)$ symmetry, thus symmetries of the cohomology problem,
On the other hand, odd permutations accompanied by the sign flips of all 
$\psi_m$'s and $\phi^m$'s are part of $SU(4)\times SU(2)_L$ symmetry 
which leave $Q$ invariant, thus being symmetries of the cohomology problem. 
To construct a better representative of this cohomology, consider the following operator 
obtained by permuting $(X,\psi_1)\leftrightarrow(Y,\psi_2)$ and flipping signs of all 
$\phi^m,\psi_m$ on (\ref{n=0-old}):
\begin{eqnarray} 
  O_0^{\prime\prime}&=&(X\cdot \psi_1-Y\cdot \psi_2)(Y\cdot \psi_3)(\psi_1\cdot \psi_2\times\psi_2)
  +(Y\cdot \psi_2-Z\cdot \psi_3)(Z\cdot \psi_1)(\psi_2\cdot \psi_3\times\psi_3)\nonumber\\
  &&+(Z\cdot \psi_3-X\cdot \psi_1)(X\cdot \psi_2)(\psi_3\cdot \psi_1\times\psi_1)\ .
\end{eqnarray}
One can show 
\begin{eqnarray}\label{n=0}
  O_0^\prime-O_0^{\prime\prime}&=&-2Q[(\psi_1\cdot\psi_2)(\psi_2\cdot\psi_3)(\psi_3\cdot\psi_1)]
  \ ,\\
  O_0&\equiv&-5(O_0^\prime+O_0^{\prime\prime})=
  \epsilon^{p_1p_2p_3}v^{m}_{\ \ p_1}v^{n}_{\ \ p_2}(\psi_m\cdot\psi_n\times \psi_{p_3})\ ,
  \nonumber
\end{eqnarray}
where 
\begin{equation}
  v^m_{\ \ n}\equiv (\phi^m\cdot\psi_n)-{\textstyle \frac{1}{3}}\delta^m_n(\phi^p\cdot\psi_p)
\end{equation}
are graviton cohomologies in the $S_2$ multiplet. $O_0$ is manifestly an $SU(3)$ singlet. 
Note that the second term of $v$ proportional to $\delta^m_n$ drops out when $v$ is
inserted into (\ref{n=0}), because of the symmetry of $\psi_m\cdot\psi_n\times\psi_{p_3}$ 
and the antisymmetry of $\epsilon^{p_1p_2p_3}$. So we can write
\begin{equation}\label{n=0-alternative}
  O_0=\epsilon^{p_1p_2p_3}(\phi^m\cdot \psi_{p_1})(\phi^n\cdot \psi_{p_2})
  (\psi_m\cdot\psi_n\times \psi_{p_3})\ .
\end{equation}

To show that $O_0$ is a black hole cohomology, one should check 
that it is $Q$-closed, not $Q$-exact, and not of graviton type. 
The first and third are trivial. $O_0$ is not graviton-like 
because it consists of seven (odd) letters: since $SU(2)$ gravitons are 
made of operators in $S_2$, they always have an even number of letters. 
To check $Q$-closedness, first note that $Q$ acts only 
on $\psi_m\cdot\psi_n\times \psi_{p_3}$ because 
$v^m_{\ \ n}$ are $Q$-closed. One finds
\begin{equation}\label{Q-psi-3}
  Q(\psi_m\cdot \psi_n\times \psi_p)={\textstyle \frac{3}{2}}
  \epsilon_{(m|qr}(\phi^q\times \phi^r)\cdot(\psi_{|n}\times\psi_{p)})
  =3\epsilon_{(m|qr}(\phi^q\cdot\psi_{|n})(\phi^r\cdot \psi_{p)})=
  3\epsilon_{(m|qr}v^{q}_{\ \ |n}v^{r}_{\ \ p)}\ .
\end{equation}
At the last step, the second term of $\phi^q\cdot\psi_n=v^q_{\ \ n}+\delta^q_n(\cdots)$ 
etc. does not survive after the index contractions. Inserting it to $QO_0$ and
replacing the product of two $\epsilon$'s by three $\delta$'s, $QO_0$ is given by 
various row/column contractions of 
four $3\times 3$ traceless matrices $v^m_{\ \ n}$. Possible terms are
${\rm tr}(v^4)$ and ${\rm tr}(v^2){\rm tr}(v^2)$, but the fermionic nature of 
$v$ and the cyclicity of trace ensure that they are all zero. 
So $QO_0$ is zero because there are no nonzero terms that can contribute.

The non-$Q$-exactness was originally shown after a calculation using computer
\cite{Chang:2022mjp,Choi:2022caq}.
Here we provide an analytic argument
by studying the $SU(1|3)$ descendants obtained by acting $Q_+^a Q_+^b$. 
For instance, one obtains
\begin{eqnarray}
  &&Q_+^2Q_+^1 O_0^\prime=\\
  &&-(Y\cdot f+\psi_3\cdot\psi_1)^2\psi_3\cdot(\psi_2\times\psi_2)
 -(X\cdot f+\psi_2\cdot\psi_3)(Z\cdot f+\psi_1 \cdot \psi_2)\psi_1\cdot(\psi_3\times\psi_3)\nonumber\\
 &&-(X\cdot f+\psi_2\cdot\psi_3)(X\cdot \psi_3)f \cdot(\psi_1\times\psi_1)
 +2(Y\cdot\psi_2-Z\cdot\psi_3)(Y\cdot f+\psi_3\cdot\psi_1)\psi_3\cdot(\psi_2\times f)
 \nonumber\\
  &&-2(Y\cdot f+\psi_3\cdot \psi_1)(X\cdot \psi_3)\psi_2\cdot(\psi_1\times f)
-(Z\cdot\psi_3-X\cdot\psi_1)(Z\cdot f+\psi_1\cdot \psi_2)f\cdot(\psi_3\times\psi_3)
\nonumber
\end{eqnarray}
which contains uncanceled $\phi^0\psi^7$ terms on the second line.
Since acting $Q$ always creates one or more $\phi$ factors, 
these terms cannot be $Q$-exact. Since a descendant of $O_0^\prime$
is not $Q$-exact, $O_0$ cannot be $Q$-exact either, providing a simple proof. 
Or alternatively, one can prove non-$Q$-exactness by acting three $Q_+$'s to $O_0^\prime$ 
and check that it contains nonzero term at $f\psi^6$ order,
\begin{eqnarray}\label{QQQ-on-O0}
  &&Q^1_+ Q^2_+ Q^3_+ O_{0}^\prime=Q^1_+Q^2_+Q^3_+O_0^{\prime\prime}=\\
  &&(X\cdot f+\psi_2\cdot\psi_3)^2 f \cdot(\psi_1\times\psi_1)
  +2(X\cdot f+\psi_2\cdot\psi_3)(Y\cdot f+\psi_3\cdot\psi_1) f\cdot(\psi_1\times\psi_2) 
  \nonumber\\
  &&+(1,2,3\rightarrow 2,3,1)+(1,2,3\rightarrow 3,1,2)=
  G^mG^n f\cdot(\psi_m\times\psi_n)\nonumber
\end{eqnarray}
where $G^m\equiv\phi^m\cdot f+\frac{1}{2}\epsilon^{mnp}\psi_n\cdot\psi_p$.
Proof of this sort will sometimes be useful later. For instance, one can show that 
$(Z\cdot f+\psi_1\cdot\psi_2)O_0^\prime$ is not $Q$-exact, since 
its descendant
\begin{equation}\label{QQ-on-t36}
  Q^2_+Q^1_+\left[(Z\cdot f+\psi_1\cdot\psi_2)O_0^\prime\right]
  =(Z\cdot f+\psi_1\cdot\psi_2)Q^2_+Q^1_+O_0^\prime
\end{equation}
contains a term at $\phi^0\psi^9$ order.

\hspace*{-0.65cm}\underline{\bf $O_1$ operator at $t^{36}$}

\hspace*{-0.65cm}Now we construct the cohomology which accounts for the $-t^{36}$ term 
of (\ref{index-core-primary}). It should be fermionic, has charge 
$j=6(R+J)=36$, and should be an $SU(2)_R\times SU(3)$ singlet because we expect unique cohomology
(unless there is a cancellation at this order which obscures the true degeneracy). 
We call this  operator $O_1$.
From the last condition, we set three $R_I$ equal and two $J_i$ equal.
Still, we do not know the individual $R$ and $J$ so we should make a guess. 
Our first guess was to add extra $\Delta J=2$ to the charges $R=\frac{3}{2}$, $J=\frac{5}{2}$ 
of $O_0$. We listed all operators in this sector and found the cohomology 
by computer.
Then we made several trials until we found the following $SU(3)$-invariant 
representative:
\begin{eqnarray}
  O_1&=&(f\cdot f) \epsilon^{c_1c_2c_3}(\phi^a \cdot\psi_{c_1})(\phi^b \cdot \psi_{c_2})(\psi_a\cdot \psi_b\times\psi_{c_3})\\
  &&+\epsilon^{b_1b_2b_3}\epsilon^{c_1c_2c_3}(f\cdot \psi_{b_1})(\phi^a\cdot \psi_{c_1})(\psi_{b_2}\cdot \psi_{c_2})
  (\psi_a\cdot \psi_{b_3}\times \psi_{c_3})\nonumber\\
  &&-{\textstyle \frac{1}{72}}\epsilon^{a_1a_2a_3}\epsilon^{b_1b_2b_3}\epsilon^{c_1c_2c_3}
  (\psi_{a_1}\cdot \psi_{b_1}\times \psi_{c_1})(\psi_{a_2}\cdot \psi_{b_2}\times \psi_{c_2})
  (\psi_{a_3}\cdot \psi_{b_3}\times \psi_{c_3})\ .\nonumber
\end{eqnarray}
It is not graviton type since it is made of 
nine (odd) letters. One can also easily check that it is not $Q$-exact. 
This is because the last term contains no scalars. 
Since $Q$ transformations (\ref{Q-bmn-SU(2)}) always yield
scalars, the last term cannot be made $Q$-exact. So $O_1$ is not $Q$-exact. 

Now we discuss the $Q$-closedness. $O_1$ takes the form of
\begin{equation}
  O_1=(f\cdot f)O_0+f\cdot \xi+\chi\ ,
\end{equation}
where the $SU(2)$ triplet $\vec{\xi}$ and the singlet $\chi$ are given by
\begin{eqnarray}
  \vec{\xi}&=&\epsilon^{b_1b_2b_3}\epsilon^{c_1c_2c_3}\vec{\psi}_{b_1}
  (\phi^a\cdot \psi_{c_1})(\psi_{b_2}\cdot \psi_{c_2})
  (\psi_a\cdot \psi_{b_3}\times\psi_{c_3})\nonumber\\
  \chi&=&-{\textstyle \frac{1}{72}}\epsilon^{a_1a_2a_3}\epsilon^{b_1b_2b_3}\epsilon^{c_1c_2c_3}
  (\psi_{a_1}\cdot \psi_{b_1}\times \psi_{c_1})(\psi_{a_2}\cdot \psi_{b_2}\times \psi_{c_2})
  (\psi_{a_3}\cdot \psi_{b_3}\times \psi_{c_3})\nonumber\\
  &=&-120\psi_1^1\psi_1^2\psi_1^3\psi_2^1\psi_2^2\psi_2^3\psi_3^1\psi_3^2\psi_3^3\ .
\end{eqnarray}
$Q$-closedness is equivalent to the following equations:
\begin{equation}\label{n=1-closed-summary}
  2(\vec{\phi}^m\times\vec{\psi}_m)O_0+Q_\psi\vec{\xi}=0\ \ ,\ \ \vec{\phi}^m\cdot(\vec{\psi_m}\times\vec{\xi})
  +Q_\psi \chi=0\ .
\end{equation}
 Note that $\vec\xi$ is related to 
$O_0$ by
\begin{equation}
  \vec{\xi}=-{\textstyle \frac{1}{2}}\epsilon^{mnp}\vec{\psi}_m
  \psi_n\cdot{\textstyle \frac{\partial}{\partial\phi^p}}O_0\ .
\end{equation}
So the first equation can be written as the following 
equations of $O_0$:
\begin{equation}\label{O1-O0}
  \hspace*{-0.3cm}4(\vec\phi^m\times\vec\psi_m)O_0=\left[(\vec\phi^a\times\vec\phi^b)
  \psi_a\cdot{\textstyle \frac{\partial}{\partial\phi^b}}+\vec\psi_a
 (\phi^a\times\phi^b)\cdot{\textstyle \frac{\partial}{\partial\phi^b}}
 -\vec\psi_{a}(\psi_{b}\times\phi^a)\cdot
 {\textstyle \frac{\partial}{\partial\psi_b}}+
 \vec\psi_{b}(\psi_{a}\times\phi^a)\cdot
 {\textstyle \frac{\partial}{\partial\psi_b}}\right]O_0\ .
\end{equation}
This is a property of $O_0$. 
The second/third terms cancel due to 
$\left(\phi^b\times\frac{\partial}{\partial\phi^b}+
\psi_b\times\frac{\partial}{\partial\psi_b}\right)O_0=0$, which holds 
because it is the $SU(2)$ gauge transformation on a gauge invariant operator $O_0$. 
One can further simplify (\ref{O1-O0}) using various properties of $O_0$. Obvious ones are
\begin{eqnarray}
  &&{\textstyle \phi^m\cdot\frac{\partial}{\partial\phi^m}O_0=n_{\rm B}O_0\ \ ,\ \ 
  \psi_m\cdot\frac{\partial}{\partial\psi_m}O_0=n_{\rm F}O_0}\ \ \ 
  (n_{\rm B},n_{\rm F})=(2,5)\nonumber\\
  &&\vec{\varepsilon}_a^{\ b}\left[{\textstyle \phi^a\cdot\frac{\partial}{\partial\phi^b}
  -\psi_b\cdot\frac{\partial}{\partial\psi_a}}\right]O_0=0\ \ \ (\vec\varepsilon_a^{\ a}=0)\ .
\end{eqnarray}
The first two equations count the numbers of bosonic/fermionic fields in $O_0$. 
The last equation is the $SU(3)$ invariance of $O_0$, which holds for any 
$\varepsilon$. Equivalently, one obtains
\begin{equation}
  {\textstyle \left[\phi^a\cdot\frac{\partial}{\partial\phi^b}
  -\psi_b\cdot\frac{\partial}{\partial\psi_a}\right]O_0=\frac{1}{3}
  (n_{\rm B}-n_{\rm F})}\delta^a_bO_0\ .
\end{equation}
Finally, note that $\delta_{ij}$ contracts the $SU(2)$ gauge triplet indices 
only between boson-fermion pairs in $O_0$, while fermion indices are 
contracted only with $\epsilon_{ijk}$. This effectively promotes $SU(2)\sim SO(3)$ to 
$SL(3)$ within $O_0$, where bosons/fermions transform in the fundamental and 
anti-fundamental representations, respectively. This leads to the following property:
\begin{equation}
  {\textstyle \left[\phi^a_i\cdot\frac{\partial}{\partial\phi^a_j}
  -\psi_a^j\cdot\frac{\partial}{\partial\psi_a^i}\right]O_0=\frac{1}{3}
  (n_{\rm B}-n_{\rm F})}\delta^j_iO_0\ .
\end{equation}
Using these properties, (\ref{O1-O0}) can be written as
\begin{equation}\label{O1-O0-simpler}
  (\vec{\psi}_a\times\vec{\phi}^b)(\phi^a\cdot{\textstyle \frac{\partial}{\partial\phi^b}})
  O_0=({\textstyle 4-\frac{n_{\rm F}+2n_{\rm B}}{3}})(\vec{\phi}^m\times\vec{\psi}_m)O_0=
  (\vec{\phi}^m\times\vec{\psi}_m)O_0\ .
\end{equation}

Both (\ref{O1-O0-simpler}) and the second equation of (\ref{n=1-closed-summary}) can be 
easily checked on a computer. We have no extra analytic insights on why (\ref{O1-O0-simpler}) 
this holds, except that using complicated representation analysis of $SU(2)\times SU(3)$ should 
provide the analytic proof. (We tried to simplify the equation for $O_0$ as much as possible
since they might provide insights on the generalization 
to higher $N$'s in the future.) On the other hand, one can easily prove the second equation of 
(\ref{n=1-closed-summary}). First note that 
$\psi_m\times\xi$ is an $SU(2)$ vector involving $8$ $\psi$'s. 
There are nine independent operators involving eight $\psi$'s, depending on 
which of the $9$ components is lacking. So it is proportional to
$\frac{\partial}{\partial\psi_m^i}\chi$. Since it has to form a gauge-invariant 
by contracting with two scalars $\phi^m_i$, $\phi^a_j$, one should be able to write
$\frac{\delta}{\delta\psi_m^i}\chi$ as
an object with two $SU(3)$ antifundamental and two $SU(2)$ triplet indices by multiplying 
invariant tensors. The only possible term is $\epsilon_{man}\epsilon_{ijk}\frac{\partial}{\partial\psi_n^k}\chi$. One can compute the proportionality constant by computing a 
term, e.g. at $m=1,a=2,i=1,j=2$, finding $-\frac{1}{2}$. So one obtains
\begin{equation}
  \phi^m\cdot(\psi_m\times\xi)=-{\textstyle \frac{1}{2}\phi^{m}_i\phi^a_j
  \epsilon_{man}\epsilon_{ijk}\frac{\partial\chi}{\partial\psi_n^k}}=
  -{\textstyle \frac{1}{2}\epsilon_{man}(\phi^m\times\phi^a)\cdot 
  \frac{\partial\chi}{\partial\psi_n}}
  =-Q_\psi\chi\ ,
\end{equation}
proving the second equation of (\ref{n=1-closed-summary}).

One may wonder if $O_1$ is a descendant of $O_0$, or a lower black hole operator times 
graviton operators appearing in (\ref{bh-index-bmn}). Since $O_{1}$ is at $t^{36}$ order, 
the only possible way of getting operators at this order from $O_0$ is 
$(Q^m_+Q^n_+O_0)(\phi^p \cdot f+\frac{1}{2}\epsilon^{pqr}\psi_q\cdot\psi_r)$. 
However, during our numerical construction of the cohomologies at this order, we separately 
constructed the last operator which is not cohomologous to $O_1$. See also the end of 
this subsection for an analytic proof (applicable to all $O_n$'s with $n\geq 1$).

\hspace*{-0.65cm}\underline{\bf $O_n$ operator at $t^{24+12n}$ ($n\geq 2$)}

\hspace*{-0.65cm}
We can use the structures of the operators $O_0$ and $O_1$ to analytically construct an infinite tower 
of cohomologies $O_n$ accounting for (\ref{index-core-primary}). Consider 
\begin{equation}\label{On}
  O_n\equiv (f\cdot f)^nO_0+n(f\cdot f)^{n-1}f\cdot \xi
  +{\textstyle \frac{2n^2+n}{3}}(f\cdot f)^{n-1}\chi
\end{equation}
for $n\geq 2$. At $n=1$, this is just $O_1$ that we discussed above. 
We will now show that these are new black hole like cohomologies at 
$t^{24+12n}$ order. It is again easy to show that these are not graviton type because
they are made of odd letters. It is not $Q$-exact because the last term does not 
contain scalars.

Now we derive the $Q$-closedness. Its $Q$-action is given by 
\begin{eqnarray}
  QO_n&=&(f\cdot f)^{n-1}\left[\vec{f}\cdot\left(2n(\vec{\phi}^m\times\vec{\psi}_m)O_0
  +nQ_\psi\vec{\xi}\right)+n(\vec{\phi}^m\times\vec{\psi}_m)\cdot\vec{\xi}
  +{\textstyle \frac{2n^2+n}{3}}Q_\psi \chi\right]\\
  &&+2(n^2-n)(f\cdot f)^{n-2}\vec{f}\cdot(\vec{\phi}^m\times\vec{\psi}_m)(f\cdot \xi)
  +{\textstyle \frac{2n(n-1)(2n+1)}{3}}(f\cdot f)^{n-2}\vec{f}\cdot(\vec{\phi}^m\times\vec{\psi}_m)\chi\ .
  \nonumber
\end{eqnarray}
The first two terms on the first line cancel due to the first equation of 
(\ref{n=1-closed-summary}). The last term on the second line is zero because it 
includes $10$ fermions.
Inserting the second equation of (\ref{n=1-closed-summary}) to the last term on 
the first line, one obtains
\begin{equation}\label{Q-On}
  QO_n={\textstyle \frac{2(n^2-n)}{3}}(f\cdot f)^{n-2}
  \left[-(f\cdot f)(\phi^m\times\psi_m)\cdot\xi
  +3(f\times\phi^m)\cdot\psi_m(f\cdot \xi)\right]\ .
\end{equation}
The second term contains $8$ fermions, where the fermions carry $ma$ indices for $SU(3)$ 
and three $SU(2)$ triplet indices to be contracted with 
$(f\times\phi^m)_k$, $f_i$, $\phi^a_j$. From the contraction structures 
of $\xi$, one finds that $b_1,c_1$ are antisymmetric so the 
corresponding $i,j$ indices should be symmetric. The only possible $8$-fermion 
terms satisfying these conditions are
\begin{equation}\label{Q-On-2nd}
  \epsilon_{man}\delta_{ij}\frac{\partial}{\partial \psi_n^k}\chi\ \ ,\ \ 
  \epsilon_{man}\delta_{k(i}\frac{\partial}{\partial \psi_n^{j)}}\chi\ .
\end{equation}
Explicitly computing two components in the second term of (\ref{Q-On}), 
one finds that the linear combination is
\begin{equation}
  \epsilon_{man}\left[\delta_{ij}{\textstyle \frac{\partial}{\partial \psi_n^k}}
  -{\textstyle \delta_{k(i} \frac{\partial}{\partial \psi_n^{j)}} }\right]\chi\ .
\end{equation}
Contracting this with $f_i$, $\phi^a_j$, $(f\times\phi^m)_k$, one obtains 
\begin{eqnarray} 
  &&\epsilon_{man}\left[(f\cdot\phi^a)(f\times\phi^m)\cdot
  {\textstyle \frac{\partial}{\partial \psi_n}}
  -{\textstyle \frac{1}{2}}[(f\times\phi^m)\cdot\phi^a]
  f\cdot{\textstyle \frac{\partial}{\partial \psi_n}}\right]\chi\\
  &&={\textstyle \frac{1}{2}}\epsilon_{man}\left[
  \left[f\times(f\times(\phi^m\times\phi^a))\right]\cdot
  {\textstyle \frac{\partial}{\partial \psi_n}}
  -[(f\times\phi^m)\cdot\phi^a]
  f\cdot{\textstyle \frac{\partial}{\partial \psi_n}}\right]\chi\nonumber\\
  &&=-{\textstyle \frac{1}{2}\epsilon_{man}(f\cdot f)(\phi^m\times\phi^a)\cdot 
  {\textstyle \frac{\partial}{\partial \psi_n}}}\chi=-(f\cdot f)Q_\psi\chi
  =(f\cdot f)(\phi^m\times\psi_m)\cdot\xi\ .\nonumber
\end{eqnarray}
So the second term of (\ref{Q-On}) cancels the first term, ensuring that $O_n$ is 
$Q$-closed.  So we have shown that the operator
\begin{eqnarray}\label{On-summary}
  O_n&=&(f\cdot f)^n \epsilon^{c_1c_2c_3}(\phi^a \cdot\psi_{c_1})(\phi^b \cdot \psi_{c_2})(\psi_a\cdot \psi_b\times\psi_{c_3})\\
  &&+n(f\cdot f)^{n-1}\epsilon^{b_1b_2b_3}\epsilon^{c_1c_2c_3}(f\cdot \psi_{b_1})(\phi^a\cdot \psi_{c_1})(\psi_{b_2}\cdot \psi_{c_2})
  (\psi_a\cdot \psi_{b_3}\times \psi_{c_3})\nonumber\\
  &&-{(\textstyle\frac{n}{72}+\frac{n^2-n}{108})}
  (f\cdot f)^{n-1}\epsilon^{a_1a_2a_3}\epsilon^{b_1b_2b_3}\epsilon^{c_1c_2c_3}
  (\psi_{a_1}\cdot \psi_{b_1}\times \psi_{c_1})(\psi_{a_2}\cdot \psi_{b_2}\times \psi_{c_2})
  (\psi_{a_3}\cdot \psi_{b_3}\times \psi_{c_3})\nonumber
\end{eqnarray} 
at $t^{24+12n}$ order is a black hole cohomology.

One may wonder if these are primaries captured in the first factor of (\ref{bh-index-bmn}), 
or if they are related to other $O_{n^\prime}$ with $n^\prime <n$ by acting some 
$Q^m_+$'s and/or gravitons on the third factor. One can show that the latter possibilities 
are all impossible. Suppose $O_n$ is obtained by acting acting $p$ $Q$'s on $O_{n^\prime}$ 
and multiplying $q$ gravitons. Then $p,q$ should satisfy
\begin{equation}
  2p+8q=12(n-n^\prime)\ ,\ \ p=0,1,2,3\ ,\ q\geq 0\ .
\end{equation}
Possible solutions are 
\begin{equation}
  (p,q,n-n^\prime)=(2,1,1)\ ,\ (0,3,2)\ ,\ (2,4,3)\ ,\ (0,6,4)\ ,\ 
  (2,7,5)\ ,\ (0,9,6)\ ,\ \cdots\ .
\end{equation}
The cases with even $n-n^\prime$ and $p=0$ yield operators 
at $t^{24+12n}$ order obtained by multiplying $O_{n^\prime}$ and 
$\frac{3}{2}(n-n^\prime)$ graviton operators of the form 
$\phi^m\cdot f+\frac{1}{2}\epsilon^{mnp}\psi_n\psi_p$. However, these cannot 
be cohomologous to $O_n$ because they do not have a term at 
$\mathcal{O}(f^{2n-2}\phi^0\psi^9)$ order that $O_n$ has, 
which cannot be changed by adding $Q$-exact terms. Now we consider the cases with 
odd $n-n^\prime$ and $p=2$, $q=\frac{3}{2}(n-n^\prime)-\frac{1}{2}$, and again 
consider whether the operator 
$(Q^a_+Q^b_+O_{n^\prime})(\phi\cdot f+\psi\cdot\psi)^q$ has a term 
at $f^{2n-2}\phi^0\psi^9$ order. Let us 
first study how the actions of $Q^a_+$ and $Q^b_+$ on $O_{n^\prime}$ can produce 
a term with no scalars. $Q^a_+$ either act as $\phi\rightarrow\psi$ or 
$\psi\rightarrow f$, so there are following possibilities:
\begin{equation}
  f^{2n^\prime}\phi^2\psi^5\rightarrow f^{2n^\prime}\psi^7\ ,\ 
  f^{2n^\prime-1}\phi\psi^7\rightarrow f^{2n^\prime}\psi^7\ ,\ \ 
  f^{2n^\prime-2}\psi^9\rightarrow f^{2n^\prime}\psi^7\ .
\end{equation}
In all three cases, we multiply gravitons of the form $(\phi \cdot f+\psi\cdot \psi)^q$ 
and see whether there can be a term at $f^{2n-2}\phi^0 \psi^9$ order.
This is possible only if $n=n^\prime+1$, $p=2$, $q=1$. That is, 
the only possible relations between different $O_n$'s are 
\begin{equation}\label{On-hairy?}
  O_n\stackrel{?}{\sim} \epsilon_{abc}(Q^a_+Q^b_+ O_{n-1})(\phi^c\cdot f+{\textstyle \frac{1}{2}}
  \epsilon^{cde}\psi_d\cdot\psi_e)\ ,
\end{equation}
where $\sim$ means up to a multiplicative factor and addition of $Q$-exact terms. 
We act three $Q^a_+$'s on (\ref{On-hairy?}) 
and show that this equation cannot hold. Acting $Q^1_+Q^2_+Q^3_+$ on the right hand side 
yields zero, so if this equation is true,  
$Q^1_+Q^2_+Q^3_+O_n$ should be $Q$-exact. However, this cannot be the case 
since it contains a term at $f^{2n+1}\psi^6$ order, which does not contain 
scalars so cannot be $Q$-exact. More concretely, one starts from
\begin{eqnarray}
O_n&=&(f\cdot f)^n O_{0}+\frac{20n}{3}(f\cdot f)^{n-1}\sum_{\textrm{cyclic}} 
(f\cdot \psi_3)(\psi_3\cdot \psi_2)(X\cdot\psi_2)(\psi_1\cdot \psi_1\times \psi_1)\nonumber\\
&&-\frac{10}{3}(\frac{n}{6}+\frac{n^2-n}{9})(f\cdot f)^{n-1}(\psi_1\cdot\psi_1\times\psi_1)(\psi_2\cdot\psi_2\times\psi_2)(\psi_3\cdot\psi_3\times\psi_3) 
\end{eqnarray}
where $\sum_{\textrm{cyclic}}$ means summation over the cyclic permutations 
of $(X,\psi_1)$, $(Y,\psi_2)$, $(Z,\psi_3)$. Acting $Q^1_+Q^2_+Q^3_+$, 
one obtains the following terms without scalars,
\begin{eqnarray}\label{QQQOn}
  \hspace*{-.5cm}&&Q^+_1 Q^+_2 Q^+_3 O_{0}\\
  \hspace*{-.5cm}
  &&=-10(X\cdot f+\psi_2\cdot\psi_3)^2 f \cdot(\psi_1\times\psi_1)-20(X\cdot f+\psi_2\cdot\psi_3)(Y\cdot f+\psi_3\cdot\psi_1) f\cdot(\psi_1\times\psi_2) +\textrm{cyclic}\nonumber\\
  \hspace*{-.5cm}&&\rightarrow 
  -20(\psi_2\cdot \psi_3)^2 (f\cdot \psi_1\times \psi_1) + 
  \textrm{cyclic}\nonumber\\
  \hspace*{-.5cm}
  &&Q^+_1 Q^+_2 Q^+_3 (f\cdot \psi_3)(\psi_3\cdot \psi_2)(X\cdot\psi_2)(\psi_1\cdot \psi_1\times \psi_1)+ \textrm{cyclic}\nonumber\\
  \hspace*{-.5cm}
  &&\rightarrow -3(f\cdot f)(\psi_2\cdot \psi_3)^2 (f\cdot \psi_1\times \psi_1)+6(f\cdot \psi_2)(f\cdot \psi_3)(\psi_2\cdot \psi_3)(f\cdot \psi_1\times \psi_1) + 
  \textrm{cyclic}\nonumber\\
  \hspace*{-.5cm}
  &&Q^+_1 Q^+_2 Q^+_3 (\psi_1\cdot\psi_1\times\psi_1)(\psi_2\cdot\psi_2\times\psi_2)(\psi_3\cdot\psi_3\times\psi_3) =-27 (f \cdot\psi_1\times\psi_1)(f \cdot\psi_2\times\psi_2)(f\cdot\psi_3\times\psi_3)\nonumber\\
  \hspace*{-.5cm}
  &&=18((f\cdot f)(\psi_2\cdot\psi_3)^2(f\cdot\psi_1\times \psi_1))-2(f\cdot \psi_2)(f\cdot \psi_3)(\psi_2\cdot \psi_3)(f\cdot \psi_1\times \psi_1) + \textrm{cyclic})\ .\nonumber
\end{eqnarray}
These terms at $f^{2n+1}\phi^0\psi^6$ order do not cancel, implying that 
$Q^1_+Q^2_+Q^3_+O_0$ cannot be $Q$-exact. 
So at least among the possibilities visible in the 
index (\ref{bh-index-bmn}), we have checked that different $O_n$'s are not related 
in trivial manners. 

Note also that the product of two $O_n$'s vanishes, $O_m O_n=0$. 
This is because each operator includes $5$ or more $\psi$'s, 
so the product involves $10$ or more $\psi$'s which vanishes by Fermi 
statistics.

\subsection{General sector and partial no-hair theorem}

Counting the graviton cohomologies with a computer using the eigenvalue 
setup explained in section 2, we obtained its index $Z_{\rm grav}$ till $t^{40}$ order. 
The remaining $SU(2)$ index is given by 
\begin{eqnarray}\label{su2-general-bh}
  Z-Z_{\rm grav}&=&\left[-t^{24}-\chi_{(1,3)}t^{32}
  -(\chi_{(1,\bar{3})}+\chi_{(3,6)})t^{34}-\chi_{(2,3)}t^{35}
  +(\chi_{(3,1)}+\chi_{(3,8)})t^{36}\right.\nonumber\\
  &&\left.-(\chi_{(2,\bar{3})}+\chi_{(4,6)})t^{37}+\chi_{(5,3)}t^{38}
  +(\chi_{(2,1)}+2\chi_{(4,1)}+\chi_{(4,8)})t^{39}\right.\nonumber\\
  &&\left.-(2\chi_{(1,6)}+\chi_{(3,\bar{3})}+\chi_{(5,\bar{3})}
  +\chi_{(5,6)})t^{40}\right]\chi_{\rm D}+\mathcal{O}(t^{41})\ .
\end{eqnarray}
The $SU(2)_R\times SU(3)$ characters and the factor
$\chi_{D}$ are given by
\begin{eqnarray}\label{characters-t40}
  \chi_{(2J^\prime+1,{\bf{\rm R}})}\!&\!\equiv\!&\!
  \chi^{SU(2)_R}_{J^\prime}(p)\chi^{SU(3)}_{\bf\rm R}(x,y)\\
  \chi_{D}\!&\!\equiv\!&\! \frac{(1-\!t^2z_1)(1-\frac{t^2}{z_2})(1-\!\frac{t^2z_2}{z_1})
  \cdot(1-\frac{tp}{z_1})(1-\!\frac{t}{pz_1})(1-tz_2p)(1-\frac{tz_2}{p})
  (1-\frac{tz_1p}{z_2})(1-\frac{tz_1}{z_2p})}{(1-t^3p)(1-\frac{t^3}{p})}
  \nonumber
\end{eqnarray}
where $t^6=e^{-\Delta_1-\Delta_2-\Delta_3}=e^{-\omega_1-\omega_2}$, 
$z_1=e^{\frac{-2\Delta_1+\Delta_2+\Delta_3}{3}}$, $z_2^{-1}=
e^{\frac{\Delta_1-2\Delta_2+\Delta_3}{3}}$, 
$p=e^{\frac{-\omega_1+\omega_2}{2}}$.
The function $\chi_{D}$ is factored out for later convenience (but we do not 
necessarily assume that only a particular type of representation appears).

Since descendant operators of $PSU(1,2|3)$ are not really  new, 
our interest is the new possible $\frac{1}{16}$-BPS primaries contained 
in  (\ref{su2-general-bh}).
Constructing all the cohomologies order by order as done in \cite{Chang:2022mjp} 
will clarify this in principle. However, we shall not comprehensively do this 
job in this paper.  Rather, we shall study the possible superconformal representation structures of the $\frac{1}{16}$-BPS states compatible with this index, finding many illuminating structures. 
As emphasized, we may miss some BPS states in case their multiplets completely cancel in the index.

The index can be written as a sum over the short $\mathcal{N}=4$
representations. Equivalently, it can be written as a sum over 
$\frac{1}{16}$-BPS multiplets of $PSU(1,2|3)\subset PSU(2,2|4)$. 
The last multiplets are embedded in the 
short representations of $PSU(2,2|4)$ in canonical manners: see appendix B. 
Knowing this representation sum is knowing the primary contents.
We will study this expansion order by order in $t$.  
As already mentioned in section 3.1 (and in the introduction), there are two classes 
of black hole cohomologies: those which can be written as products of other black hole 
cohomologies and gravitons which we call `hairy' and 
the rest which we call `core.'

We start by studying the black hole cohomologies that we identified in 
section 3.1. Among these, two of them $O_0$, $O_1$ appear within the $t^{40}$ order. We can show that all $O_n$'s are core black hole primaries of 
the $\frac{1}{16}$-BPS multiplets. The coreness
of $O_n$ is already shown in section 3.1, at least within the states 
visible in the index (\ref{bh-index-bmn}), since it 
suffices to show this within the BMN sector. We only need to show that they are 
$\frac{1}{16}$-BPS primaries 
in their full $PSU(1,2|3)$ representations. $O_0$ is clearly a $\frac{1}{16}$-BPS  primary 
since it is the lowest black hole cohomology. Since $j\equiv J_1+J_2=5$ is too large, $O_0$ can only belong to the $\mathcal{N}=4$ multiplet 
$A_1\overline{L}[4;0]^{[2,0,0]}_{9}$. The primary $O_0$ of the 
$\frac{1}{16}$-BPS multiplet is obtained by acting $Q^\prime\equiv Q^4_+$ on 
a primary of this $\mathcal{N}=4$ multiplet. The index over this multiplet is
\begin{equation}
  \chi_{24}\equiv-t^{24}\chi_D(t,x,y,p)\ ,
\end{equation}
where $\chi_D$ is defined in (\ref{characters-t40}) and in appendix B. 
So the first term $-t^{24}$ in the square bracket of (\ref{su2-general-bh}) 
corresponds to the contribution of this multiplet. 
Next we consider other $O_n$'s. We can prove that they are also primaries 
by showing that acting any of the nine $Q$'s in $PSU(1,2|3)$ yields nontrivial 
and independent cohomologies. (This is because $O_n$ does not contain derivatives and cannot be 
a conformal descendant.) We have shown in section 3.1 that the action of any 
$Q^m_+$ on $O_n$ is nontrivial and independent because acting all three of them yields a nontrivial cohomology. One can also show that
$\overline{Q}_{m\dot\alpha}O_n$ are all nontrivial and independent. 
It suffices to show that the six $\overline{Q}_{m\dot\alpha}$'s 
acting on $Q^1_+Q^2_+Q^3_+O_n$ are independent. 
This is easily shown by studying the terms obtained 
by acting $\overline{Q}_{m\dot\alpha}$ on the $\mathcal{O}(f^{2n+1}\phi^0\psi^6)$ 
order terms of $Q^1_+Q^2_+Q^3_+O_n$ in (\ref{QQQOn}). 
In particular, one obtains terms at $f^{2n}\phi^0\psi^6D\psi$ by acting 
$\overline{Q}_{m\dot\alpha}$ on $f$. These terms cannot be $Q$-exact 
since it involves neither $\bar\phi^m$ or $\bar\lambda_{\dot\alpha}$. 
This proves that all $6$ operators $\overline{Q}_{m\dot\alpha}Q^1_+Q^2_+Q^3_+O_n$ are nontrivial. They are also
independent since their $SU(2)_R\times SU(3)$ quantum numbers are different.
This shows that $O_{n\geq 1}$ are $\frac{1}{16}$-BPS primaries. 
$O_n$ belongs to the $\mathcal{N}=4$ multiplet 
$A_1\overline{L}[4+4n;0]^{[2,0,0]}_{9+4n}$, 
which contributes to to the index as $-t^{24+12n}\chi_D(t,x,y,p)$.

Now with the nature of $O_n$ understood, we come back to study the series 
(\ref{su2-general-bh}) till $t^{40}$ order, trying to better characterize 
other cohomologies order by order in $t$. Once the lowest operator $O_0$ is identified, 
all the states in its $\frac{1}{16}$-BPS multiplet are not really new operators. 
So we subtract $\chi_{24}$ from $Z-Z_{\rm grav}$ and see what are left: 
\begin{eqnarray}\label{grav-thres-deficit}
  Z-Z_{\rm gra}-\chi_{24}&=&\left[-\chi_{(1,3)}t^{32}
  -(\chi_{(1,\bar{3})}+\chi_{(3,6)})t^{34}-\chi_{(2,3)}t^{35}
  +(\chi_{(3,1)}+\chi_{(3,8)})t^{36}\right.\nonumber\\
  &&-(\chi_{(2,\bar{3})}+\chi_{(4,6)})t^{37}+\chi_{(5,3)}t^{38}
  +(\chi_{(2,1)}+2\chi_{(4,1)}+\chi_{(4,8)})t^{39}\nonumber\\
  &&\left.-(2\chi_{(1,6)}+\chi_{(3,\bar{3})}+\chi_{(5,\bar{3})}
  +\chi_{(5,6)})t^{40}\right]\chi_D+\mathcal{O}(t^{41})\ .
\end{eqnarray}
Somewhat surprisingly, after subtracting the multiplet of $O_0$, one finds that 
the remaining index starts from $t^{32}$ order. Namely, in the range $t^{25}\sim t^{31}$, 
the index does not capture any new black hole cohomologies except the trivial descendants 
of $O_0$. At first sight this may look like a boring result, but the triviality of the 
index in this range has a nontrivial implication.

Recall that cohomologies multiply to yield new cohomologies. This is because 
of the Leibniz rule of the classical $Q$ acting on product operators.
So apparently, one can multiply light graviton cohomologies 
to $O_0$ or its descendants to obtain many new cohomologies in the range 
$t^{25}\sim t^{31}$.
The possible product cohomologies of $O_0$ and gravitons below $t^{32}$ order are
\begin{eqnarray}\label{no-hair-till-31}
  &&O_0(\bar\phi^{(m}\cdot\bar\phi^{n)})\ ,\ \ 
  O_0(\bar\phi^m\cdot\bar\lambda_{\dot\alpha})\ ,\ \ 
  O_0(\bar\lambda_{\dot{+}}\cdot\bar\lambda_{\dot{-}})\ ,\nonumber\\
  &&O_0(\bar\phi^m\cdot\psi_{n+}-{\textstyle \frac{1}{3}}\delta^m_n
  \bar\phi^p\cdot\psi_{p+})\ ,\ \ 
  O_0(\bar\lambda_{\dot\alpha}\cdot\psi_{m+}
  -{\textstyle \frac{1}{2}}\epsilon_{mnp}\phi^n\cdot D_{+\dot\alpha}\phi^p)\ ,\nonumber\\
  &&O_0\partial_{+\dot{\alpha}}(\bar\phi^{(m}\cdot\bar\phi^{n)})\ .
\end{eqnarray}
Other possible products below $t^{32}$ 
involving the descendants of $O_0$ are  
\begin{eqnarray}\label{no-hair-2}
  \overline{Q}O_0&\times&
  (\bar\phi^m\cdot \bar\phi^n\ ,\ 
  \bar\phi^m\cdot\bar\lambda_{\dot\alpha}\ ,\ 
  \bar\lambda_{\dot{+}}\cdot\bar\lambda_{\dot{-}}\ ,\ 
  \bar\phi^m\cdot\psi_{n+}-{\textstyle \frac{1}{3}}\delta^m_n
  \bar\phi^p\cdot\psi_{p+})\ ,
  \nonumber\\
  (Q,\overline{Q}\overline{Q})O_0&\times&
  (\bar\phi^m\cdot \bar\phi^n\ ,\ 
  \bar\phi^m\cdot\bar\lambda_{\dot\alpha})\ ,
  \nonumber\\
  (Q\overline{Q},\overline{Q}\overline{Q}\overline{Q},\partial)O_0&\times&
  (\bar\phi^m\cdot\bar\phi^n)\ .
\end{eqnarray}
The triviality of the index (\ref{grav-thres-deficit}) in 
this range implies two possibilities for these product cohomologies. 
The first possibility is that these product cohomologies are $Q$-exact, 
i.e. absent in the BPS spectrum. 
Another possibility is that these product cohomologies are nontrivial but there 
are cancellations in the index, either among themselves or with new 
core black hole cohomologies.\footnote{We have checked that cancellations 
cannot happen within the product cohomologies listed above.  
It is logically possible (although a bit unnatural) that some new core black
hole primaries appear in this range, precisely canceling with some of the product operators
above if they are not $Q$-exact. Although in different contexts, certain black holes 
are known not to appear in the index. For instance, asymptotically flat 
multi-center BPS black holes or BPS black rings are not captured by the index 
\cite{Dabholkar:2009dq}.}
Among (\ref{no-hair-till-31}) and (\ref{no-hair-2}), we have explicitly shown after 
very nontrivial numerical/analytic calculations that 
\begin{equation}
  O_0(\bar\phi^m\cdot \bar\phi^n)\ ,\ O_0(\bar\phi^m\cdot\bar\lambda_{\dot\alpha})\ ,\ 
 O_0(\bar\phi^m\cdot \psi_{n+}-{\textstyle \frac{1}{3}}\delta^m_n\bar\phi^p\cdot\psi_{p+})
\end{equation}
are all $Q$-exact. See Appendix D for some details. We did not manage to prove the $Q$-exactness 
of other operators. Since these operators do not appear at all in the index, all of them may be 
$Q$-exact till $t^{31}$ order. More robustly/modestly, we can say that our index 
exhibits a no-hair behavior for $O_0$ till $t^{31}$ order. It will be interesting 
to clarify this issue in the future.

The $Q$-exactness of these product operators implies that $O_0$ abhors 
the dressings by certain gravitons, reminiscent of the black hole no-hair theorem. 
Especially,  $(\bar\phi^m\cdot\bar\phi^n)\sim {\rm tr}(\bar\phi^m\bar\phi^n)$ multiplied to 
$O_0$ are $Q$-exact. This is interesting because these operators correspond to 
bulk scalar fields which have been discussed in the context of hairy 
AdS$_5$ black holes \cite{Bhattacharyya:2010yg,Markeviciute:2018yal,Markeviciute:2018cqs}.
More precisely, it is the `s-wave' modes of these scalars that have been used 
to construct hairy black holes, precisely dual to the conformal primary operator 
${\rm tr}(\bar\phi^m\bar\phi^n)$. Here, note that the BPS limits of the hairy black holes 
constructed this way all exhibit substantial back reactions to the core black holes, 
at least near the horizon, no matter how small the hair parameter is
\cite{Markeviciute:2018yal,Markeviciute:2018cqs}. 
In section 4, we shall revisit these aspects in the gravity dual and 
more carefully discuss the possible forms of hairy BPS black holes.

Now we consider the lowest term $-\chi_{(1,3)}t^{32}$  of (\ref{grav-thres-deficit}). 
In fact, this term comes from the following product of $O_0$ and gravitons:
\begin{equation}\label{t32-non-Q-exact}
  O_0 (\bar\phi^m\cdot f_{++}+{\textstyle \frac{1}{2}}\epsilon^{mnp}\psi_{n+}\cdot\psi_{p+})\ .
\end{equation}
It is easy to show that this is not $Q$-exact, e.g. by acting two 
$Q^m_+$ as shown in (\ref{QQ-on-t36}). These operators contain 
terms at $f^0\phi^0\psi^9$ order, which cannot be $Q$-exact. 
So the operators (\ref{t32-non-Q-exact}) themselves are not $Q$-exact either. 
Therefore, the no-hair interpretation that we made so far holds 
only for certain low-lying gravitons, at best. See section 4 for more detailed 
discussions on examples of `allowed' hairs. Among the conformal primaries of $S_2$ listed 
in Appendix A, these three gravitons are the only ones which explicitly appear in the 
index when multiplied to $O_0$. At this stage, it may seem that two more gravitons 
$f_{++} \cdot \bar\lambda_{\dot\alpha}+{\textstyle\frac{2}{3}}
  \psi_{m+}\cdot D_{+\dot\alpha}\bar\phi^m
  -{\textstyle \frac{1}{3}}\bar\phi^m \cdot D_{+\dot\alpha}\psi_{m+}$ at $\mathcal{O}(t^9)$ might multiply $O_0$ to 
show up at $t^{33}$ order, but we will see below that the index does not capture them. 
Therefore, out of the $32$ particle species of conformal primary particles 
in the $S_2$ multiplet, $29$ gravitons  except $\bar\phi^m\cdot f_{++}+
{\textstyle \frac{1}{2}}\epsilon^{mnp}\psi_{n+}\cdot\psi_{p+}$
do not appear in the index when they multiply $O_0$. 
In the BMN sector, our studies in section 3.1 imply a similar theorem for 
all $O_n$, at least as seen by the index. Among the $17$ particle species of gravitons 
in the BMN sector, all $14$ particles except  $\bar\phi^m\cdot f_{++}+
{\textstyle \frac{1}{2}}\epsilon^{mnp}\psi_{n+}\cdot\psi_{p+}$
do not appear in the index when they multiply any $O_n$.

The $3$ product coholomologies at $t^{32}$ order violating the no-hair theorem 
should be the primaries of $PSU(1,2|3)$. This is again contained in a short multiplet 
of $A_1\overline{L}$ type, whose contribution to the index is given by 
$\chi_{32}=t^{32}\chi_{(1,3)}\chi_{ D}$. 
We subtract this from $Z-Z_{\rm grav}-\chi_{24}$, and study the remaining cohomologies. 
We can then try to interpret the lowest order term of the remainder and judge whether it
comes from new core black hole primaries or products of already known core primaries and
gravitons. If one can clarify the nature of the cohomologies at this lowest order, one 
can again subtract the characters of their supermultiplets and keep exploring even higher
orders. Since it becomes more and more difficult to judge the $Q$-exactness of the 
possible product operators, we shall only make much simpler and structural studies 
till the $t^{40}$ order. Namely, we shall try to see if the surviving index can be 
explained as the products of known gravitons and core primaries $O_n$, without the 
need of any new core black hole primaries. Studies we made so far showed that 
this is possible till $t^{32}$ order. Namely, the index till this order is compatible 
with having no more new core primaries and only three more product cohomologies 
(\ref{t32-non-Q-exact}). We shall show that the graviton spectrum is such that 
new core black hole primaries should appear at $t^{39}$ order at the latest. This 
not only proves from the index the existence of new core black hole primaries, 
but will also show scenarios of possible hairy black holes (to be further discussed 
in section 4).

After eliminating the contribution of the multiplet $\chi_{32}$ to the index, the 
remaining index vanishes at $t^{33}$ order. In principle, there are two possible 
product operators which completely cancel in the index if they are not $Q$-exact, given by
\begin{equation}
  O_0\partial_{+\dot{\alpha}}(\bar\lambda_{\dot\beta}\cdot\bar\lambda^{\dot\beta})\ \ ,\ \ 
  O_0\left(f_{++}\cdot\bar\lambda_{\dot\alpha}+{\textstyle\frac{2}{3}}
  \psi_{m+}\cdot D_{+\dot\alpha}\bar\phi^m
  -{\textstyle \frac{1}{3}}\bar\phi^m \cdot D_{+\dot\alpha}\psi_{m+}\right)\ .
\end{equation}
So these product operators, even if they exist, do not appear in the index. 
The lowest nonzero term of $Z-Z_{\rm grav}-\chi_{24}-\chi_{32}$ is 
$-(\chi_{(1,\bar{3})}+\chi_{(3,6)})t^{34}$. 
The only possible product operators which may account for this term, if they are 
not $Q$-exact, are 
\begin{equation}
  O_0{\partial_+}^{\dot\alpha}\left(\bar\lambda_{\dot\alpha}\cdot\psi_{m+}
  -\textstyle{\frac{1}{2}}\epsilon_{mnp}\bar\phi^n\cdot D_{+\dot\alpha}\bar\phi^p\right)
  \ \ ,\ \ 
  O_0\partial_{+\dot\alpha}\partial_{+{\dot{\beta}}}(\bar\phi^{m}\cdot\bar\phi^{n})\ .
\end{equation}
If they are nontrivial, they are in the $\mathcal{N}=4$ representations 
$A_1\overline{L}[6;2]^{[2,2,0]}_{13}$ and $A_1\overline{L}[6;0]^{[3,0,1]}_{13}$
representations, respectively. Assuming that they are both non-$Q$-exact, this order 
can be accounted for by hairy black hole operators. Their multiplets
will contribute $-(\chi_{(1,\bar{3})}+\chi_{(3,6)})t^{34}\chi_D$ to the index. Subtracting them, 
the leading term is $-\chi_{(2,3)}t^{35}$. The only possible product cohomologies which 
can account for this term are
\begin{equation}
  O_0\partial_{+\dot\alpha}(f_{++}\cdot\bar\phi^m+{\textstyle \frac{1}{2}}
  \epsilon^{mnp}\psi_{n+}\cdot\psi_{p+})\ ,
\end{equation} 
if they are not $Q$-exact. In this case, its multiplet is again $A_1\overline{L}$ type 
and contributes $-\chi_{(2,3)}t^{35}\chi_D$ to the index. Subtracting this, 
the lowest term is $+(\chi_{(3,1)}+\chi_{(3,8)})t^{36}$. Since there is one 
fermionic black hole primary $O_1$, we study whether the product cohomologies may 
account for $+(1+\chi_{(3,1)}+\chi_{(3,8)})t^{36}$. 
The only possible set is  
\begin{equation}
  O_0\partial_{+\dot\alpha}(f_{++}\cdot\bar\lambda_{\dot\beta}+{\textstyle\frac{2}{3}}
  \psi_{m+}\cdot D_{+\dot\beta}\bar\phi^m-{\textstyle \frac{1}{3}}\bar\phi^m \cdot 
  D_{+\dot\beta}\psi_{m+})\ \ ,\ \ 
  O_0\partial_{+\dot\alpha}\partial_{+\dot\beta}(\bar\phi^m\cdot\psi_{n+}-
  {\textstyle \frac{1}{3}}\delta^m_n\bar\phi^p\cdot\psi_{p+})\ .
\end{equation} 
Subtracting the contributions of these multiplets if not $Q$-exact, again in the 
$A_1\overline{L}$ multiplets, the lowest term is $-(\chi_{(2,\bar{3})}+\chi_{(4,6)})t^{37}$.
The only possible product cohomologies that can account for this term are 
\begin{equation}
  O_0\partial_{+\dot\alpha} \partial_+^{\ \ \dot\beta}
  (\bar\lambda_{\dot\beta}\cdot\psi_{m+}-\textstyle{\frac{1}{2}}
  \epsilon_{mnp}\bar\phi^n\cdot D_{+\dot\beta}\bar\phi^p)\ \ ,\ \ 
  O_0\partial_{+\dot\alpha}\partial_{+\dot\beta}\partial_{+\dot\gamma}
  (\bar\phi^m\cdot\bar\phi^n)\ .
\end{equation}
Further processing to subtract the contributions of their multiplets, the lowest term 
is $+\chi_{(5,3)}t^{38}$.  one possible set of product cohomologies which can account for 
this is 
\begin{equation}
  O_0\partial_{+(\dot\alpha}\partial_{+\dot\beta}\partial_{+\dot\gamma}
  (\bar\lambda_{\dot\delta)}\cdot\bar\phi^m)\ .
\end{equation}
Apart from these, the following two sets of product cohomologies  
\begin{equation}
  O_0\partial_{+\dot\alpha}\partial_{+\dot\beta}
  {\partial_+}^{\dot\gamma}(\bar\lambda_{\dot\gamma}\cdot\bar\phi^m)\ \ ,\ \ 
  O_0\partial_{+\dot\alpha}\partial_{+\dot\beta}
  (f_{++}\cdot\bar\phi^m+{\textstyle \frac{1}{2}}\epsilon^{mnp}\psi_{n+}\cdot\psi_{p+})
\end{equation}
exactly cancel in the index, so there are two possible ways in which product hairy 
cohomologies can account for this order. In either case, they are all in the 
$A_1\overline{L}$ type multiplets. 

Subtracting the last multiplets, the lowest term is 
$+(\chi_{(2,1)}+2\chi_{(4,1)}+\chi_{(4,8)})t^{39}$. All possible 
product cohomologies at this order are 
\begin{eqnarray}
  ({\bf4},{\bf1})^\textrm{F}&:&O_0\partial_{+\dot\alpha}\partial_{+\dot\beta}\partial_{+\dot\gamma}
  (\bar\lambda_{\dot\delta}\cdot\bar\lambda^{\dot\delta})\ ,\\
  ({\bf2},{\bf1})^\textrm{B}&:&O_0\partial_{+\dot\alpha}\partial_+^{\ \ \dot\beta}
  (f_{++}\cdot\bar\lambda_{\dot\beta}+{\textstyle\frac{2}{3}}
  \psi_{m+}\cdot D_{+\dot\beta}\bar\phi^m
  -{\textstyle \frac{1}{3}}\bar\phi^m \cdot D_{+\dot\beta}\psi_{m+})\ ,\nonumber\\
  ({\bf4},{\bf1})^\textrm{B}&:&O_0\partial_{+(\dot\alpha}\partial_{+\dot\beta}
  (f_{++}\cdot\bar\lambda_{\dot\gamma)}+{\textstyle\frac{2}{3}}
  \psi_{m+}\cdot D_{+\dot\gamma)}\bar\phi^m-{\textstyle \frac{1}{3}}\bar\phi^m \cdot D_{+\dot\gamma)}\bar\psi_{m})\ ,\nonumber\\
  ({\bf4},{\bf8})^\textrm{B}&:&O_0\partial_{+\dot\alpha}\partial_{+\dot\beta}\partial_{+\dot\gamma}
  (\bar\phi^m\cdot\psi_{n+}-{\textstyle \frac{1}{3}}
  \delta^m_n\bar\phi^p\cdot\psi_{p+})\ .\nonumber
\end{eqnarray}
We have shown the superscripts B/F for their bosonic/fermionic statistics, respectively. 
With these candidates, we find that the closest one can get to the index at this order 
is the case in which all three classes of bosonic operators are nontrivial while the 
fermionic operators are $Q$-exact. In this case, their contribution at this order 
is maximal and becomes $+(\chi_{(2,1)}+\chi_{(4,1)}+\chi_{(4,8)})t^{39}$. Therefore, 
there should be at least $4$ core black hole primaries to account for the remaining 
$+\chi_{(4,1)}t^{39}$. Of course this is the latest order in which new 
core black hole primaries should appear, and it may well appear at the lowest orders 
by some non-$Q$-exactness assumptions we made for product cohomologies being invalid.

So we have shown that, from the index data till $t^{40}$ order, there should exist 
more core primary operators except $O_n$ in the BMN sector. This conclusion is obtained
by supposing otherwise, and trying to explain the index as product cohomologies of $O_n$ 
and gravitons but finding a contradiction at $t^{39}$. We should also emphasize that 
the structure of the index admits natural explanations in terms of hairy product operators 
in a wide range $t^{33}\sim t^{38}$. Note also that most of the gravitons appearing in this 
range are conformal descendants in the $S_2$ multiplet. We shall find an interpretation 
of this in the next section.

\section{Comments on hairy black holes}

In section 3.2, we explored possible hairy black hole states in the $SU(2)$ theory. 
To better address this concept, we should understand the ways in which 
hairy BPS black holes can form in AdS.
% (SL added)
In this section, we study scalar hairs of the BPS black holes in both wave and probe particle
perspectives and present a qualitative picture in the bulk gravity as to which scalar hairs
may dress the BPS black holes.

The simplest form of hairy black holes has been 
explored in \cite{Basu:2010uz,Bhattacharyya:2010yg}, called the `non-interacting mix' 
of small black holes and dilute graviton hairs. The idea is that 
the back-reaction of the black hole to the hair is weak when the black hole size $q$ is small, 
and vice versa when the hair density $\varepsilon$ is low.
In the limit of small $q$ and $\varepsilon$, the leading hairy 
black hole solution is a superposition of 
the non-hairy black hole and the graviton wavefunction in vacuum AdS. Systematic 
back-reactions to each other can be computed as a series in small $q,\varepsilon$, 
rendering a nonlinear superposition.
Nonlinear superposition is naturally expected in the BPS sector, even 
at finite $q,\varepsilon$, where the cancellation of mutual forces is often
ensured by supersymmetry. In fact, our product cohomologies realize
such superpositions of BPS states in some sense. Although the true BPS states do not 
generally admit product forms, there are product representatives of the cohomologies.
Such superposition representations were crucial in \cite{Grant:2008sk} 
for gaining a deeper understanding of multi-graviton states beyond the low-energy limit.

Meanwhile, some numerical solutions for hairy BPS black holes
are known from \cite{Markeviciute:2018yal,Markeviciute:2018cqs}.
%As for the hairy BPS black holes, some numerical solutions 
%are known from \cite{Markeviciute:2018yal,Markeviciute:2018cqs}.
These solutions 
appear to be somewhat different from the non-interacting mix 
picture of \cite{Basu:2010uz,Bhattacharyya:2010yg} in the following sense. 
No matter how small the hair density $\varepsilon$ is, it is shown
that the hair back-reacts substantially to the black hole near the horizon. 
In fact, the back-reaction is such that a (mild) singularity is created 
at the  horizon. On the other hand, we would like to interpret 
our hairy black hole operators of section 3 as realizing the non-interacting mix 
picture of \cite{Basu:2010uz,Bhattacharyya:2010yg}, as superpositions are 
realized in the cohomology setup. In this section,  we will try to 
better understand these two notions of hairy BPS black holes and explore 
a setup in which BPS solutions along the line of \cite{Basu:2010uz,Bhattacharyya:2010yg} 
may exist.

One  no-hair theorem proved in appendix D is 
the absence of $(\bar\phi^m\cdot\bar\phi^n)O_0$.
On the other hand, in \cite{Bhattacharyya:2010yg,Markeviciute:2018yal,Markeviciute:2018cqs}, 
the bulk scalar dual to  ${\rm tr}(\bar\phi^m\bar\phi^n)$ 
was discussed as a hair. We shall first review our current knowledge on 
hairy BPS black holes in $AdS_5\times S^5$ with this scalar. 
We shall try to clarify the relations of these results with our findings in section 3
along the way.

Attempts made in the literature were to find continuous 
deformations of non-hairy black holes. The simplest non-hairy BPS black holes 
are those carrying equal R-charges and angular momenta, 
$R\equiv R_1=R_2=R_3$, $J\equiv J_1=J_2$ \cite{Gutowski:2004ez}. 
They carry one free parameter $q$. (This is related to the three 
$\mu_I$ parameters of \cite{Gutowski:2004yv} by $\mu_1=\mu_2=\mu_3=q$.)
We shall follow the notation of \cite{Liu:2007rv} for later convenience.
The metric and the gauge field are given by
\begin{eqnarray}\label{background-bh}
  ds^2&=&-f^2(dt+w\sigma_3)^2+f^{-1}h_{mn}dx^mdx^n\nonumber\\
  &=&-f^2\left(dt+w\sigma_3\right)^2+f^{-1}\left[\frac{dx^2}{4xh}
  +\frac{x}{4}\left(\sigma_1^2+\sigma_2^2+h\sigma_3^2\right)\right]\nonumber\\
  A&=&-f(dt+w\sigma_3)+U\sigma_3
\end{eqnarray}
where the 1-forms $\sigma_i$ satisfying 
$d\sigma_1=-\sigma_2\wedge \sigma_3$, etc. are given by
%\begin{equation}
%  -\frac{i}{2}\tau_i\sigma_i=U^{-1}dU\ \ ,\ \ 
%  U=e^{J_3\phi}e^{J_2\theta}e^{J_3(\psi+\frac{\pi}{2})}\ \ ,
%  \ \ J_i=-\frac{i}{2}\tau_i\ ,
%\end{equation}
\begin{equation}
  \sigma_1=\cos\psi d\theta + \sin\psi \sin\theta d\phi\ ,\ \ 
  \sigma_2=-\sin\psi d\theta+\cos\psi \sin\theta d\phi\ ,\ \ 
  \sigma_3=d\psi+\cos\theta d\phi\ ,
\end{equation}
and various functions are given by
\begin{equation}\label{background-ftn}
  H\equiv f^{-1}=1+\frac{q}{x}\ \ ,\ \ U
  %(=U_3\frac{e^3}{\sigma_3})
  =\frac{x}{2\ell}+\frac{q}{\ell}\ \ ,\ \ 
  h=1+\frac{x+3q}{\ell^2}\ \ ,\ \ w=\frac{x+3q}{2\ell}+\frac{3q^2}{4\ell x}\ .
\end{equation}
$\ell$ is the radius of AdS$_5$. The following 4d vierbeins
\begin{equation}\label{4d-vierbein}
  {\textstyle e^{\hat{x}}=\frac{dx}{2\sqrt{xh}}\ \ ,\ \ 
  e^{\hat{1},\hat{2}}=\frac{\sqrt{x}}{2}\sigma_{1,2}
  \ \ ,\ \  e^{\hat{3}}=\frac{\sqrt{xh}}{2}\sigma_3\ .}
\end{equation}
for the metric $h_{mn}$ will be useful below. 
The boundary $x\rightarrow\infty$ is asymptotically AdS$_5$, whose
canonical time and angle coordinates $t^\prime , \psi^\prime$ are related to those above by 
$t^\prime=t$, $\psi^\prime=\psi+2t$. The unprimed coordinate system rotates  
with the event horizon, located at $x=0$. The mass and charges are given by
\begin{equation}\label{two-charges}
  {\textstyle M=\frac{3R+2J}{\ell}
  \ \ ,\ \ R=\frac{N^2}{2}\left(\frac{q}{\ell^2}+\frac{q^2}{2\ell^2}\right)
  \ \ ,\ \ J=\frac{N^2}{2}\left(\frac{3q^2}{2\ell^4}+\frac{q^3}{\ell^6}\right)\ .}
\end{equation}
%and the mass is given by $M=\frac{3R+2J}{\ell}$.
The solution has one free parameter $q$ but two distinct charges $R,J$. 
So $R,J$ should satisfy a relation on the solution space, which is given by 
\cite{Choi:2018hmj}
\begin{equation}\label{charge-relation}
  {\textstyle \mathcal{G}(R,J)\equiv R^3+\frac{N^2}{2}J^2-\left( 3R+\frac{N^2}{2} \right)
  \left(3R^2-N^2J\right)=0\ .}
\end{equation}

Hairy BPS black holes have been studied as a 1-parameter deformation of 
the solution above. A real scalar field $\varphi$ was added, only depending on the 
radial variable $x$ and preserving the $SU(2)_R\times U(1)\subset SO(4)$ 
isometry. The functions $f(x),U(x), h(x), w(x),\varphi(x)$ 
now satisfy  coupled ODE's in $x$. \cite{Markeviciute:2018yal,Markeviciute:2018cqs} 
solved these equations numerically first in the 
non-BPS case and then reduced the mass to get close to the BPS limit. 
One important feature reported in \cite{Markeviciute:2018yal,Markeviciute:2018cqs} is that 
the solutions seem to exhibit a singularity which replaces
the event horizon. The limiting solution has an extra continuous parameter 
apart from $q$. We would like to first see whether BPS black holes 
in this setup can be understood from the non-interacting mix picture, in the 
sense of superposing the solution. In particular, we set the hair condensate 
$\varphi\sim \varepsilon$ to be small, keeping the black hole size general. 
We shall try to keep $\varphi$ as a small perturbation around fixed black hole 
background (\ref{background-bh}), trying to realize  the spirit of 
\cite{Basu:2010uz,Bhattacharyya:2010yg}.
(The failure of this procedure will be related to the unique nature of the solutions 
of \cite{Markeviciute:2018yal,Markeviciute:2018cqs}.)

We study the BPS equations for a real scalar $\varphi$ in the notation of \cite{Liu:2007rv}. 
Their scalar is related to the scalar $\Phi$ of \cite{Bhattacharyya:2010yg}  
by $\Phi=2\sinh\varphi$. The functions $f(x),U(x),w(x),h(x),\varphi(x)$ satisfy
eqn.(4.7) of  \cite{Liu:2007rv}. For small $\varphi\sim\varepsilon$, $f,U,w,h$ are given by 
(\ref{background-ftn}) at the leading $\mathcal{O}(\varepsilon^0)$ order. 
At the next-to-leading $\mathcal{O}(\varepsilon^1)$ order, $\varphi$ 
should satisfy
\begin{equation}\label{scalar-eqn-BPS}
  \varphi^\prime(x)=-\frac{2U(x)}{\ell xh(x)}\sinh\varphi(x)\ .
\end{equation}
At this order, we  plug in the zeroth order solutions (\ref{background-ftn}) 
for $U,h$. Since we keep the hair small, $\varphi\sim\mathcal{\varepsilon}$, 
we take $\sinh\varphi\approx\varphi$ and consider the linear
differential equation for $\varphi$.
Normally, if we solve the general equation of motion in the AdS black hole 
background, one studies the second order differential equation with the 
normalizable boundary condition at the boundary $x\rightarrow\infty$ and 
the infalling boundary condition at the horizon $x\rightarrow 0$. This  
yields the quasinormal modes at discrete choices of frequency
$\omega_n$. These frequencies will typically have nonzero imaginary parts, either 
implying that the modes fall into the horizon (${\rm Im}(\omega_n)<0$) or  
that they trigger superradiant instability (${\rm Im}(\omega_n)>0$).
However, in the BPS equation, we only expect stationary modes with 
${\rm Im}(\omega_n)=0$. (In the time coordinate chosen above, $\varphi$ will be 
$t$-independent and real.) (\ref{scalar-eqn-BPS}) can be 
integrated with the integration constant being its overall amplitude $\varepsilon$. 
So one has no free parameter left to match either normalizability at $x=\infty$ or the 
infalling (or in the BPS case, stationary normalizable) condition at $x=0$. We just find 
either acceptable or pathological solution.

Plugging in (\ref{background-ftn}) to (\ref{scalar-eqn-BPS}), the solution for 
$\varphi$ is given by
\begin{equation}
  \tanh\frac{\varphi(x)}{2}=\varepsilon
  x^{-\frac{2q/\ell^2}{1+3q/\ell^2}}
  \left(1+\frac{3q}{\ell^2}+\frac{x}{\ell^2}\right)^{-\frac{1+q/\ell^2}
  {1+3q/\ell^2}}\ .
\end{equation}
The left hand side should be understood as $\approx \frac{\varphi(x)}{2}$ in 
our small hair approximation. 
At the boundary $x \to \infty$, it behaves like the proper normalizable mode $\sim x^{-1}$ 
for the scalar dual to the Yang-Mills operator ${\rm tr}(X^2+Y^2+Z^2)$ 
\cite{Bhattacharyya:2010yg}. However, the solution diverges at $x=0$ as
\begin{equation}\label{s-wave-singular}
  \varphi(x)\sim \varepsilon x^{-\frac{2q/\ell^2}{1+3q/\ell^2}}\ .
\end{equation}
So our perturbative approach with small $\varphi$ breaks down. 
Also, for small black hole size $q$, this solution provides the first correction to 
the graviton wavefunction in vacuum AdS ($q=0$) due to the small black hole ($q \neq 0$), which 
again diverges. So the iteration procedure of \cite{Basu:2010uz,Bhattacharyya:2010yg} for the
nonlinear superposition would not work even for small $\varepsilon$, $q$ 
(say, without finite temperature regularization). 
The exact numerical results of
\cite{Markeviciute:2018yal,Markeviciute:2018cqs} report divergence not with the scalar 
but with certain tidal forces (geometry), very different from what we find above. 
Since the nature of the background black hole solution changes by turning on small 
$\varepsilon$, the superposition picture of \cite{Basu:2010uz,Bhattacharyya:2010yg} does 
not seem to apply straightforwardly in this setup.

% From here

To improve the situation, we  first show  that $\varphi$ 
beyond the s-wave ansatz may provide smooth solutions in the perturbative regime.  
To find these solutions, we should keep a complex scalar including $\varphi$ 
and consider the BPS equations given by PDE's in AdS$_5$. We study a small complex scalar sitting in a hypermultiplet and identify the complex 
scalar which contains $\varphi$. The general BPS equation for 
the scalars in the hypermultiplet is given in \cite{Bellorin:2007yp}, eqn. (3.12) and (3.27). 
(3.12) demands $t$-independence of the scalar, while (3.27) is a spatial differential 
equation given by
\begin{equation}\label{hyper-eqn}
  D_m q^X={(\Phi^r)_m}^{n}D_nq^Y{(J^r)_Y}^X\ .
\end{equation}
Here, $q^X$ is the scalars in the hypermultiplet, with $X,Y=1,\cdots,4$, 
$(\Phi^r)_m^{\ n}$ with $r=1,2,3$ are the three complex structures of the 
base space and $(J^r)_X^{\ Y}$ are the three complex structures of the scalar target space. 
They satisfy 
\begin{equation}\label{3-complex}
  {(\Phi^1)_m}^n{(\Phi^2)_n}^p={(\Phi^3)_m}^p\ ,\ 
  {(\Phi^2)_m}^n{(\Phi^3)_n}^p={(\Phi^1)_m}^p\ ,\ 
  {(\Phi^3)_m}^n{(\Phi^1)_n}^p={(\Phi^2)_m}^p\ ,\ 
\end{equation}
and similar relations hold for $J^r$.
$D_m$ is the covariant derivative involving our $U(1)$ gauge field 
$A_\mu$. $\Phi^r$ can be taken to be  
\begin{equation}\label{phi3}
  \Phi^1=e^{\hat{x}}\wedge e^{\hat{2}}-e^{\hat{3}}\wedge e^{\hat{1}}\ ,\ 
  \Phi^2=e^{\hat{2}}\wedge e^{\hat{3}}-e^{\hat{x}}\wedge e^{\hat{1}}\ ,\ 
  \Phi^3=e^{\hat{x}}\wedge e^{\hat{3}}-e^{\hat{1}}\wedge e^{\hat{2}}
\end{equation}
on the black hole background (after some permutations of $r=1,2,3$: e.g. 
$\Phi^3$ given above is called $J=\Phi^2$ in \cite{Liu:2007rv} and $J^1$ in \cite{Gutowski:2004ez}).
In this setup, introducing the following inverse-vierbeins $E_{\hat{x}},E_{\hat{1},\hat{2},\hat{3}}$ 
for (\ref{4d-vierbein}) on the 4d base space,
\begin{eqnarray}
  &&{\textstyle E_{\hat{x}}=2\sqrt{xh}\frac{\partial}{\partial x}\ \ ,\ \ 
  E_{\hat{3}}=\frac{2}{\sqrt{xh}}\frac{\partial}{\partial\psi}}\\
  &&{\textstyle 
  E_{\hat{1}}=\frac{2}{\sqrt{x}}(\cos\psi\frac{\partial}{\partial\theta}
  +\frac{\sin\psi}{\sin\theta}(\frac{\partial}{\partial\phi}
  -\cos\theta\frac{\partial}{\partial\psi}))\ ,\ \ 
  E_{\hat{2}}=\frac{2}{\sqrt{x}}(-\sin\psi\frac{\partial}{\partial\theta}
  +\frac{\cos\psi}{\sin\theta}(\frac{\partial}{\partial\phi}
  -\cos\theta\frac{\partial}{\partial\psi}))}\ ,
  \nonumber
\end{eqnarray}
one can show 
\begin{equation}
  \mathcal{E}_{i}^m {(\Phi^r)_m}^n=-i{(\sigma^r)_{i}}^{j}
  \mathcal{E}_{j}^n
  \ \ ,\ \ i,j=1,2
\end{equation}
with $\mathcal{E}_{i}^m\equiv((E_{\hat{x}}+iE_{\hat{3}})^m,(E_{\hat{1}}+iE_{\hat{2}})^m)$, 
where $\sigma^r$ are the Pauli matrices. Defining 
$\tilde{\mathcal{E}}^m_i\equiv((E_{\hat{1}}-iE_{\hat{2}})^m,-(E_{\hat{x}}-iE_{\hat{3}})^m)$, 
it also satisfies 
\begin{equation}
  \tilde{\mathcal{E}}^m_i{(\Phi^r)_m}^n=-i{(\sigma^r)_i}^j\tilde{\mathcal{E}}_j^n\ .
\end{equation}
The $\Phi^r$ in (\ref{phi3}) are chosen such that the right 
hand side is proportional to $\sigma^r$. (The overall factor 
in front of $\sigma^r$ should be $-i$ for the relation 
(\ref{3-complex}) to hold.)
One can similarly introduce a complex basis in the field space. 
For small scalar fluctuations, we can approximate the hypermultiplet target space to be flat.  
Combining four real scalars $q^X$ to two complex scalars $Q^a=(Q,\tilde{Q}^\ast)$, 
with $a=1,2$, $q^Y{(J^r)_Y}^X$ can be rewritten as
$-iQ^b{(\sigma^r)_b}^a$.  As we shall see below, with the gauging 
$D_\mu Q^a\equiv (\partial_\mu-2iA_\mu)Q^a$, $Q$ will become the complex scalar 
which contains $\varphi$.

Now contracting the equation (\ref{hyper-eqn}) with 
$\mathcal{E}_{i}^m$ or $\tilde{\mathcal{E}}_i^m$, and changing the real index $X$
to the complex index $a$, we study the BPS equation for $Q^a$.
Using ${(\sigma^r)_i}^j{(\sigma^r)_b}^a=\frac{3}{2}\delta^a_i\delta_b^j
-\frac{1}{2}{(\sigma^s)_i}^a{(\sigma^s)_b}^j$, these equations reduce to
\begin{equation}
  \mathcal{E}_i^m D_mQ^i=0\ ,\ \tilde{\mathcal{E}}_i^m D_m Q^i=0\ .
\end{equation}
In terms of the component scalars, 
these equations are 
\begin{equation}\label{BPS-all}
  (E_{\hat{x}}+iE_{\hat{3}})^mD_mQ=-(E_{\hat{1}}+iE_{\hat{2}})^mD_m\tilde{Q}^\ast\ ,\ 
  (E_{\hat{1}}-iE_{\hat{2}})^mD_mQ=(E_{\hat{x}}-iE_{\hat{3}})^mD_m\tilde{Q}^\ast\ .
\end{equation}
Keeping only one complex scalar  $Q$ while turning off the other $\tilde{Q}=0$,
one obtains  
\begin{equation}\label{hyper-BPS}
  (E_{\hat{x}}+iE_{\hat{3}})^mD_mQ=0\ \ ,\ \ 
  (E_{\hat{1}}-iE_{\hat{2}})^mD_mQ=0\ .
\end{equation} 
This is the BPS equation for the complex scalar $Q$ that we wish to study.\footnote{On the other hand, turning off $Q=0$, $\tilde{Q}$ satisfies similar equations with $D_\mu \tilde{Q}\equiv (\partial_\mu+2iA_\mu)\tilde{Q}$. 
Solving these equations following our studies in the next paragraph, 
one finds that there are no normalizable 
solutions at $x\rightarrow\infty$. In general, we think  
the normalizability of the solutions of (\ref{BPS-all}) 
at $x\rightarrow\infty$ should demand $\tilde{Q}=0$.}
We have checked that (\ref{hyper-BPS}) implies the following 
equation of motion 
\begin{equation}\label{scalar-eom}
  {\textstyle 
  \left[\frac{1}{\sqrt{-g}}D_\mu(\sqrt{-g}g^{\mu\nu}D_\nu)+\frac{4}{\ell^2}\right]Q=0}\ .
\end{equation}
Inserting the inverse-vierbeins, (\ref{hyper-BPS}) can be written as 
\begin{equation}\label{hair-1st-eqn}
  \left({\textstyle xh\frac{\partial}{\partial x}
  +i\frac{\partial}{\partial\psi}}\right)Q={\textstyle -\frac{2}{\ell}}UQ\ \ ,\ \ 
  \left({\textstyle \sin\theta\frac{\partial}{\partial\theta}
  -i(\frac{\partial}{\partial\phi}-\cos\theta\frac{\partial}{\partial\psi})}\right)Q=0
  \ .
\end{equation}
We believe that $Q$ is the complex scalar of \cite{Bhattacharyya:2010yg} appearing 
in their consistent truncation. (It carries the same $U(1)$ charge as their scalar and 
satisfies the same linearized equation of motion.)
When $Q$ depends on $x$ only, the first equation 
reduces to (\ref{scalar-eqn-BPS}) with $Q=2\sinh\varphi\approx 2\varphi$.

Now we study all smooth solutions of (\ref{hair-1st-eqn}) 
at $\mathcal{O}(\varepsilon^1)$. We take 
\begin{equation}
  Q(x,\theta,\phi,\psi)=\Phi(x)e^{im\psi}f(\theta,\phi)
\end{equation}
where $m$ is half-integral. The equations reduce to
\begin{equation}
  \Phi^\prime=\frac{m\ell^2-x-2q}{x(x+3q+\ell^2)}\Phi\ \ \ ,\ \ \ 
  z^\ast\frac{\partial f}{\partial z^\ast}=
  2m\frac{1-|z|^2}{1+|z|^2}f
\end{equation}
where $z\equiv \tan\frac{\theta}{2}e^{-i\phi}$.  
The solutions are given by
\begin{eqnarray}\label{spinning-hair}
  \Phi(x)&=&{\textstyle \varepsilon x^{\frac{m-2q/\ell^2}{1+3q/\ell^2}}
  \left(1+\frac{3q}{\ell^2}+\frac{x}{\ell^2}\right)^{-\frac{1+m+q/\ell^2}{1+3q/\ell^2}}}\\
  f(\theta,\phi)&=&{\textstyle \left(\frac{z^\ast}{(1+|z|^2)^2}\right)^{\frac{m}{2}}z^\alpha
  =(\cos\frac{\theta}{2})^{\frac{3m}{2}-\alpha}
  (\sin\frac{\theta}{2})^{\frac{m}{2}+\alpha}}e^{i(\frac{m}{2}-\alpha)\phi}\nonumber
\end{eqnarray}
where the factor $z^\alpha$ represents the integration constant for $z^\ast$ given by
arbitrary functions of $z$. 
One should take $\alpha=-\frac{m}{2},-\frac{m}{2}+1,\cdots,\frac{3m}{2}$
to have regular solutions on $S^2(\theta,\phi)$, 
\begin{equation}
  f(\theta,\phi)e^{im\psi}={\textstyle (\cos\frac{\theta}{2}e^{i\phi_1})^{m_1}
  (\sin\frac{\theta}{2}e^{i\phi_2})^{m_2}}\ \ ,\ \ (\psi,\phi)=\phi_1\pm\phi_2
\end{equation}
where $m_1+m_2=2m$ and $m_1,m_2=0,1,2,\cdots$.
The angle-dependent part is the BPS spherical harmonics. 
It is  just the usual `derivative' factor for the BPS graviton wavefunctions 
in AdS$_5$, or in the dual QFT,   
$(\partial_{+\dot{+}})^{m_1}(\partial_{+\dot{-}})^{m_2}$ acting on the 
conformal primary operators. One finds the following behaviors at infinity and on the horizon:
\begin{equation}
  \Phi(x)\sim \left\{
  \begin{array}{ll}
    x^{-1}&\textrm{for }x\rightarrow\infty\\
    x^{\frac{m-2q/\ell^2}{1+3q/\ell^2}}&\textrm{for }x\rightarrow 0
  \end{array}
  \right.\ .
\end{equation} 
We get the expected normalizable boundary condition at $x\rightarrow\infty$.
At the horizon $x\rightarrow 0$, the solution remains finite only for the modes satisfying 
\begin{equation}\label{over-rotate}
  m=\frac{m_1+m_2}{2}\geq \frac{2q}{\ell^2}\ . 
\end{equation} 

% Till here

This gives an intuitive understanding of the well-behaved perturbative BPS hairs. 
If the scalar field ($\sim$ particle) rotates with large spherical harmonics 
quantum numbers $m_1,m_2\gg 1$ in vacuum AdS, its average  
radial position will be far away from the center. On the other hand, the 
conformal primary state with $m_1,m_2=0$ prefers to be at the central region 
of AdS. This can be seen from $\Phi(x)$ of (\ref{spinning-hair}) 
at $q=0$ (no black hole):
\begin{equation}
  \Phi(x)\rightarrow\frac{\varepsilon  x^m}{(1+x/\ell^2)^{m+1}}\ .
\end{equation}
The radial wavefunction $\Phi(x)$ is peaked at $x=x_\star\equiv m\ell^2$, but the peak is not sharp.
This is basically because the damping after the peak $x_\star$ is
slow, related to its slow asymptotic decay $\Phi\sim x^{-1}$ for the 
mass $M^2\ell^2=-4$. Anyway, the average 
radial position moves outward for larger $m=\frac{m_1+m_2}{2}$.
Now imagine placing the same field ($\sim$ particle) in the black hole 
background. The central region of AdS is now occupied by the black hole, behind the 
event horizon. Therefore, the fields ($\sim$ particles) with lower $m_1,m_2$ 
quantum numbers will be more likely to be swallowed by the black hole, or to back-react 
more strongly to it. (\ref{spinning-hair}) shows that, below the threshold 
$\frac{2q}{\ell^2}$ for $m$, this expectation is manifested by the 
exterior wavefunction exhibiting a divergence at the event horizon.
To have reasonable hairy black holes below this threshold,  the 
hair should give a large back-reaction to the background black hole.

Our studies so far intrinsically used the wave picture.
Let us also consider the particle picture.
Note that $\Phi(x)$ in (\ref{spinning-hair}) is never peaked around its maximum at 
any choice of $m,q$, basically due to the slow fall-down $\sim x^{-1}$ 
at large $x$.
However, if one considers a semi-classical Kaluza-Klein graviton with large mass
and large orbital angular momentum quantum number $m$,
one can have a sharply peaked wavefunction in AdS.
%However, if one considers Kaluza-Klein gravitons with large enough masses,
%one can have sharply peaked wave functions in AdS if the orbital 
%angular momentum quantum number $m$ is also large.
We shall reconsider the 
perturbative hair as a particle probe in this case, assuming that it is realized 
by a sharply peaked wavepacket of a field with large mass and angular momenta. 
This viewpoint will provide new understandings of 
(\ref{over-rotate}). For instance, consider a spinless graviton 
particle dual to the chiral primary (\ref{scalar-primary}) with $n\gg 1$.
The scaling dimension $\Delta=n$ and the charge $\delta R=\frac{n}{3}$ of this 
operator are dual to the following mass and charge for the graviton:
\begin{equation}
  (M\ell)^2=n(n-4)\approx n^2\ ,\ \ \delta R={\textstyle \frac{n}{3}}\ .
\end{equation}
The bosonic probe particle action in the black hole background 
(\ref{background-bh}) is given by 
\begin{equation}
  S={\textstyle \frac{n}{\ell}}\int d\tau\left[
  -\sqrt{-g_{\mu\nu}\dot{x}^\mu\dot{x}^\nu}
  -{\textstyle \frac{1}{3}} A_\mu\dot{x}^\mu\right]\ .
\end{equation}
We try to find the rotating BPS solutions for the time-like worldline with 
$t(\tau)=\tau$, outside the horizon $x>0$. Denoting by $\delta J_1,\delta J_2$ 
the Noether charges of the particle for the rotations on $\phi_1,\phi_2$, respectively,
one finds the following stationary radius $x$ as a function of charges:
\begin{equation}\label{bound-particle}
  x=\ell^2\left(\frac{\delta J_1+\delta J_2}{3\delta R}-\frac{2q}{\ell^2}\right)
  \equiv \ell^2\left(\frac{2\delta J}{3\delta R}-\frac{2q}{\ell^2}\right)>0
  \ \rightarrow\ \delta J>\frac{nq}{\ell^2}\ .
\end{equation}
The particle can orbit outside the event horizon when the particle 
is rotating fast enough. This is morally in accordance with (\ref{over-rotate}). 
In particular, blindly inserting  $n=2$ and $\delta J=m$ for the 
quantum particle of the field $\Phi$, one happens to recover the bound (\ref{over-rotate}).

When the bound (\ref{bound-particle}) is 
saturated, the particle is precisely at the horizon.
Then it will be ambiguous to distinguish whether this is a hair 
or part of the black hole, at least in the point particle approximation. 
Indeed one finds an interesting signature of this ambiguity. To explain 
this, recall that the background black hole satisfies a charge relation 
(\ref{charge-relation}).
Absorbing a probe hair into the black hole will shift its charges by
$R \to R+\delta R$ and $J \to J+\delta J$,
%Adding a probe hair will give extra contributions to 
%the charges, $R+\delta R$, $J+\delta J$,
where $\delta R,\delta J$ from the 
probe should be much smaller than the background charges. Since we 
take the background charges $R,J$ to be at order $N^2$, 
the hair charges should also satisfy $\delta R\ll N^2$, $\delta J\ll N^2$. 
In this limit, the shift in polynomial $\mathcal{G}$ in (\ref{charge-relation}) is given approximately by
\begin{equation}\label{charge-relation-violation}
  \delta \mathcal{G}(R,J)\approx \frac{N^4}{2}
  \left(1+\frac{q}{\ell^2}\right)^3\left(\delta J-\frac{3q}{\ell^2}\delta R\right)\ .
\end{equation}
If the probe exists outside the horizon, the hair violates the charge relation 
(\ref{charge-relation}) in an over-rotating manner, $\mathcal{G}>0$. 
If the probe particle is exactly at the horizon, one finds from (\ref{bound-particle}) 
that the charge relation is respected. So adding such `hairs' moves the black hole 
within the charge sector of the non-hairy black holes, as long as their 
charges are concerned. An outside observer unable to 
see such particles exactly at the horizon, only seeing the total charges, 
may regard the net system as a non-hairy black hole satisfying 
the charge relation. This is just a naive probe particle picture. 
In the wave picture, the solution (\ref{spinning-hair}) has 
finite nonzero value $\Phi(0)$ at the horizon. Perhaps it might be interesting 
to study its interaction with the strong-coupling modes of the near-horizon 
AdS$_2$ \cite{Boruch:2022tno} and see if there are quantum lessons to be 
learned.\footnote{Also, since the divergence (\ref{s-wave-singular}) at $m_1,m_2=0$ 
below the threshold is absent in the finite temperature regularization
\cite{Markeviciute:2018yal,Markeviciute:2018cqs} of AdS$_2$, this may also be related 
to the coupling to the near-horizon modes. 
We thank Shiraz Minwalla and Gustavo J. Turiaci for pointing it out to us.}

To conclude, we found a picture on why the conformal primary states 
(with $m=0$ or $\delta J=0$) of a graviton cannot be the BPS black hole hairs 
while respecting the superposition picture. This is in accordance with the QFT 
operator spectrum that we found in section 3 for $SU(2)$. Our studies in this section 
also imply that particles with fast enough orbital motions may provide black hole hairs 
along the strategy of \cite{Basu:2010uz,Bhattacharyya:2010yg}. This is why we had in mind 
the hairs from conformal descendant gravitons in the range $t^{34}\sim t^{39}$, beyond 
just a logical possibility.
Although the results in our sections 3 and 4 are in the opposite extreme, 
$N=2$ vs. $\infty$, we think this picture is natural. It will be 
exciting to further explore this issue by finding new black hole cohomologies at higher $N$, 
and also by exploring hairy black holes more comprehensively.

\section{Conclusion and remarks}

In this paper, we studied the classical cohomologies of local BPS 
operators in $\mathcal{N}=4$ Yang-Mills theory with $SU(N)$ gauge group. 
Even for low $N$'s, we distinguish the cohomologies for the gravitons (subject 
to the stringy exclusion principle) and the rest. The latter have a chance to 
describe quantum black hole microstates in AdS/CFT at finite Newton constant 
if $N$ is finite. In the $SU(2)$ theory, we constructed an infinite family 
$O_n$ of non-graviton cohomologies. We studied the partial no-hair behavior of 
these cohomologies when graviton cohomologies are multiplied. We first showed that 
almost all conformal primary gravitons multiplied to our new cohomologies do not appear 
in the superconformal index, providing in a sense a partial no-hair theorem. 
One natural explanation is that the product cohomologies are trivial, 
i.e. $Q$-exact. For a selection of products involving $O_0$ and simple conformal 
primary gravitons, we explicitly showed the $Q$-exactness. On the other hand, 
the index allows a chance that conformal 
descendant gravitons may dress our new cohomologies. We provided a gravity 
interpretation of these phenomena by studying when the graviton hairs can be
perturbatively superposed with non-hairy black holes.

We would like to discuss various issues and further interesting directions.

We should be able to construct new cohomologies 
at larger $N$ and large charges. This will allow us to address dual black 
hole physics more quantitatively. Considering that this is a completely well 
defined and classical combinatorics problem, it is amazing to confront the computational 
complexity towards its solutions. In this paper we mostly relied on analytic insights to 
construct a small subset of these new cohomologies, and this strategy may continue 
to work to a certain extent for larger charges and $N$. However, to understand the 
full landscape of these cohomologies, probably one should rely on efficient 
computerized calculations. It will also be valuable to understand the time and 
resources required for this calculation, which by itself may shed light 
on the nature of the black hole microstates.

Since constructing all  cohomologies is cumbersome, 
we employed a streamlined strategy. We computed the index over finite $N$ 
gravitons and subtracted it from the full index to first notice which 
charge sectors host new cohomologies. Although finite $N$ gravitons 
are completely well defined, enumerating them 
without overcounting is tricky due to the trace relations. In fact, computing the 
finite $N$ graviton index has been a major bottleneck of this project.
Since the basic ingredients are diagonal matrices, we hope that this 
combinatoric problem can be solved without too much difficulty.

To count the finite $N$ gravitons more easily, we extensively 
studied the so-called BMN sector of the cohomology problem. Counting graviton 
cohomologies in this sector is much easier since one just needs to consider 
diagonal matrices without any derivatives. For $SU(2)$ theory, 
we could analytically count the BMN gravitons completely. After subtracting 
the graviton index from the full index, we constructed all the cohomologies that saturate 
the remaining index. We hope that similar calculations in the BMN sector will be not 
too difficult for higher $SU(N)$ groups. We expect the BMN spectrum to be 
much more interesting at higher $N$. For instance, having in mind applying 
the large $N$ techniques of \cite{Choi:2021lbk}, we envision that the 
`small black hole' like scaling of the entropy $S\sim\frac{j^{3/2}}{N}$
will be discovered when  $\frac{j}{N^2}$ is given by a small number independent of 
$N$.

For higher $N$, it is important to find  the `threshold' cohomology, 
the lowest operator not of graviton type. For $N=2$, it has scaling dimension 
$E=\frac{19}{2}$ which is quite larger than $N^2=4$. However, the threshold for 
large $N$ should be much smaller than $N^2$. This is because
small black holes exist at charges given by $\epsilon N^2$ with an arbitrary 
small number $\epsilon$ independent of $N$. To accommodate such black holes, the threshold 
energy level should not scale like $N^2$. 
The situation is in contrast to the black hole threshold 
in AdS$_3$. BTZ black holes start to appear at an energy proportional to
the central charge, which is a parameter analogous to $N^2$, 
without having a small black hole branch with negative specific heat. 

Perhaps the infinite tower of new cohomologies $O_n$ in section 3.1 is the 
most surprising discovery of this paper. This strongly signals that the 
black hole cohomologies may have unexpected emergent structures. It will be interesting 
to see if similar towers can be found for higher $SU(N)$ theories. Also, it may 
be interesting to generalize these findings to other $\mathcal{N}=1$ SCFTs with 
or without gravity duals. In particular, since the operator $O_0$ involves a
baryon-like factor $\psi_{m+}\cdot(\psi_{n+}\times\psi_{p+})$ of the $SO(3)$ 
vector-valued fields, it may be possible to understand and generalize them 
from the viewpoint of Regge trajectories.

We also emphasize that we provided a rather operational criterion to distinguish 
graviton and black hole cohomologies, although we think it is natural. It should 
be valuable to establish more intrinsic (maybe information theoretic) criteria 
to distinguish the two.

We would like to comment on aspects of hairy black holes as seen by our cohomologies. 

One feature of non-hairy BPS black holes is the charge relation. 
In AdS$_5$, this relation is given by (\ref{charge-relation}). Why 
this relation holds, if any, and how hairy black holes violate it are still 
not well understood microscopically. Here note that recent studies of BPS black holes 
were often made using the index which cannot see one charge. 
So the charge relation cannot be addressed with the index. On the other hand, once we 
construct the cohomologies, all charge information is available and the charge 
relation can be studied in principle.

We also try to classify the possible patterns of hairs. 
We find three possibilities. In a sense, all three possibilities 
are realized in our new cohomologies, at least morally. 
\begin{itemize}

\item The first class of hairs is discussed in detail in this paper,  
`superposing' already existing black holes and gravitons. In gravity, 
this generalizes the spirit of \cite{Basu:2010uz,Bhattacharyya:2010yg} to the BPS 
sector, and the back-reaction to each other is controlled and suppressed as 
one tune the parameters of the solutions. We probed this class of hairs 
by studying the perturbative scalar beyond s-waves in section 4. In 
the QFT dual, non-$Q$-exact product cohomologies of black holes and gravitons 
are interpreted as hairy black hole operators in this class.

\item Second, there may be hairy black holes in which the back-reaction of 
the hair to the core black hole is essential. The numerical hairy black holes 
of \cite{Markeviciute:2018yal,Markeviciute:2018cqs} seem to belong to 
this class, in that even very small hairs demand completely new core black holes with 
new near-horizon structures. Such cores are 
possible only by the back-reaction of the hair. Somewhat curiously, 
we know a black hole cohomology that behaves like this. It is our $SU(2)$ threshold 
operator $O_0$ (\ref{n=0}). There, two on-shell (i.e. $Q$-closed) graviton cohomologies 
$v^m_{\ n}$ are multiplied to an off-shell core of the form 
$\psi_{m+}\cdot (\psi_{n+}\times \psi_{p+})$. The latter could be made on-shell 
only with the aid, or back-reactions, of the gravitons.

\item Third, it is a logical possibility that the back-reaction of the core black hole to 
the gravitons may allow qualitatively new hairs. This may have been 
realized in our black hole cohomologies $O_{n\geq 1}$, (\ref{On}) or (\ref{On-summary}).
The first term of this operator takes the form of an on-shell black hole operator 
$O_0$ times an off-shell matter $(f\cdot f)^n$. After adding two more terms as 
a result of the back-reactions, this could be promoted to an on-shell operator. 
It will be interesting to clarify whether this signals a more exotic form of hairs or 
otherwise this is better viewed as representing internal structures of black holes.

\end{itemize}

\vskip 0.5cm

\hspace*{-0.8cm} {\bf\large Acknowledgements}
\vskip 0.2cm

\hspace*{-0.75cm} 
We thank Francesco Benini, Jaehyeok Choi, Nakwoo Kim, Seungkyu Kim, Ki-Hong Lee, Kimyeong Lee, Shiraz Minwalla, 
Silviu S. Pufu, Jaewon Song and Gustavo J. Turiaci for helpful discussions. This work is supported in 
part by a KIAS Individual Grant PG081602 at Korea Institute for Advanced Study (SC), 
the DoE grant DE-SC0007859 (SL) and the NRF grants 2021R1A2C2012350 (SK, EL), 2021R1A6A1A10042944
(JP), 2021R1A2C1012440 (JP).

\appendix

\section{Supersymmetry and supergravitons}

In this section, we list symmetry transformations for the classical Poincare supercharges 
in the $\mathcal{N}=4$ SYM theory.
The $\mathcal{N} = 4$ SYM fields comprise a gauge field $A_{\alpha\dot{\beta}}$, six scalars $\Phi_{ij}$ 
and fermions $\Psi_{i\alpha}$, $\overline{\Psi}^{j\dot{\alpha}}$.
The transformations of the fields generated by some Poincare supercharges are given by \cite{Biswas:2006tj,Grant:2008sk}
\begin{align}
  [Q^i_\alpha,\Phi_{jk}]&=\delta^i_j\Psi_{k\alpha}-\delta^i_k\Psi_{j\alpha}
  \nonumber\\ 
  \{Q^i_\alpha, \Psi_{j\beta}\}&=-2i\delta^i_jf_{\alpha\beta}
  -\epsilon_{\alpha\beta}[\overline{\Phi}^{ik},\Phi_{kj}]
  \nonumber\\
  \{Q^i_\alpha,\overline{\Psi}^j_{\dot\beta}\}&=
  2iD_{\alpha\dot\beta}\overline{\Phi}^{ij}
  \nonumber\\
  [Q^i_\alpha,A_{\beta\dot{\gamma}}]&=-\epsilon_{\alpha\beta}
  \overline{\Psi}^i_{\dot\gamma},
\end{align}
and the action of $\overline{Q}_{m\dot{\alpha}}$ is given similarly.
Our focus will be on the $\frac{1}{16}$-BPS operators that are annihilated by both 
$Q\equiv Q^4_-$ and $S\equiv S^-_4$.
The BPS fields are given by
\begin{align}\label{bpsletterappendix}
\bar{\phi}^n\equiv \overline{\Phi}^{4m}, \quad \psi_{n+}\equiv-i\Psi_{n+}, \quad n=1,2,3, \quad \bar{\lambda}_{\dot{\alpha}}\equiv\bar{\Psi}^4_{\dot{\alpha}},\quad f_{++},
\end{align}
 together with the covariant derivatives
 \begin{align}
     D_{+\dot{\alpha}}=\partial_{+\dot{\alpha}}-i[A_{+\dot{\alpha}},\ \cdot\ ]\ ,
 \end{align}
where $[D_{+\dot{\alpha}},D_{+\dot{\beta}}]=\epsilon_{\dot{\alpha}\dot{\beta}}f_{++}$. 
The letters are also constrained by the equation of motion $D_{+\dot{[\alpha}}\bar{\lambda}_{\dot{\beta]}}=\epsilon_{\dot{\alpha}\dot{\beta}}[\psi_{n+},\bar{\phi}^n]$.

The transformation rules of BPS fields under $Q$ are given by
\begin{align}
[Q,\bar{\phi}^n]&=0\nonumber\\
\{Q,\psi_{n+}\}&=-i\epsilon_{nmp}[\bar{\phi}^m,\bar{\phi}^p]\nonumber\\
\{Q,\bar{\lambda}_{\dot{\beta}}\}&=0\nonumber\\
[Q,f_{++}]&=-i[\psi_{m+},\bar{\phi}^m]\nonumber\\
[Q,D_{+\dot{\alpha}}]&=-i[\bar{\lambda}_{\dot{\alpha}},\ \ \}\ .
\end{align}
The supercharges $Q^m_{+}$ and $\bar{Q}_{m\dot{\alpha}}$ in $PSU(1,2|3)$, which commute with 
$Q$, act on the BPS fields as follows 
\begin{align}
[Q^m_{+},\bar{\phi}^n]&=i\epsilon^{mnp}\psi_{p+}\nonumber\\
\{Q^m_{+},\psi_{n+}\}&=-2i\delta^m_n f_{++}\nonumber\\
\{Q^m_{+},\bar{\lambda}_{\dot{\alpha}}\}&=-2iD_{+\dot{\alpha}}\bar{\phi}^m\nonumber\\
[Q^m_{+},D_{\dot{+\alpha}}]&=0\nonumber\\
[Q^m_{+},f_{++}]&=0\nonumber\\
[\bar{Q}_{m\dot{\alpha}},\bar\phi^n]&=-\delta^{n}_{m}\bar{\lambda}_{\dot{\alpha}}\nonumber\\
\{\bar{Q}_{m\dot{\alpha}},\psi_{n+}\}&=-2\epsilon_{mnp}D_{+\dot{\alpha}}\bar\phi^p\nonumber\\
\{\bar{Q}_{m\dot{\alpha}},\bar{\lambda}_{\dot{\beta}}\}&=\epsilon_{\dot{\alpha}\dot{\beta}}\epsilon_{mnp}[\bar\phi^n,\bar\phi^p]\nonumber\\
[\bar{Q}_{m\dot{\alpha}}, D_{+\dot{\beta}}]&=-\epsilon_{\dot{\alpha}\dot{\beta}}[\psi_{m+},\ \ \}\nonumber\\
[\bar{Q}_{m\dot{\alpha}},f]&=D_{+\dot{\alpha}}\psi_{m+}.
\end{align}
The cohomologies in $S_2$ can be taken to be 
\begin{eqnarray}
  &&{\rm tr}(\bar\phi^{(m}\bar\phi^{n)})\ \ ,\ \  
  {\rm tr}(\bar\phi^m\bar\lambda_{\dot\alpha})\ \ ,\ \ 
  {\rm tr}(\bar\lambda_{\dot{+}}\bar\lambda_{\dot{-}}),\\
  &&{\rm tr}(\bar\phi^m\psi_{n+})-{\textstyle \frac{1}{3}}\delta^m_n
  {\rm tr}(\bar\phi^l\psi_{l+})\ \ ,\ \ 
  {\rm tr}(\bar\lambda_{\dot\alpha}\psi_{m+}
  -\epsilon_{mnp}\bar\phi^nD_{+\dot\alpha}\bar\phi^p),\nonumber\\
  &&{\rm tr}(\bar\phi^m f_{++}-{\textstyle\frac{1}{4}}\epsilon^{mnp}\psi_{n+}\psi_{p+})
  \ \ ,\ \ {\rm tr}(\bar\lambda_{\dot\alpha}f_{++}-{\textstyle\frac{2}{3}}
  \psi_{m+}D_{+\dot\alpha}\bar\phi^m+{\textstyle\frac{1}{3}}\bar\phi^m D_{+\dot\alpha}\psi_{m+}).
  \nonumber
\end{eqnarray}
as well as arbitrary number of derivatives $\partial_{+\dot\alpha}$ acting on them.

We can represent the $SU(2)$ adjoint fields in the basis of Pauli matrices:
\begin{align}
\bar{\phi}^1&=\textstyle{\frac{1}{\sqrt{2}}}X\cdot \vec{\sigma}, \quad \bar{\phi}^2=\textstyle{\frac{1}{\sqrt{2}}}Y\cdot \vec{\sigma}, \quad \bar{\phi}^3=\textstyle{\frac{1}{\sqrt{2}}}Z\cdot \vec{\sigma}, \quad \bar{\lambda}_{\dot{\alpha}}=\lambda_{\alpha}\cdot \vec{\sigma}\nonumber\\
{\psi}_{1+}&=\psi_1\cdot \vec{\sigma}, \quad \psi_{2+}=\psi_2\cdot \vec{\sigma}, \quad \psi_{3+}=\psi_3\cdot \vec{\sigma}, \quad f_{++}=-\textstyle{\frac{1}{\sqrt{2}}}f\cdot\vec{\sigma}
\end{align}
where $\vec{\sigma}\equiv(\textstyle{\frac{\sigma_1}{2}},\textstyle{\frac{\sigma_2}{2}},\textstyle{\frac{\sigma_3}{2}})$ and $[\sigma_i,\sigma_j]=2i\epsilon_{ijk}\sigma_{k}$.
$\phi^m=(X,Y,Z)$, $\psi_m$, $\lambda_{\alpha}, f$ denote 3-dimensional vectors. 
We abbreviate bars and dots for vectors. In this notation, inner and outer products will be used instead of trace and commutators.
The $Q$-transformations are given by
\begin{equation}
  Q\phi^m=0\ ,\ \ Q\psi_m={\textstyle \frac{1}{2}}\epsilon_{mnp}\phi^n\times \phi^p
  \ ,\ Qf=\phi^m\times\psi_m\ ,\ [Q,D_{{\alpha}}]=\lambda_{{\alpha}}\times
\end{equation}
where the field strength is written as $\epsilon_{\alpha\beta}f\times=-\sqrt{2}[D_{\alpha},D_{\beta}]$ and the equation of motion is written as $-\sqrt{2}D_{[\alpha}\lambda_{\beta]}=\epsilon_{\alpha\beta}\phi^m\times \psi_m$ 
In the vector notation, the cohomologies in $S_2$ are written as
\begin{eqnarray}
  &&\phi^{(m}\cdot\phi^{n)}\ ,\ \ 
  \phi^m\cdot\lambda_{\alpha}\ ,\ \ 
  \lambda_{{+}}\cdot\lambda_{{-}}\ ,\nonumber\\
  &&\phi^m\cdot\psi_{n}-{\textstyle \frac{1}{3}}\delta^m_n
  \phi^p\cdot\psi_{p}\ ,\ \ 
  \lambda_{\alpha}\cdot\psi_{m}-
  \frac{1}{2}\epsilon_{mnp}\phi^n\cdot D_{\alpha}\phi^p\ ,\nonumber\\
  && {\phi}^m\cdot f+{\textstyle \frac{1}{2}}\epsilon^{mnp}\psi_n\cdot\psi_p\ ,\ \ 
  \lambda_{\alpha}\cdot f+{\textstyle\frac{2}{3}}
  \psi_{m}\cdot D_{\alpha}\phi^m-{\textstyle \frac{1}{3}}\phi^m \cdot D_{\alpha}\psi_{m}\ .
\end{eqnarray}
When we use vector notation to act $Q^m_+$ on $S_2$ multiplet, a set of numerical factors emerges. For instance, $Q^m_{+}\bar{\phi}^n$ equals $\sqrt{2}i\epsilon^{mnp}\psi_p$. However, we can simplify matters by defining the action of $Q^m_+$ in a way that eliminates these numerical factors: 
\begin{align}
    Q^m_+ \phi^n &\equiv\epsilon^{mnp} \psi_p, \nonumber\\
    Q^m_+ \psi_n &\equiv \delta^{m}_n f, \nonumber\\
    Q^m_+ {\lambda}_{{\alpha}} &\equiv -D_{+{\alpha}}\phi^m.
\end{align}
According to this definition, 
\begin{align}
    Q^n_+(\phi^m\cdot\psi_{n})={\phi}^m\cdot f+{\textstyle \frac{1}{2}}\epsilon^{mlp}\psi_l\cdot\psi_p \quad (m\neq n),
\end{align}
where $n$ is not subject to any summation.

\section{Representations of $PSU(2,2|4)$ and $\frac{1}{16}$-BPS states}

In this section we review the structures of representations of the $\mathcal{N}=4$
superconformal group $PSU(2,2|4)$, and of its subgroup $PSU(1,2|3)$ that commutes 
with $Q$. We refer the readers to \cite{Cordova:2016emh} for more details.

Superconformal group of 4-dimensional $\mathcal{N}=4$ theory is $PSU(2,2|4)$,
whose bosonic subgroup consists of the 4d conformal group $SO(2,4)$
and the R-symmetry $SU(4)$.
States, or weights, of its representations are often labeled with scaling dimension $E$,
Dynkin labels $j$ and $\bar{j}$ for the Lorentz group $SO(4) = SU(2) \times SU(2)$,
and Dynkin labels $[r_1, r_2, r_3]$ for the R-symmetry group $SU(4)$. (We hope that 
$j$ appearing in the representation theory here, $j=J_1+J_2$, is not confused 
with $j=6(R+J)$ that often appears in the series expansion in the main part of 
this paper.)
We often refer to scaling dimension $E$ as energy with radial quantization in mind.
A representation of the bosonic subalgebra is often expressed as
\begin{eqnarray}
[j; \bar{j}]^{[r_1,r_2,r_3]}_E~,
\end{eqnarray}
using its highest weights.
It is also conventional to define angular momenta
\begin{eqnarray}\label{appdefJ}
J_1 = \frac{j + \bar{j}}{2}~, \qquad J_2 = \frac{j - \bar{j}}{2} 
\end{eqnarray}
and R-symmetry charges
\begin{eqnarray}\label{appdefR}
R_1 = r_2 + \frac{r_1+r_3}{2}~, \qquad
R_2 = \frac{r_1+r_3}{2}~, \qquad
R_3 = \frac{r_1-r_3}{2}~,
\end{eqnarray}
as charges for rotations on the orthogonal 2-planes.
Dynkin labels $j$, $\bar{j}$, $r_1$, $r_2$ and $r_3$ are all integers so that $j \pm \bar{j}$
is even for bosons and odd for fermions.
It follows that $J_1$ and $J_2$ are integers for bosons and half-integers for fermions.

A representation of the superconformal group $PSU(2,2|4)$ is completely determined by its
superconformal primary.
Given a superconformal primary that is annihilated by conformal supercharges,
superconformal descendants are obtained by the action of supercharges,
of which there are 16 in the 4d $\mathcal{N}=4$ theory.
There are a finite number of such descendants due to fermion statistics,
and following each descendant is an infinite number of conformal descendants,
obtained by acting any number of $P^\mu$.

By unitarity, the scaling dimension of a superconformal primary is bounded from below.
Let $[j; \bar{j}]^{[r_1,r_2,r_3]}_E$ be the bosonic subalgebra representation of the primary.
For generic $j$, $E$ must be not smaller than certain value determined by its quantum numbers
and in the exceptional case of $j=0$, an isolated value of $E$ which is exactly 2 below the bound is allowed.
To summarize,
\begin{eqnarray}\label{appunibound}
E &\geq& 2 + j + \frac12 \left( 3r_1 + 2r_2 + r_3 \right) = 2 + J_1 + J_2 + R_1 + R_2 + R_3~, \nonumber \\
{\rm or} \quad E &=& \frac12 \left( 3r_1 + 2r_2 + r_3 \right) = R_1 + R_2 + R_3 \qquad {\rm and} \qquad j=0~.
\end{eqnarray}
If the inequality is not saturated, the superconformal representation that follows from the primary
is \emph{long}: the primary is not annihilated by any of supercharges.
The representation is said to be of $L$-type.
If the inequality is saturated, the representation is \emph{short}.
If further $j \neq 0$ then a supercharge annihilates the primary,
and the representation is said to be of $A_1$-type.
If instead $j = 0$ then successive action of two supercharges is needed to annihilate the primary,
and the representation is said to be of $A_2$-type.
Finally, in the exceptional case of the second line of (\ref{appunibound}),
the primary is annihilated by some supercharges
and the representation is said to be of $B_1$-type.

Similar unitarity bounds and classification of representations
according to $\bar{j}$ in place of $j$ and $r_1 + 2r_2 + 3r_3$ in place of $3r_1 + 2r_2 + r_3$
apply independently, although this classification is less relevant to the following discussion.
So for instance, the superconformal representation with primary 
$[j; \bar{j}]^{[r_1,r_2,r_3]}_E$ that belongs to $A_1$-type and $\bar{L}$-type with 
respective definitions above, will be denoted as 
$$A_1 \bar{L} [j; \bar{j}]^{[r_1,r_2,r_3]}_E~.$$
For more details, see \cite{Cordova:2016emh}.

We turn to a subject more relevant to the body of this paper:
$\frac{1}{16}$-BPS contents of these representations.
We choose a $\frac{1}{16}$-BPS sector that is annihilated by a supercharge $Q\equiv Q^4_-$
with quantum numbers
\begin{eqnarray}\label{appQqn}
{\textstyle 
(E, j, \bar{j}, r_1, r_2, r_3 ) = (\frac12, -1, 0, 1, 0, 0)
\quad \leftrightarrow \quad (E,J_1,J_2,R_1,R_2,R_3) = (\frac12, -\frac12, -\frac12, 
\frac12, \frac12, \frac12)~.}
\end{eqnarray}
Subgroup of $PSU(2,2|4)$ that commutes with the chosen supercharge $Q$ is $PSU(1,2|3)$.
The Dynkin labels, or charges, associated with this subgroup have the same scaling dimension $E$
and $\bar{j}$ for the Lorentz group, as well as $[r_2, r_3]$ for the R-symmetry group. 
It is possible to decompose representations of the larger group $PSU(2,2|4)$ into multiple representations of the subgroup $PSU(1,2|3)$.

Unitarity puts a lower bound on the scaling dimension of any states in the theory:
\begin{eqnarray}\label{appBPSrelation}
E &\geq& j + \frac12 \left( 3r_1 + 2r_2 + r_3 \right) = J_1 + J_2 + R_1 + R_2 + R_3~.
\end{eqnarray}
A state that is annihilated by the chosen supercharge $Q$, in other words
a $\frac{1}{16}$-BPS state,
is characterized by saturation of the unitarity bound (\ref{appBPSrelation}).

The linearity of this relation has an important implication.
Suppose there are two representations of the bosonic subalgebra of $PSU(2,2|4)$,
say $[j ; \bar{j}]^{[r_1,r_2,r_3]}_{E_1}$ and $[k ; \bar{k}]^{[s_1,s_2,s_3]}_{E_2}$,
both of which saturate (\ref{appBPSrelation}).
The highest weights of both representations are $\frac{1}{16}$-BPS.
Now take the direct product of these two representations, and expand as a sum over irreducible representations.
Due to the linearity of the unitarity bound, only the one where Dynkin labels simply add up,
$[j+k ; \bar{j}+ \bar{k}]^{[r_1+s_1,r_2+s_2,r_3+s_3]}_{E_1+E_2}$,
contains $\frac{1}{16}$-BPS state.

Of 16 supercharges, there is a single supercharge $Q^\prime\equiv Q^4_+$ 
whose quantum numbers satisfy $E - J_1 - J_2 - R_1 - R_2 - R_3 = -2$.
For all other supercharges, $E - J_1 - J_2 - R_1 - R_2 - R_3 \geq 0$.
In particular, 9 supercharges that belong to $PSU(1,2|3)$ as well as our
$\frac{1}{16}$-BPS supercharge $Q$
have quantum numbers such that $E - J_1 - J_2 - R_1 - R_2 - R_3 = 0$.

With this information, let us now examine the $\frac{1}{16}$-BPS contents of superconformal representations.

First consider the $B$-type representations: when the superconformal primary $[j; \bar{j}]^{[r_1,r_2,r_3]}_E$
has $j=0$ and satisfies $E = \frac12 \left( 3r_1 + 2r_2 + r_3 \right) = R_1 + R_2 + R_3$.
Clearly the highest weight of this primary saturates (\ref{appBPSrelation}), it is a $\frac{1}{16}$-BPS state. It is annihilated by the supercharge $Q$ (because it saturates the 
bound (\ref{appBPSrelation})) and $Q^\prime$ (otherwise it will create a state 
violating  (\ref{appBPSrelation})), among others.
States obtained from the highest weight using lowering operators of the commuting subgroup $PSU(1,2|3)$
are also annihilated by $Q$, and are $\frac{1}{16}$-BPS states.
They belong to a representation $[\bar{j}]^{[r_2,r_3]}_E$ under bosonic subalgebra of 
$PSU(1,2|3)$. All states in the $PSU(1,2|3)$ representation that starts with the primary $[\bar{j}]^{[r_2,r_3]}_E$ are annihilated by $Q$ as the primary is.
Meanwhile, these are the only states in the original representation of $PSU(2,2|4)$
that are $\frac{1}{16}$-BPS.
This is because for a state in the $PSU(2,2|4)$ representation to not belong to the $PSU(1,2|3)$ representation, some operator that does not commute with $Q$ must be applied to the primary.
However, such an operator has $E - J_1 - J_2 - R_1 - R_2 - R_3 > 0$ so no weights in 
the product representation between the primary and such an operator can saturate the 
unitarity bound.

Next consider the $A$-type representations, either $A_1$ or $A_2$:
when the superconformal primary $[j; \bar{j}]^{[r_1,r_2,r_3]}_E$
satisfies $E = 2+ j+ \frac12 \left( 3r_1 + 2r_2 + r_3 \right) = 2 + J_1 + J_2 + R_1 + R_2 + R_3$.
No weights in the primary saturate the unitarity bound (\ref{appBPSrelation}).
For a state in the representation to be $\frac{1}{16}$-BPS,
it needs an agent that raises its charges more than its energy, and 
$Q^\prime$ is precisely the one.
Among direct product between the primary $[j; \bar{j}]^{[r_1,r_2,r_3]}_E$
and $Q^\prime = [1; 0]^{[1,0,0]}_{\frac12}$,
a single irreducible representation of the bosonic subalgebra $[j+1 ; \bar{j}]^{[r_1+1,r_2,r_3]}_{E+\frac12}$
contains weights that are $\frac{1}{16}$-BPS.
Such states are annihilated by $Q$ because now they saturate the bound 
(\ref{appBPSrelation}).
From this point, the analysis is the same as for $B$-type multiplets.
A $PSU(1,2|3)$ representation that starts from the primary $[\bar{j}]^{[r_2,r_3]}_{E+\frac12}$
of its bosonic subalgebra is the $\frac{1}{16}$-BPS contents of the original representation.

$L$-type representations do not contain any $\frac{1}{16}$-BPS states.

Before we conclude, we present some examples.

Free graviton multiplets are defined as a trace over an arbitrary number of symmetrized free scalars,
and their descendants.
From this principle, it is clear that the superconformal primary must be $[0;0]^{[0,n,0]}_n$.
$n$ is the number of scalars inside trace, and is restricted to $2 \leq n \leq N$ for the $SU(N)$ theory
due to trace relations.
This is of type $B \bar{B}$ according to the aforementioned classification.
It was named $S_n$ in \cite{Kinney:2005ej}. Therefore,
\begin{eqnarray}
S_n &=& B \bar{B}[0;0]^{[0,n,0]}_n ~.
\end{eqnarray}
Its $\frac{1}{16}$-BPS contents have appeared in Table \ref{graviton-multiplet}.

Also appearing with importance are $A_1 \bar{L}$-type multiplets.
For example, the black hole threshold operator with
$(E,j,\bar{j},r_1,r_2,r_3) = (\frac{19}{2},5,0,3,0,0)$
belongs to $A_1 \bar{L}[4;0]^{[2,0,0]}_9$. 
All the other BPS operators that we encounter in section 3 belong the
$A_1\overline{L}$ type multiplets, except for the pure BPS graviton multiplets 
(in $B\overline{B}$).

Here let us first explain why $A_1$-type multiplets appear so often 
in our discussions on new $\frac{1}{16}$-BPS states.
As explained above, the primary of an $A_1$-type multiplet have $j\neq 0$.
$\frac{1}{16}$-BPS states within this multiplet form a $PSU(1,2|3)$ multiplet.
Its primary is obtained by acting $Q^\prime$ on the primary of the $A_1$-type multiplet,
it therefore carries $j> 1$.
On the other hand, since the primary of an $A_2$-type multiplet carries $j=0$,
the $PSU(1,2|3)$ primary of the $\frac{1}{16}$-BPS states should carry precisely $j=1$. 
Finally, the primary of a $B$-type multiplet has $j=0$ and is also the primary of
an $PSU(1,2|3)$ multiplet that consists of $\frac{1}{16}$-BPS states.
Therefore, all the $PSU(1,2|3)$ primaries belonging to either $A_2$ or $B$ 
type multiplet should have $j=1$ or $0$. At this point, recall from the discussion below
(\ref{appBPSrelation}) that when taking a product of two representations that contain 
$\frac{1}{16}$-BPS states, only the irreducible representation whose Dynkin labels are 
the sum of those of both representations remain $\frac{1}{16}$-BPS. Therefore, when
constructing a $\frac{1}{16}$-BPS operator out of the BPS letters in the free theory 
($\bar\phi^m$, $\psi_{m+}$, $\bar\lambda_{\dot\alpha}$, $f_{++}$ and derivatives 
$D_{+\dot\alpha}$ on them), the quantum number $j\geq 0$ of these letters 
will only add and the cases with $j=0$ or $1$ will be rather rare.

First, the net quantum number $j$ of a $PSU(1,2|3)$ primary can be zero only if 
the operator is made only of $\bar\phi^m,\bar\lambda_{\dot\alpha}$, without any 
derivatives. The cohomologies in 
this sector are strongly believed to be well understood \cite{Kinney:2005ej}. 
They are the $\frac{1}{8}$-BPS chiral rings made of commuting or anticommuting 
(in the case of two $\bar\lambda_{\dot\alpha}$'s) letters within the trace, 
annihilated by $Q^{4}_+$ as well as our  $Q\equiv Q^4_-$. 
If they are made of scalars only, they are the primaries of $B\overline{B}$ for 
the gravitons, because in order for Dynkin labels to add up the scalars
must appear symmetrized.
If they contain $\bar\lambda_{\dot\alpha}$'s, they are descendants 
obtained by acting $\overline{Q}_{m\dot\alpha}$'s on the scalar primaries. So the 
case with $j=0$ is the graviton multiplets.

The $PSU(1,2|3)$ multiplet with $j=1$ for the primaries should belong to 
$A_2$. The $PSU(1,2|3)$ primaries in this class can contain arbitrary numbers of 
$\bar\phi^m$ and $\bar\lambda_{\dot\alpha}$ fields, together with
only one $\psi_{m+}$ or $D_{+\dot\alpha}$ to meet $j=1$.

With these structures in mind, we can understand some of the setups of section 3. 
Namely, we considered the core black hole primaries $O_n$ which contain five or more 
$\psi$ fields in section 3.1, and also considered the possible hairy black hole 
primaries which are products of gravitons and $O_n$. These operators have more than 
two $\psi$ fields, having $j>1$. For these, it suffices to consider the $A_1$ 
multiplets only. 

We can also understand why the BPS operators in section 3 belong to $A_1\overline{L}$-type
multiplets, rather than $A_1\overline{B}$, $A_1\overline{A}_2$ or $A_1\overline{A}_1$.
First, primaries of $A_1\overline{B}$-type multiplets of $PSU(2,2|4)$ satisfy 
$J_1=J_2$, $R_3+J_2+1=0$ and $E=1+J_1+R_1+R_2$. (See Table 21 of \cite{Cordova:2016emh},
knowing that their $R_{1,2,3}$ correspond to lowercase $r_{1,2,3}$ in (\ref{appdefR}).) 
The $\frac{1}{16}$-BPS primaries
obtained by acting $Q^\prime$ on them satisfy 
$J_1=J_2$, $R_3+J_2=0$ and
$E=R_1+R_2+J_1$. These states are enhanced 
$\frac{1}{8}$-BPS states, preserving extra supersymmetry $\overline{Q}_{3\dot{-}}$.
They are the so-called Schur operators.
Second, primaries of $A_1\overline{A}_2$- or $A_1\overline{A}_1$-type multiplets
satisfy $R_3+J_2=0$.
Acting $Q^\prime$, the $\frac{1}{16}$-BPS primaries in the multiplets satisfy $R_3+J_2=1$.
Thus, for all the $\frac{1}{16}$-BPS states in these three classes, one finds that 
$R_3+J_2=0$ or $1$.
On the other hand, in section 3, we either considered the 
$\frac{1}{16}$-BPS multiplets of $O_n$ or graviton multiplied by $O_n$. 
For $O_n$, $R_3+J_2=4+2n\geq 4$ so they cannot belong to these multiplets. 
Multiplying BPS gravitons to $O_n$ never decreases the value of this charge, 
since all $\frac{1}{16}$-BPS operators carry non-negative values of $R_3+J_2$. 
This implies that all the possible short multiplets discussed in section 3 should 
belong to $A_1\overline{L}$.\footnote{This does not mean that the other three short 
multiplets discussed in this paragraph are irrelevant to the black hole microstates. 
For instance, the $\frac{1}{8}$-BPS states in $A_1\overline{B}$ are believed to host 
exotic black hole like entropy growth, from the study of the Macdonald index
\cite{Choi:2018hmj}.}

With the relevance of $A_1\overline{L}$-type multiplets in our problems better understood, 
let us now explain their structures. 
Given an $A_1 \bar{L}$-type representation with primary $[j; \bar{j}]^{[r_1,r_2,r_3]}_E$,
we have seen above that the first $\frac{1}{16}$-BPS state in this representation is
$[\bar{j}]^{[r_2,r_3]}_{E + \frac12} \subset [j+1; \bar{j}]^{[r_1+1,r_2,r_3]}_{E + \frac12}$.
The $A_1 \bar{L}$-type representation is generic enough,
in some sense \emph{minimally} $\frac{1}{16}$-BPS,
so that all 9 supercharges $\in SU(1,2|3)$ can be applied without annihilating.
They are
\begin{eqnarray}\label{appQcommute}
Q^m_+ &=& [0]^{[1,0]}_{\frac12} ~~ \text{and} ~~ j=1~, \nonumber \\
\bar{Q}_{m \dot{\pm}} &=& [1]^{[0,1]}_{\frac12} ~~ \text{and} ~~ j=0~.
\end{eqnarray}
That is, there are {\bf 3} $Q$'s that raise $j$ by 1 and $E$ by $\frac12$,
and {\bf 2} $\times$ {\bf 3} $\bar{Q}$'s that only raise $E$ by $\frac12$.
Note that during the construction of the descendants, it is convenient 
to regard all $9$ supercharges as anti-commuting operators. Namely, despite some pairs 
yielding nontrivial anticommutators $\{Q^m_+,\overline{Q}_{n\dot\alpha}\}\sim 
\delta^m_n P_{+\dot\alpha}$, we dismiss the right-hand side for a moment since 
the conformal descendants will be supplemented later. Then each of the 9 supercharges 
can be applied only once. The states obtained this way forms the conformal primaries. 
Then, one can act arbitrary numbers of $P_{+\dot\alpha}\sim \partial_{+\dot\alpha}$ 
to construct the conformal descendants.

In terms of unrefined index ${\rm Tr}~t^{2E+j}$ that we often use,
the {\bf 3} $Q$'s each contribute by a factor of $-t^2$, 
the {\bf 2} $\times$ {\bf 3} $\bar{Q}$'s by $-t$,
and each of two derivatives by $t^3$.
Therefore, the $\frac{1}{16}$-BPS character of the representation
$A_1 \bar{L} [j; \bar{j}]^{[r_1,r_2,r_3]}_E$ is given by
\begin{eqnarray}\label{appALchar}
\chi_{BPS} \left[ A_1 \bar{L} [j; \bar{j}]^{[r_1,r_2,r_3]}_E \right] &=&
(-1)^{2E+1} d_{\bar{j}} d_{[r_2,r_3]} t^{2E+j+2} \cdot \frac{(1-t)^6 (1-t^2)^3}{(1-t^3)^2}~,
\end{eqnarray}
where $d_{\bar{j}} = \bar{j}+1$ and $d_{[r_2,r_3]} = \frac12 (r_2+1)(r_3+1)(r_2+r_3+2)$
multiply to dimension of the first $\frac{1}{16}$-BPS states $[\bar{j}]^{[r_2,r_3]}_{E + \frac12}$.
The refined version where $p$, $z_1$ and $z_2$ are used as fugacities for
$SU(2)_R \times SU(3) \subset PSU(1,2|3)$ (see below (\ref{su2-general-bh}) for convention),
is
\begin{eqnarray}\label{appALcharref}
\chi_{BPS} \left[ A_1 \bar{L} [j; \bar{j}]^{[r_1,r_2,r_3]}_E \right] &=&
(-1)^{2E+1} t^{2E+j+2} \cdot \chi_{(\bar{j}+1,[r_2,r_3])} \cdot \chi_{ D}~,
\end{eqnarray}
with the characters defined as (\ref{characters-t40}).

\section{Counting BMN gravitons for $SU(2)$}

In this section, we count graviton-type cohomologies in the BMN sector of the
$SU(2)$ theory. In terms of eigenvalues, BPS graviton polynomials are
arbitrary products of
\begin{eqnarray}\label{BMNsggrav}
  {\bf 6}&:& x^2\ ,\ y^2\ ,\ z^2\ ,\ xy\ ,\ yz\ ,\ zx\\
  {\bf 8}&:&\psi_1\cdot(y,z)\ ,\ \psi_2\cdot(z,x)\ ,\ \psi_3\cdot(x,y)\ ,\ 
  \psi_1 x-\psi_2 y\ ,\ \psi_2 y-\psi_3 z\nonumber\\
  {\bf 3}&:&xf-\textstyle{\frac{1}{2}}\psi_2\psi_3\ ,\ yf-\textstyle{\frac{1}{2}}\psi_3\psi_1\ ,\ zf-\textstyle{\frac{1}{2}}\psi_1\psi_2\ .
  \nonumber
\end{eqnarray}
and the goal of this section is to count independent polynomials among them.
$\psi_{1,2,3}$ are Grassmann variables while $x$, $y$, $z$, $f$ are bosonic. 

In the third line of (\ref{BMNsggrav}), $xf,yf,zf$ are accompanied by two-fermion terms,
but for the purpose of counting independent graviton polynomials,
these terms can be omitted, as we prove now.

Let $\mathfrak{V}$ be the infinite set of all possible products of $6+8=14$ polynomials
in the first two lines of (\ref{BMNsggrav}).
Define two series of vector spaces $V_k$ and $\tilde{V}_k$ as
\begin{eqnarray}
    V_k &=& {\rm span} \left\{ \mathfrak{v} \times (xf)^a(yf)^b(zf)^c
    ~|~ \mathfrak{v} \in \mathfrak{V}~,~ a+b+c \leq k \right\}~, \\
    \tilde{V}_k &=& {\rm span} \left\{ \mathfrak{v} \times
    (xf-\textstyle{\frac{1}{2}}\psi_2\psi_3)^a(yf-\textstyle{\frac{1}{2}}\psi_3\psi_1)^b
    (zf-\textstyle{\frac{1}{2}}\psi_1\psi_2)^c
    ~|~ \mathfrak{v} \in \mathfrak{V}~,~ a+b+c \leq k \right\}~. \nonumber
\end{eqnarray}
We want to show that rank of $V_\infty$ and rank of $\tilde{V}_\infty$ are equal.
We do this by induction.
Clearly ${\rm rank}(V_0) = {\rm rank}(\tilde{V}_0)$.
Now, suppose that ${\rm rank}(V_{k-1}) = {\rm rank}(\tilde{V}_{k-1})$
and let us show that ${\rm rank}(V_k) = {\rm rank}(\tilde{V}_k)$.
The equivalent statement is the following:
\begin{itemize}
\item
Consider a pair of polynomials
\begin{eqnarray}
    v &=& \sum_{i=1}^n r_i (xf)^{a_i} (yf)^{b_i} (zf)^{c_i}~, \nonumber \\
    \tilde{v} &=& \sum_{i=1}^n r_i (xf-\textstyle{\frac{1}{2}}\psi_2\psi_3)^{a_i}
(yf-\textstyle{\frac{1}{2}}\psi_3\psi_1)^{b_i}(zf-\textstyle{\frac{1}{2}}\psi_1\psi_2)^{c_i}~, \nonumber
\end{eqnarray}
where $r_i \in V_0 = \tilde{V}_0$ and $a_i + b_i + c_i = k$ for all $i$
so that $v \in V_k$ and $\tilde{v} \in \tilde{V}_k$.

Then $v \in V_{k-1}$ if and only if $\tilde{v} \in \tilde{V}_{k-1}$.
\end{itemize}

The $\leftarrow$ part is easy. If $\tilde{v} \in \tilde{V}_{k-1}$,
then $\tilde{v}$ equals a linear combination of polynomials that are at most of degree $k-1$
in $xf-\textstyle{\frac{1}{2}}\psi_2\psi_3$ and the likes.
Collecting terms with degree $k$ in $f$, the equality becomes $v=0 \in V_{k-1}$.

To show the $\rightarrow$ part, first note that $v \in V_{k-1}$ implies $v=0$,
since $v$ is homogeneous in $f$ with degree $k$. Now,
\begin{eqnarray}
    \tilde{v} ~=~ - \frac12 \sum_i^n r_i
    \Big[ & a_i \psi_2 \psi_3 (xf-\textstyle{\frac{1}{2}}\psi_2\psi_3)^{a_i-1}
    (yf-\textstyle{\frac{1}{2}}\psi_3\psi_1)^{b_i} (zf-\textstyle{\frac{1}{2}}\psi_1\psi_2)^{c_i}& \\
    & +b_i \psi_3 \psi_1 (xf-\textstyle{\frac{1}{2}}\psi_2\psi_3)^{a_i}
    (yf-\textstyle{\frac{1}{2}}\psi_3\psi_1)^{b_i-1} (zf-\textstyle{\frac{1}{2}}\psi_1\psi_2)^{c_i}& \nonumber \\
    & +c_i \psi_1 \psi_2 (xf-\textstyle{\frac{1}{2}}\psi_2\psi_3)^{a_i}
    (yf-\textstyle{\frac{1}{2}}\psi_3\psi_1)^{b_i} (zf-\textstyle{\frac{1}{2}}\psi_1\psi_2)^{c_i-1}& \Big]~. \nonumber
\end{eqnarray}
If $r_i \psi_j \psi_{j+1}$ for all $i$ and $j=1,2,3$ all belong to $V_0 = \tilde{V}_0$,
it will establish $\tilde{v} = \tilde{V}_{k-1}$.
Indeed, if $r_i$, which is a product of 14 polynomials in the first two lines of (\ref{BMNsggrav}),
contains any of the {\bf 6} in the first line, 
this factor can combine with two $\psi$'s and $r_i \psi_j \psi_{j+1} \in V_0$.
For example, $y^2 \psi_2\psi_3 = (\psi_2 y - \psi_3 z) (\psi_3 y)$.
On the other hand, if $r_i$ contains two or more factors of the {\bf 8} in the second line,
after multiplication by two $\psi$'s it will vanish due to Grassmannian nature of $\psi$,
so automatically $r_i \psi_j \psi_{j+1}= 0 \in V_0$.

Therefore the only possibility that remains in concern is when $r_i$ is precisely one of the {\bf 8}.
This leaves only a finite number of exceptions that one can explicitly work out.
That is, if $v = 0$ with the eight $r_i$ (they cannot mix with other $r_i$ due to homogeneity)
with appropriate numerical coefficients $\alpha_i$:
\begin{eqnarray}
r_1 = \alpha_1 \psi_1 y~,~ r_2 = \alpha_2 \psi_1 z~,~ \cdots
~,~ r_8 = \alpha_8 (\psi_2 y-\psi_3 z)~, \nonumber
\end{eqnarray}
it follows that $\tilde{v} = 0$ as well.
This completes the proof that ${\rm rank}(V_k) = {\rm rank}(\tilde{V}_k)$
given ${\rm rank}(V_{k-1}) = {\rm rank}(\tilde{V}_{k-1})$,
and by induction the number of independent products of (\ref{BMNsggrav})
is not affected by the $\psi \psi$ terms in the third line.

With this rule established, we now count the number of independent graviton polynomials in the BMN sector.
This task is greatly simplified by the fact that all $6+8+3 = 17$ but only two single-graviton generators
are monomials, because linear independence between monomials is rather transparent.
Our strategy will be to order the counting problem carefully 
so that we can work with the monomial basis as far as possible,
and treat the contribution from the two polynomial generators later.

Since there are 3 Grassmann variables, it is convenient to classify 
the graviton operators into $2^3 = 8$ sectors according to their Grassmannian contents.

\hspace*{-.65cm}\underline{\bf 0-fermion sector}

\hspace*{-.65cm}We first focus on the 0-fermion sector: graviton operators that do not contain 
any $\psi$'s.
It is clear that such operators are created by multiplying bosonic single-gravitons on 
the first and third lines of (\ref{BMNsggrav}).
Since all of them are monomials, we may simply write down a list of distinct monomials that can be
obtained by multiplying bosonic single-gravitons, then their linear independence is guaranteed.
The first six single-gravitons can be used to create any monomial $x^ay^bz^c$,
where $a,b,c$ are non-negative integers and $a+b+c$ is even.
Including $xf$, $yf$, $zf$, an eligible monomial may contain any number of $f$
as long as it is supported by at least as many $x$, $y$, or $z$.
Therefore, multi-gravitons in the 0-fermion sector are precisely described as
\begin{eqnarray}\label{BMNgrav0f}
G_0 &=& \{ x^ay^bz^cf^d ~|~ a,b,c,d \in \mathbb{Z}^{\geq 0}~,~a+b+c \geq d~,~ a+b+c+d = 0~(\text{mod } 2) \}~.
\end{eqnarray}

Because we can attribute to each of $x$, $y$, $z$ and $f$ a unit of their own quantum numbers,
the partition function for $G_0$ can be simply defined by the sum over monomials,
\begin{eqnarray}\label{BMNdefgf}
Z_0(x,y,z,f) &=& \sum_{g \in G_0} g~.
\end{eqnarray}
It can be computed as follows.
If there were no restrictions to $a,b,c,d$ except being non-negative integers,
the generating function would be $\frac{1}{(1-x)(1-y)(1-z)(1-f)}$.
From this, we subtract the sum of monomials for which $d > a+b+c$, which is
\begin{eqnarray}\label{BMNgrav0f1}
\frac{1}{(1-xf)(1-yf)(1-zf)} \cdot \frac{f}{1-f}~.
\end{eqnarray}
Then we project to the even part under $(x,y,z,f) \to (-x,-y,-z,-f)$, obtaining 
\begin{eqnarray}\label{BMNgrav0f2}
Z_0 &=& \left[\frac{1}{(1-x)(1-y)(1-z)(1-f)}- \frac{1}{(1-xf)(1-yf)(1-zf)} \cdot \frac{f}{1-f} 
\right]_{\rm even} \nonumber \\
&=& \left[ \frac{1-f(xy+yz+zx-xyz)+f^2xyz}{(1-x)(1-y)(1-z)(1-xf)(1-yf)(1-zf)} 
\right]_{\rm even} \nonumber \\
&=& \frac{1+\chi_2+f (\chi_3 - \chi_1 \chi_2) + f^2 (\chi_3^2+\chi_1 \chi_3)}{(1-x^2)(1-y^2)(1-z^2)(1-xf)(1-yf)(1-zf)}~.
\end{eqnarray}
Abbreviations for cyclic polynomials
\begin{eqnarray}\label{defchi}
\chi_1 = x+y+z~, \qquad
\chi_2 = xy+yz+zx~, \qquad
\chi_3 = xyz~,
\end{eqnarray}
will be used from now on.

\hspace*{-.65cm}\underline{\bf 1-fermion sector}

\hspace*{-.65cm}Now we list (independent) operators with one fermion, 
either $\psi_1$, $\psi_2$ or $\psi_3$. These are obtained by multiplying any operator in 
0-fermion sector $G_0$ by a generator on the second line of (\ref{BMNsggrav}). As mentioned
earlier, the last two of these may create non-monomial operators, so let us first
proceed without them. 

Operators with one $\psi_1$ can only be obtained by multiplying operators in $G_0$
by either $y\psi_1$ or $z\psi_1$. As a result, the list of such operators is simply the following monomials:
\begin{equation}\label{BMNgrav1f1}
\{ x^ay^bz^cf^d \psi_1 ~|~ a,b,c,d \in \mathbb{Z}^{\geq 0}~,~b+c \geq 1~,~a+b+c-1 \geq d~,~ a+b+c+d = 1~(\text{mod } 2) \}~.
\end{equation}
Operators containing one $\psi_2$ or one $\psi_3$ can be listed by 
cyclic permutations of letters.

Next, we ask what new operators arise when multiplying an operator in the 0-fermion sector 
$G_0$ by $x\psi_1-y\psi_2$.
If $x\psi_1-y\psi_2$ multiplies $x^ay^bz^cf^d \in G_0$ such that (i) $c \geq 1$ 
or (ii) $a \geq 1$ and $b \geq 1$,
both monomials $x^{a+1}y^bz^cf^d \psi_1$ and $x^ay^{b+1}z^cf^d \psi_2$ that appear in the product
are already counted in (\ref{BMNgrav1f1}) and corresponding $\psi_2$ sector respectively. 
So no new independent operators arise. Therefore, new operators that are obtained using
$x\psi_1-y\psi_2$ are classified as follows:
\begin{enumerate}
\item $(x^{a \geq 1}y^0z^0f^d)\cdot (x\psi_1-y\psi_2)$: In this case, the second monomial $x^ay^1z^0f^d \psi_2$
is already counted in $\psi_2$ sector corresponding to (\ref{BMNgrav1f1}),
while the first monomial is not counted in the $\psi_1$ sector.
Therefore, these can be regarded new monomials $x^{a+1}y^0z^0f^d \psi_1$ in $\psi_1$ sector.

\item $(x^0y^{b \geq 1}z^0f^d)\cdot (x\psi_1-y\psi_2)$: In this case, the first monomial $x^1y^bz^0f^d \psi_1$
is already counted in $\psi_1$ sector (\ref{BMNgrav1f1}),
while the second monomial is not counted in the $\psi_2$ sector.
Therefore, these can be regarded new monomials $x^0y^{b+1}z^0f^d \psi_2$ in $\psi_2$ sector.

\item $(1) \cdot (x\psi_1-y\psi_2)$: In this case, both monomials $x\psi_1$ and $y\psi_2$ have not been counted
in respective sectors. Therefore, this cannot be regarded as a new monomial in one of $\psi_1$ or $\psi_2$ sector.
Instead, this should be understood as an exceptional non-monomial operator. 
\end{enumerate}
Similar arguments can be made for multiplication by $y\psi_2 - z\psi_3$.

As a result, the list of monomials in $\psi_1$ sector is now extended to
\begin{equation}\label{BMNgrav1f2}
G_{\psi_1} = \{ x^ay^bz^cf^d \psi_1 ~|~ a,b,c,d \in \mathbb{Z}^{\geq 0}~,~a+b+c-1 \geq d~,~ a+b+c+d = 1~(\text{mod } 2) \}
\backslash \{x \psi_1 \}~.
\end{equation}
List of monomials $G_{\psi_2}$ in $\psi_2$ sector and $G_{\psi_3}$ in $\psi_3$ sector 
are defined by cyclicity. In addition, there are two exceptional operators 
$x\psi_1-y\psi_2$ and $y\psi_2 - z\psi_3$ that are not monomials 
and do not belong to any of $G_{\psi_m}$. So the whole set $G_1$ of 1-fermion 
BPS gravitons is given by
\begin{eqnarray}\label{BMNgrav1f3}
G_1 &=& G_{\psi_1} \cup G_{\psi_2} \cup G_{\psi_3} \cup \{ x\psi_1-y\psi_2~,~y\psi_2 - z\psi_3\}~.
\end{eqnarray}
Alternatively, one can take $G_{\psi_1}$ to \emph{not} exclude $x \psi_1$,
similarly $G_{\psi_2}$ and $G_{\psi_3}$ to \emph{not} exclude $y \psi_2$ and $z \psi_3$ respectively,
but instead exclude just  $x \psi_1 + y \psi_2 + z\psi_3$ at the end.

The existence of such non-monomial operators forbids us from attributing individual quantum numbers to $\psi$'s.
Instead, they carry a negative unit of respective scalar quantum numbers,
and a positive unit of overall $\psi$-number:
\begin{eqnarray}\label{BMNdefqn}
x \to [x]~, y \to [y]~, z \to [z]~, f \to [f]~, \psi_1 \to \frac{[\psi]}{[x]}~, \psi_2 \to \frac{[\psi]}{[y]}~, \psi_3 \to \frac{[\psi]}{[z]}~.
\end{eqnarray}
The partition function of the 1-fermion sector is given by a function of $x$, $y$, $z$, $f$ 
and $\psi$.
The partition function in $\psi_1$ sector (and of the rest of the 1-fermion sector) can be computed
analogously to the 0-fermion sector.
Starting from $\frac{1}{(1-x)(1-y)(1-z)(1-f)} \cdot \frac{\psi}{x}$, we implement the restriction $a+b+c-1 \geq d$
by subtracting its complement, extract the odd part under $(x,y,z,f) \to (-x,-y,-z,-f)$, and further subtract $x \psi_1 \to \psi$.
\begin{eqnarray}\label{BMNgrav1f4}
Z_{\psi_1} &=& \left[\frac{1}{(1-x)(1-y)(1-z)(1-f)}- \frac{1}{(1-xf)(1-yf)(1-zf)} \cdot \frac{1}{1-f} \right]_{\rm odd}
\cdot \frac{\psi}{x} - \psi \nonumber \\
&=& \left[ \frac{x+y+z - (xy+yz+zx)(1+f) + xyz(1+f+f^2)}{(1-x)(1-y)(1-z)(1-xf)(1-yf)(1-zf)} 
\right]_{\rm odd}
\cdot \frac{\psi}{x} - \psi \nonumber \\
&=& \frac{\chi_1 + \chi_3 -f (\chi_2 +\chi_2^2 - \chi_1 \chi_3 - \chi_3^2) + f^2 \chi_3 (1+ \chi_2)}
{(1-x^2)(1-y^2)(1-z^2)(1-xf)(1-yf)(1-zf)} \cdot \frac{\psi}{x} - \psi~.
\end{eqnarray}
Note that $Z_{\psi_2}$ and $Z_{\psi_3}$ can be computed similarly.
Further including $x\psi_1-y\psi_2~,~y\psi_2 - z\psi_3$, one obtains the following 
partition function for $G_1$:
\begin{eqnarray}\label{BMNgrav1f5}
Z_{1}  &=& \frac{\chi_1 + \chi_3 -f (\chi_2 +\chi_2^2 - \chi_1 \chi_3 - \chi_3^2) + f^2 \chi_3 (1+ \chi_2)}
{(1-x^2)(1-y^2)(1-z^2)(1-xf)(1-yf)(1-zf)} \cdot \frac{\chi_2}{\chi_3} \cdot \psi - \psi~.
\end{eqnarray}

\hspace*{-.65cm}\underline{\bf 2-fermion sector}

\hspace*{-.65cm}We consider operators that contain two of three $\psi$'s.
These are obtained by multiplying a generator on the second line of (\ref{BMNsggrav})
to an operator in $G_1$.
Focusing on the $\psi_1 \psi_2$ sector, we first note there are three ways to obtain an operator in this sector.
\begin{enumerate}
\item Multiply either $x\psi_2$ or $z\psi_2$ to an operator in $G_{\psi_1}$ (\ref{BMNgrav1f2}).
Such a set of operators are
\begin{equation}\label{BMNgrav2f1}
\{ x^ay^bz^cf^d \psi_1\psi_2 ~|~ a,b,c,d \in \mathbb{Z}^{\geq 0},~a+c \geq 1,~a+b+c-2 \geq 
d,~ a+b+c+d = 0~(\text{mod } 2) \}
\backslash \{x^2 \psi_1 \psi_2 \}~.
\end{equation}

\item Multiply either $y\psi_1$ or $z\psi_1$ to an operator in $G_{\psi_2}$, analogous to (\ref{BMNgrav1f2}):
\begin{equation}\label{BMNgrav2f2}
\{ x^ay^bz^cf^d \psi_1\psi_2 ~|~ a,b,c,d \in \mathbb{Z}^{\geq 0},~b+c \geq 1,~a+b+c-2 \geq 
d,~ a+b+c+d = 0~(\text{mod } 2) \}
\backslash \{y^2 \psi_1 \psi_2 \}~.
\end{equation}

\item Multiply $x\psi_2$, $z\psi_2$, $y\psi_1$ or $z\psi_1$ to $x\psi_1 - y\psi_2$.
These supplement $x^2 \psi_1 \psi_2$ and $y^2 \psi_1 \psi_2$ excluded in (\ref{BMNgrav2f1}) and (\ref{BMNgrav2f2}).
\end{enumerate}
Taking the union of the three sets above, we arrive at
\begin{equation}\label{BMNgrav2f3}
G_{\psi_1\psi_2} = \{ x^ay^bz^cf^d \psi_1 \psi_2 ~|~ a,b,c,d \in \mathbb{Z}^{\geq 0}~,~a+b+c-2 \geq d~,~
a+b+c+d = 0~(\text{mod } 2) \}~,
\end{equation}
and similarly for $\psi_2\psi_3$ and $\psi_3\psi_1$ sectors.

Note that we have not explicitly considered multiplying, for example, 
$x\psi_3$ or $y\psi_3$ to $x\psi_1-y\psi_2$. Both monomials obtained this way are already
included in $G_{\psi_3\psi_1}$ and $G_{\psi_2 \psi_3}$,
so they do not add any new independent operators.
Furthermore, there is a possibility of multiplying $x\psi_1-y\psi_2$ or $y\psi_2-z\psi_3$ to the operators in
the 1-fermion sector.
These may give rise to
\begin{eqnarray}
(x\psi_1-y\psi_2)(x\psi_1-y\psi_2) &\sim& xy \psi_1 \psi_2~,\nonumber \\
(y\psi_2-z\psi_3)(y\psi_2-z\psi_3) &\sim& yz \psi_2 \psi_3~, \nonumber \\
(x\psi_1-y\psi_2)(y\psi_2-z\psi_3) &\sim& xy \psi_1 \psi_2 + yz \psi_2 \psi_3 + zx \psi_3 \psi_1~,
\end{eqnarray}
but again, all of the monomials are already counted in respective 2-fermion sectors.
Therefore, we conclude that the 2-fermion sectors can be written completely in monomial basis,
by (\ref{BMNgrav2f3}) and its cyclic versions:
\begin{eqnarray}\label{BMNgrav2f4}
G_2 &=& G_{\psi_1\psi_2} \cup G_{\psi_2\psi_3} \cup G_{\psi_3\psi_1}~.
\end{eqnarray}

The partition function of 2-fermion sector can be computed as before.
The result is:
\begin{eqnarray}\label{BMNgrav2f5}
Z_{\psi_1\psi_2} &=& \frac{\chi_1^2-\chi_2 -\chi_2^2 +2 \chi_1 \chi_3+\chi_3^2 +f(\chi_3 -\chi_1\chi_2)
+f^2 \chi_3 (\chi_1+\chi_3)}{(1-x^2)(1-y^2)(1-z^2)(1-xf)(1-yf)(1-zf)} \cdot \frac{\psi^2}{xy}~,
\end{eqnarray}
for the individual sector, and
\begin{eqnarray}\label{BMNgrav2f6}
Z_{2} &=& Z_{\psi_1\psi_2} +Z_{\psi_2\psi_3} +Z_{\psi_3\psi_1} \nonumber \\
&=& \frac{\chi_1^2-\chi_2 -\chi_2^2 +2 \chi_1 \chi_3+\chi_3^2 +f(\chi_3 -\chi_1\chi_2)
+f^2 \chi_3 (\chi_1+\chi_3)}{(1-x^2)(1-y^2)(1-z^2)(1-xf)(1-yf)(1-zf)} \cdot \frac{\chi_1}{\chi_3} \cdot \psi^2~,
\end{eqnarray}
for the entire 2-fermion sector.

\hspace*{-.65cm}\underline{\bf 3-fermion sector}

\hspace*{-.65cm}We finally investigate the 3-fermion sector, i.e. operators 
that contain all $\psi_1$, $\psi_2$ and $\psi_3$.
One way to obtain 3-fermion operators is to multiply $x\psi_3$ or $y\psi_3$ to the $\psi_1\psi_2$-sector
(\ref{BMNgrav2f3}). Set of such operators is
\begin{equation}\label{BMNgrav3f1}
\{ x^ay^bz^cf^d \psi_1 \psi_2 \psi_3 ~|~ a,b,c,d \in \mathbb{Z}^{\geq 0}~,~a+b \geq 1~,~a+b+c-3 \geq d~,~
a+b+c+d = 1~(\text{mod } 2) \}~.
\end{equation}
By cyclicity, there are two more sets of 3-fermion operators that are obtained by $x \to y \to z \to x$ from
(\ref{BMNgrav3f1}). Their union is,
\begin{equation}\label{BMNgrav3f2}
G_{\psi_1 \psi_2 \psi_3} = \{ x^ay^bz^cf^d \psi_1 \psi_2 \psi_3 ~|~ a,b,c,d \in \mathbb{Z}^{\geq 0}~,~a+b+c-3 \geq d~,~
a+b+c+d = 1~(\text{mod } 2) \}~.
\end{equation}
One can easily check that multiplying non-monomial blocks $x\psi_1-y\psi_2$ or $y\psi_2-z\psi_3$
to 2-fermion sector does not produce any new operator.

Partition function of the 3-fermion sector (\ref{BMNgrav3f1}) is
\begin{equation}\hspace{-0.5cm}\label{BMNgrav3f3}
Z_{3} = \left[ \frac{-1+\!\chi_1^2-\!2\chi_2-\!\chi_2^2+\!2\chi_1\chi_3+\!\chi_3^2+\!f(\chi_1+\!\chi_3)
-\!f^2 (\chi_2+\!\chi_2^2-\!\chi_1\chi_3-\!\chi_3^2)+\!f^3 \chi_3 
(1+\!\chi_2)}{(1-x^2)(1-y^2)(1-z^2)(1-xf)(1-yf)(1-zf)} +\!1\right]\frac{\psi^3}{f \chi_3}~.
\end{equation}

\hspace*{-.65cm}\underline{\bf The index}

\hspace*{-.65cm}The complete list of BPS multi-graviton operators in BMN sector of 
the $SU(2)$ theory is given by (\ref{BMNgrav0f}), (\ref{BMNgrav1f3}), (\ref{BMNgrav2f4}) 
and (\ref{BMNgrav3f2}). Corresponding partition function is $Z_0 + Z_1 + Z_2 + Z_3$,
each of which is presented in (\ref{BMNgrav0f2}), (\ref{BMNgrav1f5}), (\ref{BMNgrav2f6}) and (\ref{BMNgrav3f3}). Attributing minus sign to the fermion number $\psi$ in the partition function and further setting $\psi,f\rightarrow xyz$ will yield the index, 
where $(x,y,z)=(e^{-\Delta_1},e^{-\Delta_2},e^{-\Delta_3})$.

To facilitate comparison with the other parts of this paper, 
we compute the unrefined index of the graviton partition function.
This is obtained simply by substituting
\begin{eqnarray}\label{unrindex}
x,y,z \to t^2~, \qquad f \to t^6 ~, \qquad \psi \to -t^6~.
\end{eqnarray}
in to the partition function. The result is
\begin{equation}\label{BMNgravindex}
\hspace*{-.3cm}Z_{\rm grav} = \frac{1\!+\!3t^4\!-\!8t^6\!-\!6t^{10}\!+\!10t^{12}\!+
\!9t^{14}\!-\!9t^{16}\!+\!16t^{18}\!
-\!18t^{20}\!-\!3t^{22}\!+\!t^{24}\!-\!3t^{26}\!+\!9t^{28}\!-\!2t^{30}\!+\!3t^{32}\!-\!3t^{34}}
{(1-t^4)^3 (1-t^8)^3}~.
\end{equation}
The difference $Z - Z_{\rm grav}$ will be the index that counts non-graviton operators.
We find a simple analytic formula for the difference:
\begin{eqnarray}\label{BMNindexdiff}
Z - Z_{\rm grav} &=& - \frac{t^{24}}{1-t^{12}} \cdot \frac{(1-t^2)^3}{(1-t^8)^3}~.
\end{eqnarray}
The fully refined index is given by (\ref{bmn-bh}).

\section{$Q$-exactness of some graviton hairs}

In this appendix, we report the $Q$-exactness of some product operators.

Six operators $O_0(\bar\phi^{(m}\cdot \bar\phi^{n)})$ at $t^{28}$ order are 
all $Q$-exact. An $SU(3)$ covariant expression is 
\begin{equation}
    \begin{aligned}
  O_0(\bar\phi^{(m}\cdot \bar\phi^{n)}) = -\frac{1}{14} Q [&20 \epsilon^{rs(m} (\bar\phi^{n)} \cdot \psi_{p+}) (\bar\phi^p \cdot \psi_{r+}) (\bar\phi^q \cdot \psi_{q+})(f_{++}\cdot \psi_{s+})   \\
  -& 20\epsilon^{prs} (\bar\phi^{(m} \cdot \psi_{p+}) (\bar\phi^{n)} \cdot \psi_{r+}) (\bar\phi^q \cdot \psi_{q+}) (f_{++}\cdot \psi_{s+}) \\
  +&30 \epsilon^{prs} (\bar\phi^{(m} \cdot \psi_{p+}) (\bar\phi^{n)} \cdot \psi_{r+})(\bar\phi^{q} \cdot \psi_{s+}) (f_{++}\cdot \psi_{q+})  \\
  -& 7 \epsilon^{a_1 a_2p} \epsilon^{b_1 b_2(m}  (\bar\phi^{n)} \cdot \psi_{p+}) (\bar\phi^q \cdot \psi_{q+}) (\psi_{a_1+} \cdot \psi_{a_2+}) (\psi_{b_1+} \cdot \psi_{b_2+}) \\
  +& 18 \epsilon^{a_1 a_2p} \epsilon^{b_1 b_2(m} (\bar\phi^{n)} \cdot \psi_{q+}) (\bar\phi^q \cdot \psi_{p+}) (\psi_{a_1+} \cdot \psi_{a_2+}) (\psi_{b_1+} \cdot \psi_{b_2+}) ]\ .
  \end{aligned}
\end{equation}
Six operators $O_0(\bar\phi^m\cdot\bar\lambda_{\dot\alpha})$ at $t^{29}$ order are also all $Q$-exact. 
An $SU(2)_R \times SU(3)$ covariant expression is  
\begin{equation}
    \begin{aligned}
        O_0(\bar\phi^m\cdot\bar\lambda_{\dot\alpha}) 
        = -\frac{1}{8}Q [& 40 \epsilon^{m np}  (f_{++} \cdot \psi_{q+}) (\bar\lambda_{\dot\alpha} \cdot \psi_{r+})(\bar\phi^{q} \cdot \psi_{n+}) (\bar\phi^{r} \cdot \psi_{p+})\\
        -&4 \epsilon^{ma_1a_2} \epsilon^{nb_1b_2} (\bar\lambda_{\dot\alpha} \cdot \psi_{n+}) (\bar\phi^p \cdot \psi_{p+}) (\psi_{a_1+} \cdot \psi_{a_2+}) (\psi_{b_1+} \cdot \psi_{b_2+})\\
        +&6 \epsilon^{ma_1a_2} \epsilon^{nb_1b_2} (\bar\lambda_{\dot\alpha} \cdot \psi_{p+}) (\bar\phi^p \cdot \psi_{n+}) (\psi_{a_1+} \cdot \psi_{a_2+}) (\psi_{b_1+} \cdot \psi_{b_2+})\\
        +&\epsilon^{na_1a_2} \epsilon^{pb_1b_2} (\bar\lambda_{\dot\alpha} \cdot \psi_{n+}) (\bar\phi^m \cdot \psi_{p+}) (\psi_{a_1+} \cdot \psi_{a_2+}) (\psi_{b_1+} \cdot \psi_{b_2+}) ]\ .
    \end{aligned}
\end{equation}
Eight operators $O_0\left(\bar\phi^m\cdot\psi_{n+}-{\textstyle \frac{1}{3}}\delta^m_n\bar\phi^p\cdot\psi_{p+}\right)$ at $t^{30}$ order are all $Q$-exact. An $SU(3)$ covariant expression is 
\begin{eqnarray}
  &&O_0 \left(\bar\phi^m\cdot\psi_{n+}-{\textstyle \frac{1}{3}}\delta^m_n\bar\phi^p\cdot\psi_{p+}\right)\\ &&=-\frac{1}{4} Q\left[\epsilon_{npq}\epsilon^{ra_1a_2} \epsilon^{qb_1b_2} \epsilon^{mc_1c_2}(\bar\phi^{p} \cdot \psi_{r+})  (\psi_{a_1+} \cdot \psi_{a_2+})(\psi_{b_1+} \cdot \psi_{b_2+}) (\psi_{c_1+} \cdot \psi_{c_2+}) \right]\ .\nonumber
\end{eqnarray}

\end{document}